\documentclass[reprint,prb,superscriptaddress]{revtex4-2}
\usepackage{braket,amsmath,amssymb,graphicx,float,hyperref,color,ulem,soul}
\allowdisplaybreaks
\begin{document}

\title{Unveiling the Kondo cloud: unitary RG study of the Kondo model}
\author{Anirban Mukherjee}
\email{mukherjee.anirban.anirban@gmail.com }
\affiliation{Department of Physical Sciences, Indian Institute of Science Education and Research-Kolkata, W.B. 741246, India}
\author{Abhirup Mukherjee}
\email{am18ip014@iiserkol.ac.in }
\affiliation{Department of Physical Sciences, Indian Institute of Science Education and Research-Kolkata, W.B. 741246, India}
\author{N. S. Vidhyadhiraja}
\email{raja@jncasr.ac.in}
\affiliation{Theoretical Sciences Unit, Jawaharlal Nehru Center for Advanced Scientific Research, Jakkur, Bengaluru 560064, India}
\author{A. Taraphder}
\email{arghya@phy.iitkgp.ernet.in}
\affiliation{Department of Physics, Indian Institute of Technology Kharagpur, Kharagpur 721302, India}
\author{Siddhartha Lal}
\email{slal@iiserkol.ac.in}
\affiliation{Department of Physical Sciences, Indian Institute of Science Education and Research-Kolkata, W.B. 741246, India}
\date{\today}
\begin{abstract}
We analyze the single-channel Kondo model using the recently developed unitary renormalisation group (URG) method, and obtain a comprehensive understanding of the Kondo screening cloud.
The fixed-point low-energy Hamiltonian enables the computation of a plethora of thermodynamic quantities (specific heat, susceptibility, Wilson ratio, etc.) as well as spectral functions, all of which are found to be in good agreement with known results. 
By integrating out the impurity, we obtain an effective Hamiltonian for the excitations of the electrons comprising the Kondo cloud. 
This is found to contain both \(k-\)space number-diagonal (Fermi liquid) as well off-diagonal four-fermion scattering terms. 
Our conclusions are reinforced by a URG study of the two-particle entanglement 
and many-body correlations 
among members of the Kondo cloud and impurity. 
The entanglement between the impurity and a cloud electron, as well as between any two cloud electrons, is found to increase under flow towards the singlet ground state at the strong-coupling fixed point. 
Both the number-diagonal and off-diagonal correlations within the conduction cloud are also found to increase as the impurity is screened under the flow, and the latter are found to be responsible for the macroscopic entanglement of the Kondo-singlet ground state.
The unitary RG flow enables an analytic computation of the  phase shifts suffered by the conduction electrons at the strong-coupling fixed point.
This reveals an orthogonality catastrophe between the local moment and strong-coupling ground states, and is related to a change in the Luttinger volume of the conduction bath under the crossover to strong coupling. Our results offer fresh insight on the nature of the emergent many-particle entanglement within the Kondo cloud, and pave the way for further investigations in more exotic contexts such as the fixed point of the over-screened multi-channel Kondo problem.
\end{abstract}
\maketitle
\section{Introduction}
In this work, we present a unitary renormalisation group (URG) analysis of the single channel Kondo model. 
The model involves a quantum impurity interacting with a conduction bath. It can be derived from the particle-hole symmetric single-impurity Anderson model using a Schrieffer-Wolff transformation that integrates out single-particle interactions \cite{schrieffer1966}. The model was initially developed to study metals with magnetic impurities, and was used by Jun Kondo~\cite{kondo1964resistance,Zhang2013} to explain the resistivity minimum that appears in these materials at low temperatures. 
Since then, various methods have been applied to determine the low energy behaviour of this model. 
Notable attempts include the Coulomb gas approach by Anderson and Yuval ~\cite{anderson1969exact,anderson1970exact} and perturbative renormalisation group approach (poor man's scaling) by Anderson ~\cite{anderson1970poor}. 
The latter method showed that the Kondo exchange coupling \(J\) flows to larger values as we go to lower energy scales, at least for small \(J\). 
The full crossover from the local moment phase at small \(J\) to the screened moment strong-coupling phase at large \(J\) was later obtained by the numerical renormalisation group (NRG) technique developed by Wilson \cite{wilson1975,bullaNRGreview}, the Bethe ansatz solution by Andrei and Wiegmann \cite{andreiKondoreview,tsvelickKondoreview} and the conformal field theory (CFT) approach \cite{affleck1995conformal,affleck1993exact}. 
Another important strong-coupling approach based on arguments of scattering phase shifts is that of the local Fermi liquid theory~\cite{nozieres1974fermi,nozaki2012} by Nozières. The low-temperature properties were found to be universal functions of a single energy scale, the Kondo temperature \(T_\text{K}\)~\cite{wilson1975,krishnamurthy-physrevlett.64.950,andreas_markus_2014,tadashi_1985,gass_kang_2011}.
All the predicted aspects of the Kondo effect, including the existence  of the Kondo cloud \cite{sorensen_erik_affleck_1996,affleck_ian_2001,simon_pascal_2003,martin2010,martin2019}, were observed experimentally in quantum dot systems~\cite{Goldhaber-Gordon1998,Cronenwett1998,Schmid_Weis1998,pustilnik_glazman_2004,Borzenets2020}. Scanning tunneling spectroscopy (STS) measurements have revealed that the Kondo effect often depends on the neighborhood of the impurities, in \(Cu\) and \(Co\) atoms~\cite{neel_berndt_2008,Zhao2005}.
It was also shown \cite{kaminski_nazarov2000}, using quantum dots, that the out of equilibrium Kondo effect also displays universality, the physics at low temperatures being decided by only two energy scales, the frequency and amplitude of the perturbation. 
The impurity spectral function has been calculated using NRG\cite{hrk_wilson_1980}, both at \(T=0\) \cite{sakai_osamu_shimizu,costi_hewson_1990} as well as \(T>0\) \cite{costi_kroha_wolfle}, as well as using diagrammatic methods \cite{kroha_wolfle}. 
The electrical resistivity was found to obey single-parameter scaling behavior in \(T/T_\mathrm{K}\) \cite{costi_hewson_1992}. 
Nozières went further and analyzed a more realistic model, the multi-channel Kondo problem in which multiple conduction bath channels interact with quantum impurity at the center \cite{Noz_blandin_1980}. 
Such a model was found, through methods like the Bethe ansatz, CFT and bosonization among others, to host a non-Fermi liquid low energy phase~\cite{Gan_Andrei_Coleman_1993,Noz_blandin_1980,emery_kivelson,Gan_mchannel_1994,Tsvelick_Weigmann_mchannel_1984,Tsvelick_weigmann_mchannel_1985,parcollet_olivier_large_N,kimura_taro_Su_N_kondo,PhysRevB.73.224445,cox_jarrell_two_channel_rev,affleck_1991_overscreen,Coleman_tsvelik}.
Kondo effect also occurs in light quark matter which interact with heavy quark impurities through gluon-exchange interactions; the scattering amplitude goes through a similar logarithmic divergence and renormalisation group calculations reveal a Kondo scale in such systems~\cite{Yasui_2013,hattori_2015}. Kondo effect can also be realized for other fermionic systems like graphene \cite{fritz_vojta_2013}, Dirac/Weyl semi-metals \cite{principi_2015,mitchell_lars_2015} and dense nuclear matter \cite{yasui_kazutaka_2017,yasui_shigehiro_2016,Yasui_2013}.

Obtaining the form of the effective Hamiltonian governing the low-energy physics of the Kondo cloud and the many-body singlet wavefunction remained a challenge. In this work, we have analyzed the Kondo model on a tight-binding lattice using the recently developed URG method \cite{anirbanurg1,anirbanurg2}. 
The method has already been applied to a variety of problems such as the 2D Hubbard model \cite{anirbanmott1,anirbanmott2}, the quantum kagome antiferromagnet \cite{santanukagome}, a generalized model of electrons in two spatial dimensions with attractive \(U(1)\)-symmetric interactions \cite{siddharthacpi}, the 1D Hubbard model \cite{1dhubjhep}, as well as other generalized models of electrons with repulsive interactions with and without translation invariance~\cite{anirbanurg2}. 
The method involves applying unitary transformations (see eq.~\ref{urg_map}) on the Hamiltonian that decouple high energy electrons, creating a family of unitarily-connected Hamiltonians at progressively smaller energy scales. 
The decoupled electrons lose their quantum fluctuations in the process, becoming integrals of motion (IOMs). 
This program leads to the one of the main results of this work - a unitarily-transformed strong-coupling fixed-point Hamiltonian. 
It is from this fixed-point Hamiltonian that we obtain the effective Hamiltonian for the electrons that comprise the Kondo cloud, by integrating out the impurity. 
This effective Hamiltonian is found to contain both Fermi liquid (diagonal) and off-diagonal four-fermion interaction terms, shedding new light on the nature of the Kondo cloud the interactions within.
It is seen that the flow towards the strong-coupling fixed point is concomitant with the growth of these off-diagonal terms, and this picture is made clearer by the evolution of the spectral function calculated using the effective Hamiltonians. 
For the purposes of benchmarking, we have also computed several thermodynamic quantities (e.g., the impurity susceptibility, specific heat, Wilson ratio etc.) from this effective theory. 
These are described in the appendices, and are found to be in good agreement with the results known from the literature.

Another important result of this work involves studies of the evolution of the macroscopic entanglement of the Kondo cloud with the impurity spin and the many-body correlations inside the Kondo cloud under renormalisation towards strong-coupling. Experimental studies of entanglement and correlations have been performed in double quantum dots~\cite{yoo_gwangshu_2018}.
For the purpose of studying the entanglement and many-body correlations, we employ the entanglement RG method developed recently by some of us in Refs.\cite{1dhubjhep,siddharthacpi,mukherjee2020}. 
These calculations are enabled by the fact that the RG transformations are unitary and preserve the total spectral weight. 
Thus, iterative applications of the inverse of these unitary transformations on the IR fixed-point singlet wavefunction generates a family of wavefunctions under a reversal of the RG flow towards the UV. 
The entanglement and correlations measures are then calculated from this family of wavefunctions, and represents their RG evolution. 
The correlations (diagonal and off-diagonal) as well as the mutual information (among members of the cloud as well as between the impurity and the cloud) are found to increase towards the strong-coupling fixed point, and this is consistent with the presence of the off-diagonal scattering term which generates the correlation among the members of the cloud. 
Our quantification of the many-particle entanglement through the mutual information of the Kondo screening cloud and the impurity and the study of its renormalisation is complementary to methods that have been used previously in the literature. 
This includes the impurity contribution to the entanglement of a region in the conduction bath (dubbed as the impurity entanglement entropy in Refs.\cite{sorensen2007,eriksson_2011,Lee_park_2015}), the entanglement entropy obtained by tracing out the impurity~\cite{sorensen1996,barzykin1998,barzykin1999}, and the concurrence between the impurity with the rest of the conduction bath ~\cite{yangfeigun2017}. 
Recent works have also shown, using negativity as a measure of entanglement \cite{Vidal_2002}, that the impurity is maximally entangled with the cloud \cite{bayat_2010}.

The rest of the paper is outlined as follows. 
In Sec.\ref{KondoModel}, we introduce the Hamiltonian for the single channel Kondo model. 
We perform the URG analysis on the Hamiltonian in Sec.\ref{URGkondo}. 
Sec.\ref{fixedPointKondo} constitutes results on the scaling of the Kondo coupling, description of the low-energy stable fixed point in terms of the effective Hamiltonian and ground state wavefunction, and the redistribution of spectral weight under the RG flow. 
In Sec.\ref{susc2}, we obtain the effective Hamiltonian for the Kondo cloud excitations and describe the salient features. 
In Sec.\ref{ent2}, we study the entanglement features and many-body correlations of the Kondo cloud. 
We end in Sec.\ref{ConcluKondo} with a discussion of the results and some avenues for future investigations. 
Appendix \ref{urg-deriv} shows the derivation of the RG equation using the URG method. 
In appendices \ref{lfl-kc} and \ref{phase_shift}, we obtain a real-space effective Hamiltonian for the low-energy excitations of the strong-coupling fixed point Hamiltonian and calculate the conduction electron scattering phase shifts close to the ground state. 
Appendices \ref{chi} and \ref{Cv} focus on calculating some thermodynamic quantities (impurity susceptibility, impurity specific heat and impurity thermal entropy) in order to benchmark with existing results. 
\section{The Model}\label{KondoModel}
 The Kondo model~\cite{kondo1964resistance,anderson1970poor} describes the coupling between a magnetic quantum impurity localized in real space with a bath of conduction electrons
\begin{eqnarray}
\label{KondoH}
\hat{H} = \sum_{\mathbf{k}\sigma}\epsilon_{\mathbf{k}}\hat{n}_{\mathbf{k}\sigma}+\frac{J}{2}\sum_{\mathbf{k},\mathbf{k}'}\mathbf{S}\cdot c^{\dagger}_{\mathbf{k}\alpha}\boldsymbol{\sigma}_{\alpha\beta}c_{\mathbf{k}'\beta}~.
\end{eqnarray}

We will consider a 2D electronic bath with a dispersion \(\epsilon_{\mathbf{k}}=-2t(\cos k_{x}+\cos k_{y})$ and with the Fermi energy set to $E_{F}=\mu\). Here 
\(J\) is the Kondo scattering coupling between the impurity and the conduction electrons. 
An important feature of the Kondo coupling is the two different classes of scattering processes: one involving spin-flip scattering processes for the bath electrons (\(c^{\dagger}_{\mathbf{k}\uparrow}c_{\mathbf{k}'\downarrow}+h.c.\)), and the other, non spin-flip (i.e., simple potential scattering processes).
In the antiferromagnetic regime \(J>0\), the spin-flip scattering processes generate quantum entanglement between the impurity spin and a macroscopic number of bath electrons (called the ``Kondo cloud"), resulting in the complete screening of the impurity via the formation of a singlet spin state. 
\begin{figure}
	\centering
	\includegraphics[width=0.3\textwidth]{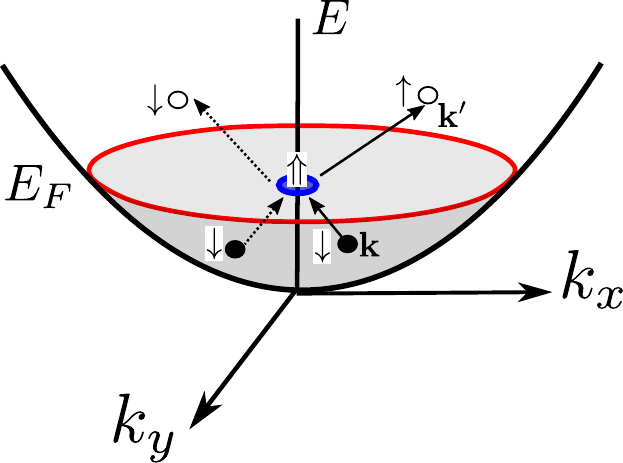}
	\caption{The Kondo model is composed of a two-dimensional conduction electron bath (Fermi liquid) coupled to a magnetic impurity via a spin-flip (solid) / non spin-flip (dashed) scattering coupling.}
\end{figure}
It is clear that the dynamical Kondo cloud corresponds to an effective single spin-1/2, such that the screening is an example of macroscopic quantum entanglement arising from electronic correlations. It is the nature of this entanglement, and the underlying quantum liquid that forms the Kondo cloud, that we seek to learn more of.
\section{URG Theory for the Kondo model}\label{URGkondo}
In constructing an effective low-energy theory for the Kondo singlet, we employ the unitary RG formalism \cite{anirbanurg1,anirbanurg2,anirbanmott1,anirbanmott2,santanukagome,siddharthacpi,1dhubjhep} to the Kondo model such that electronic states from the non-interacting conduction bath are step-wise disentangled, starting with the highest energy electrons at the bandwidth and eventually scaling towards the FS. While this aspect is similar to Anderson's poor man's scaling~\cite{anderson1970poor}, we shall see that several other aspects of the unitary RG formalism are different from those adopted in the poor man's scaling approach.

\begin{figure}
\centering
\includegraphics[width=0.25\textwidth]{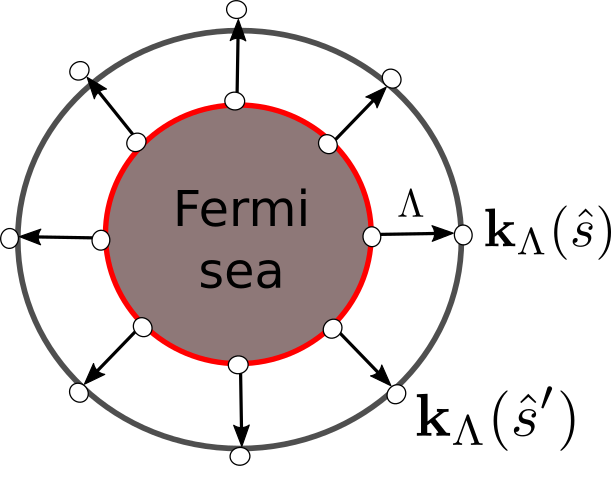}
\caption{Fermi surface geometry for a circular Fermi volume of non-interacting electrons in 2 spatial dimensions.}
\label{FSgeom}
\end{figure}

The electronic states are labeled in terms of the normal distance $\Lambda$ from the FS and the orientation unit vectors (Fig.\ref{FSgeom}) $\hat{s}$: $\mathbf{k}_{\Lambda\hat{s}}=\mathbf{k}_{F}(\hat{s})+\Lambda\hat{s}$, where $\hat{s}=\frac{\nabla\epsilon_{\mathbf{k}}}{|\nabla\epsilon_{\mathbf{k}}|}|_{\epsilon_{\mathbf{k}}=E_{F}}$. The various \(\hat s\) represent the normal directions of the Fermi surface. The states are labeled as $|j,l,\sigma\rangle = |\mathbf{k}_{\Lambda_{j}\hat{s}},\sigma\rangle, l:=(\hat{s}_{m},\sigma)$. The $\Lambda$'s are arranged as follows: $\Lambda_{N}>\Lambda_{N-1}>\ldots>0$, where the electronic states farthest from FS $\Lambda_{N}$ are disentangled first, eventually scaling towards the FS. This leads to the Hamiltonian flow equation~\cite{anirbanurg1}
\begin{equation}
\centering
\label{urg_map}
H_{(j-1)}=U_{(j)}H_{(j)}U^{\dagger}_{(j)}~,
\end{equation}
where the unitary operation $U_{(j)}$ is the unitary map at RG step $j$. 
$U_{(j)}$ disentangles all the electronic states 
$|\mathbf{k}_{\Lambda_{j}\hat{s}_{m}},\sigma\rangle$
on the isogeometric curve and has the form~\cite{anirbanmott1,anirbanurg1}
\begin{equation}
\centering U_{(j)}=\prod_{l}U_{j,l}, U_{j,l}=\frac{1}{\sqrt{2}}[1+\eta_{j,l}-\eta^{\dagger}_{j,l}]~,
\end{equation}
where $\eta_{j,l}$ are electron-hole transition operators following the algebra
\begin{equation}
\lbrace\eta_{j,l},\eta_{j,l}^{\dagger}\rbrace=1~,~\left[\eta_{j,l},\eta_{j,l}^{\dagger}\right]=1~.
\end{equation}
The transition operator can be represented in terms of the diagonal ($H^{D}$) and off-diagonal ($H^{X}$) parts of the Hamiltonian as follows 
\begin{eqnarray}
\eta_{j,l}&=&Tr_{j,l}(c^{\dagger}_{j,l}H_{j,l})c_{j,l}\frac{1}{\hat{\omega}_{j,l}-Tr_{j,l}(H_{j,l}^{D}\hat{n}_{j,l})\hat{n}_{j,l}}~.~~\label{e-TransOp}
\end{eqnarray}
We note that in the numerator of the expression for $\eta_{j,l}$, the operator $Tr_{j,l}(c^{\dagger}_{j,l}H_{j,l})c_{j,l}+h.c.$ is composed of all possible scattering vertices that modify the configuration of the electronic state $|j,l\rangle$~\cite{anirbanurg1}. The generic forms of $H^{D}_{j,l}$ and $H^{X}_{j,l}$ are as follows
\begin{equation}\begin{aligned}
H^{D}_{j,l}=&\sum_{\Lambda\hat{s},\sigma}\epsilon^{j,l}\hat{n}_{\mathbf{k}_{\Lambda\hat{s}},\sigma}+\sum_{\alpha}\Gamma_{\alpha}^{4,(j,l)}\hat{n}_{\mathbf{k}\sigma}\hat{n}_{\mathbf{k}'\sigma'}\\
	   &+\sum_{\beta}\Gamma_{\beta}^{8,(j,l)}\hat{n}_{\mathbf{k}\sigma}\hat{n}_{\mathbf{k}'\sigma'}(1-\hat{n}_{\mathbf{k}''\sigma''})+\ldots~,\\
H^{X}_{j,l}=&\sum_{\alpha}\Gamma_{\alpha}^{2}c^{\dagger}_{\mathbf{k}\sigma}c_{\mathbf{k}'\sigma'}+\sum_{\beta}\Gamma_{\beta}^{2}c^{\dagger}_{\mathbf{k}\sigma}c^{\dagger}_{\mathbf{k}'\sigma'}c_{\mathbf{k}_{1}'\sigma_{1}'}c_{\mathbf{k}_{1}\sigma_{1}}+\ldots~.
\end{aligned}\end{equation}
The indices \(\alpha\) and \(\beta\) are strings that denote the quantum numbers of the incoming and outgoing electronic states at a particular interaction vertex \(\Gamma^n_{\alpha}\) or  \(\Gamma^m_{\beta}\). The operator $\hat{\omega}_{j,l}$ accounts for the quantum fluctuations arising from the non-commutation between different parts of the renormalised Hamiltonian and has the following form~\cite{anirbanurg1}
\begin{eqnarray}
\hat{\omega}_{j,l}&=&H^{D}_{j,l}+H^{X}_{j,l}-H^{X}_{j,l-1}~.\label{qfOp}
\end{eqnarray}
Upon disentangling electronic states $\hat{s},\sigma$ along a isogeometric curve at distance $\Lambda_{j}$, the following effective Hamiltonian $H_{j,l}$ is generated 
\begin{eqnarray}
H_{j,l}=\prod_{m=1}^{l}U_{j,m}H_{(j)}[\prod_{m=1}^{l}U_{j,m}]^{\dagger}~.
\end{eqnarray}
Here, $\tau_{j,l}= n_{j,l}-\frac{1}{2}$. We note that $H_{j,2n_{j}+1}=H_{(j-1)}$ is the Hamiltonian obtained after disentangling all $2n_{j}$ number of electronic states on the isogeometric curve $j$ at distance $\Lambda_{j}$. Below, we depict the different terms generated upon successive disentanglement of the states $|\mathbf{k}_{\Lambda_{j}\hat{s}_{l}},\sigma\rangle$ on a given curve,
\begin{eqnarray}
H_{j,l+1}&=&Tr_{j,l}(H_{(j,l)})+\lbrace c^{\dagger}_{j,l}Tr_{j,l}(H_{(j,l)}c_{j,l}),\eta_{j,l}\rbrace\tau_{j,l}, \tau_{j,l}\nonumber\\
	 &=&\hat{n}_{j,l}-\frac{1}{2}\nonumber\\ 
H_{j,l+2}&=&Tr_{j,l+1}(Tr_{j,l}(H_{(j,l)}))\nonumber\\
	 &+&Tr_{j,l+1}(\lbrace c^{\dagger}_{j,l}Tr_{j,l}(H_{(j,l)}c_{j,l}),\eta_{j,l}\rbrace\tau_{j,l})\nonumber\\
&+&\lbrace c^{\dagger}_{j,l+1}Tr_{j,l+1}(Tr_{j,l}(H_{(j,l)})c_{j,l+1}),\eta_{j,l+1}\rbrace\tau_{j,l+1}\nonumber\\
&+&\lbrace c^{\dagger}_{j,l+1}Tr_{j,l+1}(\lbrace c^{\dagger}_{j,l}Tr_{j,l}(H_{(j,l)}c_{j,l}),\eta_{j,l}\rbrace c_{j,l+1}),\nonumber\\
&&\eta_{j,l+1}\rbrace\tau_{j,l}\tau_{j,l+1}~.\nonumber\\
H_{j,l+3}&=&\ldots\text{terms with } \tau_{j,l}, \tau_{j,l+1}, \tau_{j,l+2}\ldots\nonumber\\
 &+& \ldots\text{terms with } \tau_{j,l}\tau_{j,l+1}, \tau_{j,l+1}\tau_{j,l+2}, \tau_{j,l}\tau_{j,l+2}\ldots\nonumber\\
 &+&\ldots\text{terms with }\tau_{j,l}\tau_{j,l+1}\tau_{j,l+2}.
\label{2ndDisentanglement}
\end{eqnarray}

Upon disentangling multiple electronic states placed in the tangential direction on a given momentum shell at distance $\Lambda_{j}$ generates RG contribution from leading order scattering processes (terms multiplied with $\tau_{j,l}$, $\tau_{j,l+1}$, etc.) that goes as $Area/Vol=1/L$ and higher order processes like terms multiplied with $\tau_{j,l}\tau_{j,l+1}$ that goes as $(Area)^{2}/Vol^{2}=1/L^{2}$ , $\tau_{j,l}\tau_{j,l+1}\tau_{j,l+2}$ that goes as $(Area)^{3}/Vol^{3}=1/L^{3}$~\cite{anirbanurg1}. Here, each factor of area arises from decoupling an entire shell of single-particle states ($\tau_{j,l}$) at every RG step, and every factor of volume arises from the Kondo coupling. Accounting for only the leading tangential scattering processes, as well as other momentum transfer processes along the normal direction $\hat{s}$, the renormalised Hamiltonian \(H_{(j-1)}\) has the form~\cite{anirbanurg1}
\begin{equation}
\hspace*{-0.1cm}Tr_{j,(1,\ldots,2n_{j})}(H_{(j)})+\sum_{l=1}^{2n_{j}}\lbrace c^{\dagger}_{j,l}Tr_{j,l}(H_{(j)}c_{j,l}),\eta_{j,l}\rbrace\tau_{j,l}~.\label{HRG}
\end{equation}
From the effective Hamiltonian obtained at the stable fixed point $\hat{H}^{*}$, we can compute the (unnormalised) density matrix operator ($\hat{\rho} = e^{-\beta \hat{H}^{*}}$) and thence the finite-temperature partition function as
\begin{equation}
Z = \mathrm{Tr}\left[ \hat{\rho}\right] = \mathrm{Tr}\left[ e^{-\beta \hat{H}^{*}}\right] = \mathrm{Tr}\left[ U^{\dagger} e^{-\beta \hat{H}^{*}} U\right] = \mathrm{Tr}\left[ e^{-\beta \hat{H}}\right]~,
\label{partfunc}
\end{equation}
where \(\beta = 1/k_B T\), \(U = \prod_{1}^{j^{*}}U_{(j)}\), $H$ is the bare Hamiltonian and $j^{*}$ is the RG step at which the IR stable fixed point is reached. This indicates that the partition function is preserved along the RG flow as the unitary transformations preserve the eigenspectrum.
\section{Crossover from local moment to strong-coupling}\label{fixedPointKondo}
\subsection{IR fixed point Hamiltonian and wavefunction}
The unitary RG process generates the effective Hamiltonian's $\hat{H}_{(j)}(\omega)$'s for various eigendirections $|\Phi(\omega)\rangle$
of the $\hat{\omega}$ operator. Note the associated eigenvalue $\omega$ identifies a sub-spectrum in the interacting many body eigenspace. The form of $\hat{H}_{(j)}(\omega)$ is given by
\begin{equation} \begin{aligned}
	\sum_{j,l,\sigma}\epsilon_{j,l}\hat{n}_{j,l} + J^{(j)}(\omega) \mathbf{S}\cdot \mathbf{s}_< +\sum^{j,n_{j}}_{\substack{a=N,\\ m=1}}J^{(a)}S^{z}s^{z}_{a,\hat{s},m}~,
\label{compHam}
\end{aligned}\end{equation}
where the label \(a\) indexes the isoenergetic shells that have already been decoupled and hence run from the furthest shell \(a=N\) to the most recently decoupled shell \(a=j\). We also defined \(\mathbf{s}_< = \frac{1}{2}\sum_{\substack{j_{1},j_{2}<j^{*},\\ m,m'}}c^{\dagger}_{j_{1},\hat{s}_{m},\alpha}\boldsymbol{\sigma}_{\alpha\beta}c_{j_{2},\hat{s}_{m'},\beta}\) as the total spin operator for the Kondo cloud and $s^{z}_{l,\hat{s},m}=\frac{1}{2}(\hat{n}_{l,\hat{s}_{m},\uparrow}-\hat{n}_{l,\hat{s}_{m},\downarrow})$. The derivation of the above equation is presented in Appendix \ref{urg-deriv}. Here, the second term is the effective Hamiltonian for the coupling of the Kondo cloud to the impurity spin, while the third encodes the interaction between the impurity spin moment and the decoupled electronic degrees of freedom that do not belong to the Kondo cloud (lying on radial shells in momentum-space indexed by the RG step $j$). Note that the third term is an Ising coupling and involves only potential scattering of the decoupled electrons and the impurity, and hence does not cause any spin-flip scattering of the impurity spin. Importantly, we will see later that eq.\eqref{compHam} encodes the entire $T=0$ RG crossover. Employing eq.\eqref{partfunc}, this enables a computation of the entire RG crossover at finite $T$ as well. The Kondo coupling RG equation for the RG steps (Appendix \ref{urg-deriv}) has the form
\begin{eqnarray}
\frac{\Delta J^{(j)}(\omega)}{\Delta\log\frac{\Lambda{j}}{\Lambda_{0}}}=\frac{n_{j}(J^{(j)})^{2}\left[(\omega - \frac{\hbar v_{F}\Lambda_{j}}{2})\right]}{(\omega - \frac{\hbar v_{F}\Lambda_{j}}{2})^{2}-\frac{\left(J^{(j)}\right)^{2}}{16}}~,\label{RGeqn}
\end{eqnarray}
where $n_{j}=\sum_{\hat{s}}1$ is the number of states on the isogeometric shell at distance $\Lambda_{j}$ from the Fermi surface. We have assumed that the conduction bath dispersion is linear near the Fermi surface: \(\epsilon_j = \hbar v_F \Lambda_j\), \(v_F\) being the Fermi velocity. Note that the denominator $\Delta\log\frac{\Lambda{j}}{\Lambda_{0}} =1$ for the RG scale parametrisation $\Lambda_{j}=\Lambda_{0}\exp(-j)$. We now redefine Kondo coupling as a dimensionless parameter
\begin{eqnarray}
K^{(j)}=\frac{J^{(j)}}{\omega-\frac{\hbar v_{F}}{2}\Lambda_{j}}~,\label{reparametrization}
\end{eqnarray} 
and operate in the regime $\omega>\frac{\hbar v_{F}}{2}\Lambda_{j}$. With the above parametrisation of eq.\eqref{reparametrization}, we can convert the difference RG relation (eq.\eqref{RGeqn}) into a continuum RG equation

\begin{eqnarray}
\frac{d K}{d\log\frac{\Lambda}{\Lambda_{0}}}=\left(1-\frac{\omega}{\omega - \frac{1}{2}\hbar v_{F}\Lambda}\right)K+\frac{n(\Lambda)K^{2}}{1-\frac{K^{2}}{16}}~,
\end{eqnarray}
where \(n(\Lambda)\) is the continuum counterpart of \(n_j\) and represents the number of electronic states on the isogeometric shell at momentum \(\Lambda.\)

Upon approaching the Fermi surface \(\Lambda_{j}\to 0\), hence \(\left(1-\frac{\omega}{\omega-\hbar v_{F}\Lambda}\right)\to 0\) and \(n(\Lambda)\) can be replaced by density of states on the Fermi surface $n(0)$:
\begin{eqnarray}
\frac{d K}{d\log\frac{\Lambda}{\Lambda_{0}}}=\frac{n(0)K^{2}}{1-\frac{K^{2}}{16}}
\end{eqnarray}
 \begin{figure}
 \centering
 \includegraphics[width=0.45\textwidth]{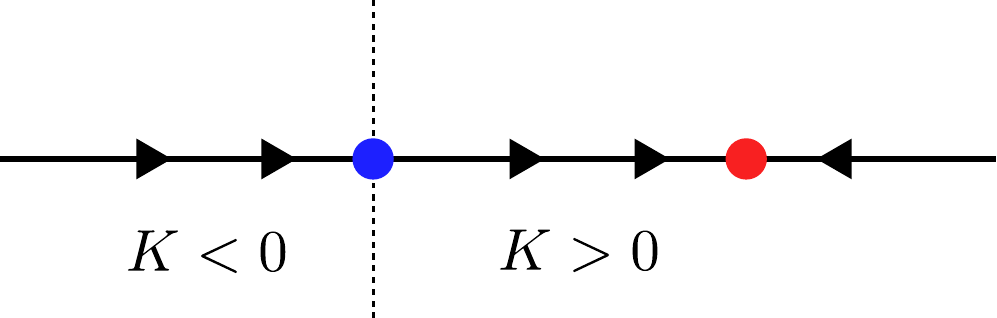}
 \caption{Schematic RG phase diagram for the Kondo problem. The red dot represents intermediate coupling fixed point at $K^{*}=4$ for the case of the antiferromagnetic Kondo coupling. The blue dot represents the critical fixed point at $K^{*}=0$ for the case of the ferromagnetic Kondo coupling.} 
 \end{figure}

At this point, we observe an important aspect of the RG equation: for $K<<1$, the RG equation reduces to the one loop form: $\frac{d K}{d\log\frac{\Lambda}{\Lambda_{0}}}=K^{2}$~\cite{anderson1970poor}. On the other hand, the non-perturbative form of the flow equation obtained from the URG formalism shows the presence of intermediate coupling fixed point at $K^{*}=4$ in the antiferromagnetic regime $K>0$. Upon integrating the RG equation and using the fixed point value $K^{*}=4$ we obtain the Kondo energy scale ($T_{K}$) and thence the effective length of the Kondo cloud ($\xi_{K}$)
\begin{align}
	\frac{1}{K_{0}}&-\frac{1}{2}+\frac{K_{0}}{16}=-n(0)\log\frac{\Lambda^{*}}{\Lambda_{0}}~,\\
	\Lambda^{*}&=\Lambda_{0}\exp\left(\frac{1}{2n(0)}-\frac{1}{n(0)K_{0}}-\frac{K_{0}}{n(0)16}\right)~,\\
	T_{K} &= \frac{\hbar v_{F}\Lambda_{0}}{k_{B}}\exp\left(\frac{1}{2n(0)}-\frac{1}{n(0)K_{0}}-\frac{K_{0}}{n(0)16}\right),~\\
	~\xi_{K} &= \frac{2\pi}{\Lambda^{*}}~.
\end{align}

\begin{figure}[!htb]
\centering
\includegraphics[width=0.45\textwidth]{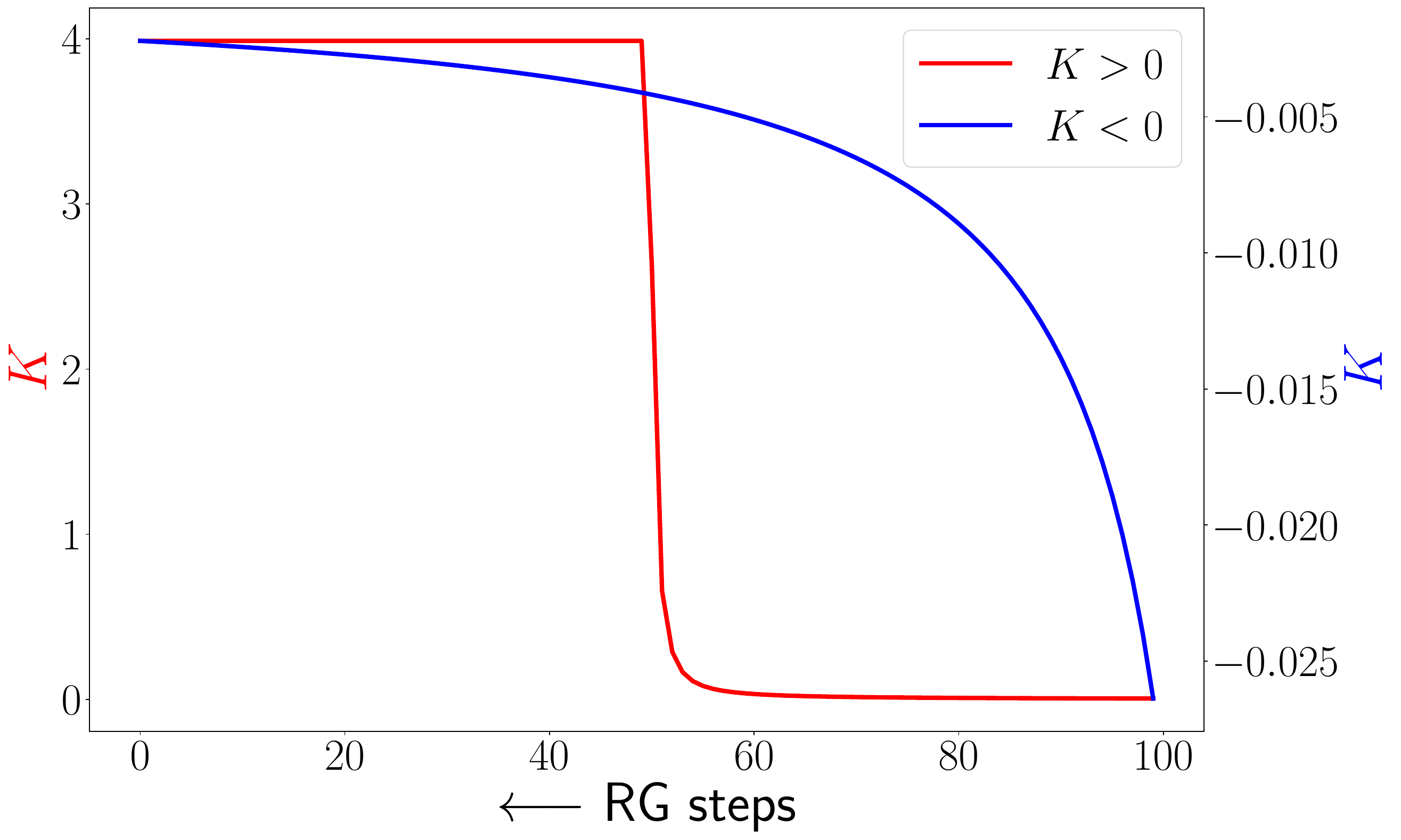}
\caption{Red: Renormalisation of the dimensionless Kondo coupling $K$ towards the strong-coupling fixed point \((K > 0)\) with RG steps ($\log\Lambda_{j}/\Lambda_{0}$). The growth of the Kondo coupling to a finite value of the intermediate coupling fixed point is evident . Blue: Decay of the dimensionless Kondo coupling $K$ to zero towards the local moment critical fixed point \((K < 0)\) with RG steps ($\log\Lambda_{j}/\Lambda_{0}$). For these plots, we chose $\omega=\hbar v_{F}\Lambda_{j}$.} \label{Kondocoupling_lm}
\end{figure}

At the IR fixed point in the antiferromagnetic regime the effective Hamiltonian is given by,
\begin{equation}\begin{aligned}
	H^{*}&=\sum_{|\Lambda|<\Lambda^{*}}\hbar v_{F}\Lambda\hat{n}_{\Lambda,\hat{s},\sigma} + J^*\mathbf{S}\cdot \mathbf{s}_< +\sum_{j'=N,m=1}^{j^{*},n_{j'}}J^{j'}S^{z}s^{z}_{j',m}~,\label{fixedPointHam}
\end{aligned}\end{equation}
In the equation, $m$ refers to the various normal directions $\hat{s}_{m}$ of the Fermi surface. In the above equation, the second term is the effective Hamiltonian for the coupling of the Kondo cloud to the impurity spin. The third encodes the Ising interaction between the impurity spin moment and the magnetic moment of the decoupled electronic degrees of freedom that do not belong to the Kondo cloud (lying on radial shells in momentum-space indexed by the RG step $j$, and corresponding to the integrals of motion (IOMs) generated during the RG flow). As reported in the appendices, the fixed point Hamiltonian has been used to compute various thermodynamic quantities like the impurity susceptibility, impurity specific heat, thermal entropy, Wilson number and the Wilson ratio. The values obtained are found to be in good agreement with those from other methods like the NRG or the Bethe ansatz \cite{andrei_rajan_1982,bullaNRGreview,wilson1975renormalization,Wilkins_oliveira_1981}.

We can now extract a zero mode from the above Hamiltonian that captures the low energy theory near the Fermi surface,
\begin{equation}\begin{aligned}
	H_{coll}&=\frac{1}{N}\sum_{|\Lambda|<\Lambda^{*}}\hbar v_{F}\Lambda\sum_{|\Lambda|<\Lambda^{*}}\hat{n}_{\Lambda,\hat{s},\sigma} + J^*\mathbf{S}\cdot \mathbf{s}_< \\
		&+ \sum_{j'=N,m=1}^{j,n_{j'}}J^{j'}S^{z}s^{z}_{j',m}\\
		&= J^*\mathbf{S}\cdot \mathbf{s}_< + \sum_{j'=N,m=1}^{j,n_{j'}}J^{j'}S^{z}s^{z}_{j',m}~.
		\label{collHam}
\end{aligned}\end{equation}
In the first step, the first term vanishes as the sum over wavevector $\Lambda$ within the symmetric window $\Lambda^{*}$ around the Fermi surface itself vanishes. Indeed, by taking the Ising coupling between the impurity spin and the IOMs ($s^{z}_{j',m}$) to be a constant, we observe that the Kondo singlet state 
\begin{equation}\begin{aligned}
\ket{\Psi^*} = \frac{1}{\sqrt{2}}\left(\ket{\uparrow_d,\Downarrow}\otimes \ket{\downarrow} - \ket{\downarrow_d,\Uparrow}\otimes\ket{\uparrow}\right)~\label{eigState}
\end{aligned}\end{equation}
corresponds to the ground state of the zero mode Hamiltonian at the IR fixed point (eq.\eqref{collHam}). In the singlet state written above, \(\ket{\uparrow_d}\) and \(\ket{\downarrow_d}\) represent the configuration of the impurity spin \(S^z\), \(\ket{\Uparrow}\) and \(\ket{\Downarrow}\) represent the configuration of the spin of the Kondo cloud \(s^z_<\) and \(\ket{\uparrow}\) and \(\ket{\downarrow}\) represent the spin of the IOMs \(s^z = \sum_{\substack{j_{1} > j^{*},\\ m}}\hat n_{j_{1},\hat{s}_{m},\alpha}\sigma_{\alpha\beta}^z\). Signatures of such a singlet ground state have been experimentally probed using NMR experiments in \(Cu\)-nuclei~\cite{boyce_slichter_1974}, using conductance measurements in carbon nanotubes~\cite{Nygrd_cobden_2000,babi_kantos_2004,chorley_galpin_2012} and using STS measurements in quantum corrals~\cite{Manoharan2000,gregory_fiete_2003,rossi_enrico_2006,Prser2011}.

\subsection{Kondo cloud size and effective Kondo coupling as functions of bare coupling $J_{0}$}
\label{cloud-size}
\begin{figure}[htpb]
	\centering
	\includegraphics[width=0.45\textwidth]{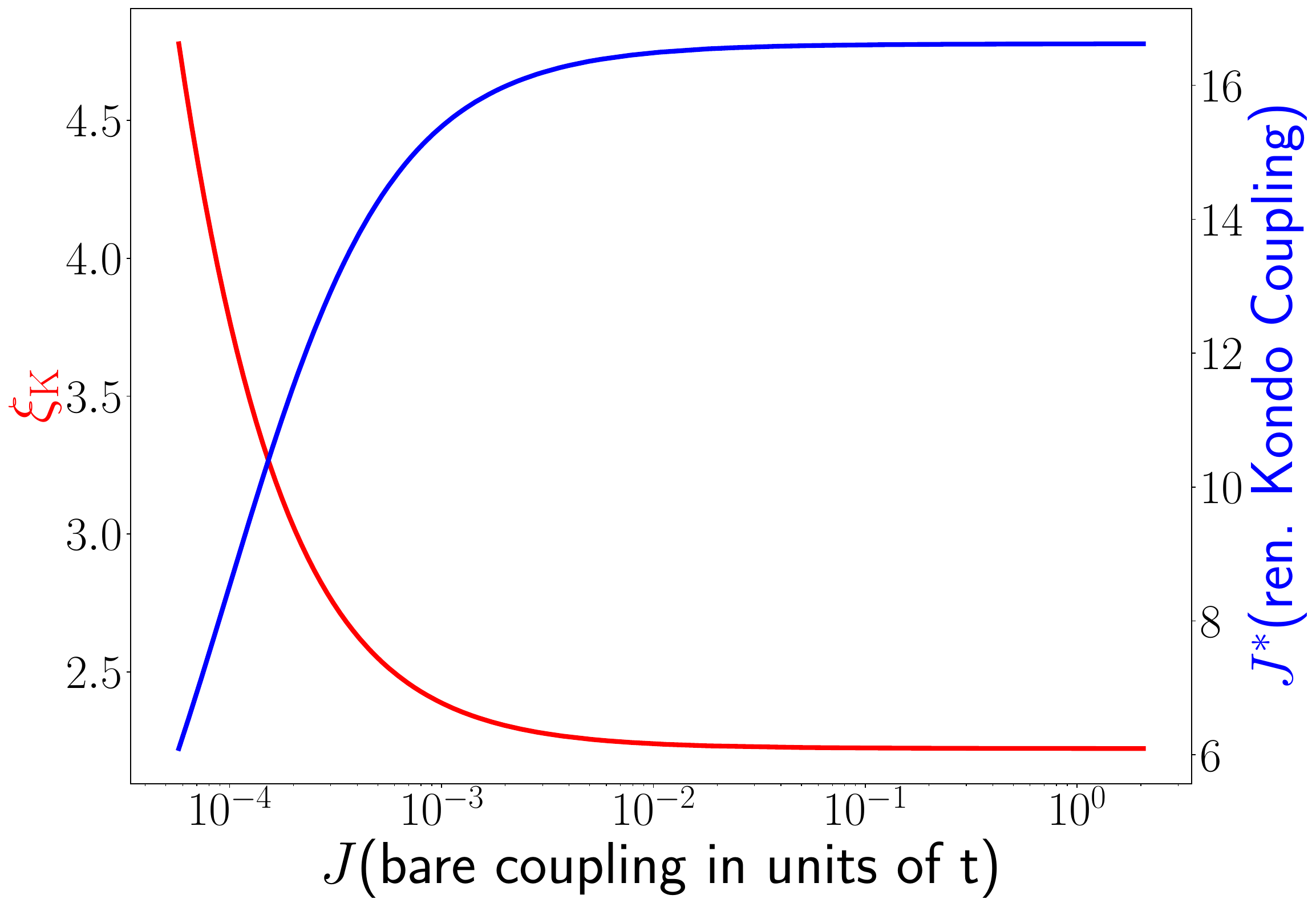}
	\caption{Red: Kondo cloud length $\xi_K$ vs.~bare Kondo coupling $J$. Blue: Renormalised Kondo coupling $J^{*}$ vs. $J$ (x-axis in log scale). The bare Kondo coupling $J$ is chosen to lie within the range $5.7\times 10^{-5}<J<5.4$.}
	\label{xi_k}
\end{figure}

\begin{figure}[!htpb]
	\centering
\includegraphics[width=0.45\textwidth]{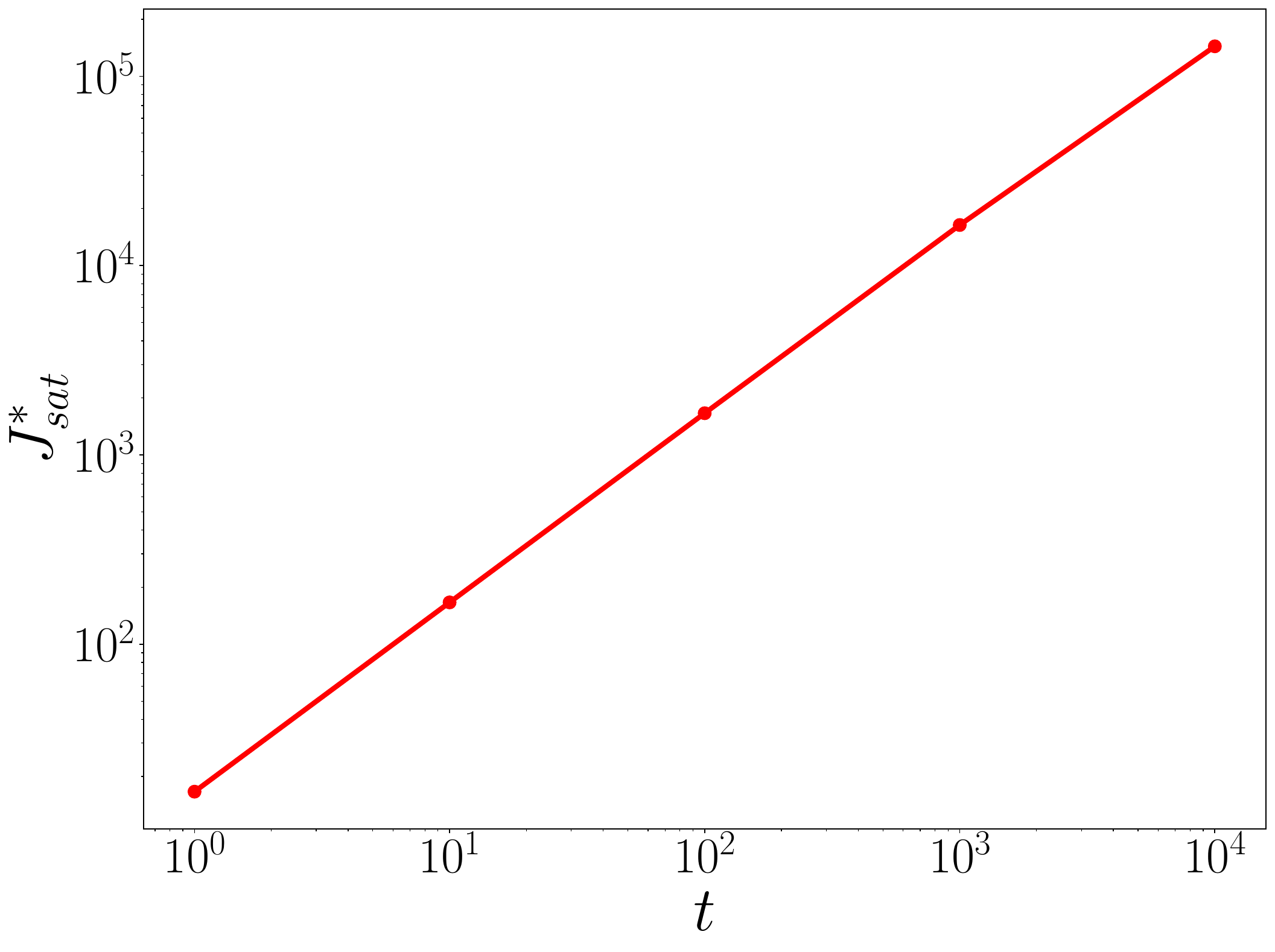}
\caption{Variation of the renormalised Kondo coupling $J^{*}$ with the hopping parameter $\mathcal{O}(1)<t<\mathcal{O}(10^{4})$ of the electronic bath (and hence the bandwidth).}
\label{infinite}
\end{figure}

In figures \ref{xi_k}, we show the variation of the Kondo cloud size $\xi_{K}$ and effective Kondo coupling $J^{*}$ as function of bare coupling $J$ (in units of $t$) in the range $5.7\times 10^{-5}<J<5.4$.
All plots below are obtained for momentum-space grid $100\times 100$ and with RG scale factor $\Lambda_{j}=b\Lambda_{j+1}$ ($b=0.9999 = 1-1/N^2~,~N=100$). The $E_{F}$ for the 2d tight binding band $-W<E_{k}=-2t(\cos k_{x}+\cos k_{y})<W$ ($W=4t$) is chosen at $E_{F}=-3.9t$, and the bare $k$-space cutoff is set at $\Lambda_{0}\simeq 2.83$. 
The variation of the renormalised Kondo coupling $J^{*}$ with the bare $J$ shown in Fig.\ref{xi_k} clearly indicates the flow under RG towards saturation at a strong-coupling value of $J^{*}_{sat}\sim 16$. Similarly, the variation of the Kondo screening length $\xi_{K}$ with $J$ in Fig.~\ref{xi_k} shows a fall to an asymptotic value of $\xi_{K}\sim 3$ lattice sites at the strong-coupling fixed point.
We recall that Wilson's NRG calculation for the Kondo problem (for a bath of conduction electrons in the continuum with linear dispersion and a very large UV cutoff $D$) shows that the renormalised Kondo coupling $J^{*}\to\infty$ under flow to strong coupling. This can be reconciled from our URG results by noting that the value of $J^{*}_{sat}$ increases upon rescaling the conduction bath bandwidth $D$ to larger values (by rescaling the nearest neighbor hopping strength $t$). This is shown in Fig. \ref{infinite}, where we see that $J^{*}_{sat}$ increases from a value of $\mathcal{O}(1)$ (in units of $t$) to $\mathcal{O}(10^{9})$ as $t$ is increased from $\mathcal{O}(1)$ to $\mathcal{O}(10^{4})$. Thus, taking the limit of $D\to\infty$ will lead to $J^{*}_{sat}\to\infty$. 
\begin{figure}[htpb!]
\centering
\includegraphics[width=0.45\textwidth]{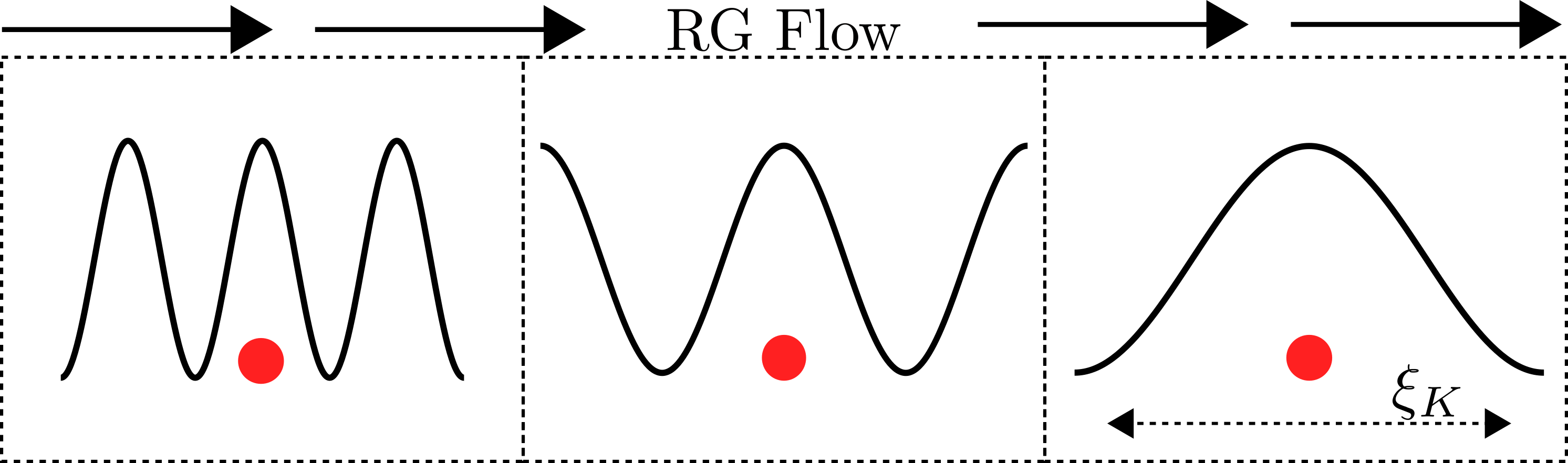}
\caption{Emergence of the Kondo length scale under URG. The red circle represents the impurity site, while the wavelike structure represents the shortest wavelength interacting with the impurity at any given point of the RG. The leftmost image is closest to the local moment fixed point while the rightmost image is closest to the strong-coupling one.}
\label{Kondo-length}
\end{figure}

There is another aspect of this method that is worth pointing out - the size of the Kondo cloud emerges as a natural length scale of the low energy theory. The URG moves forward by decoupling high momentum states. Each momentum state wavefunction \(\Psi_k\) is associated with a de Broglie wavelength \(\lambda_k \sim \frac{1}{k}\). At high energies (temperatures), all possible wavelengths are interacting with the impurity. Under RG, the shortest of these wavelengths get decoupled and the impurity interacts with only longer wavelengths. The shortest such wavelength that is still interacting with the impurity at the fixed point defines the Kondo length scale \(\xi_K\). 

The members of the Kondo cloud then involve all states with wavelengths starting from \(\xi_K\) and extending up to \(\infty\). From existing results as well as our own investigations (inset of fig.\ref{kondo_spec_func}), it is clear that the width of the spectral function increases with temperature. This width defines the relevant energy scale \(\omega(T)\) (and hence a relevant length scale \(\xi(T) \sim \omega^{-1}(T)\)) for interactions with the impurity. Higher temperatures therefore lead to reduced \(\xi(T)\). This means that at sufficiently large temperatures, the relevant length scale \(\xi(T)\) is shorter than the Kondo length scale \(\xi_K\): \(\xi(T) \ll \xi_K\), which means that the Kondo effect will not manifest at temperatures \(T \gg T_K\): the RG flow is not able to filter out the low energy physics because of the thermal fluctuations. Similar insights were obtained in \cite{mitchell_bulla_2011} using a real space renormalisation group flow.
\subsection{Redistribution of spectral weight under RG flow}
The presence of a unitarily-transformed effective Hamiltonian at each stage of the RG allows one to map the journey in terms of changes in the distribution of the total spectral weight. In this section, we will compute the impurity spectral function at various values of the running coupling \(J\), following the expression \cite{costi2000,rosch_costi2003}
\begin{equation}\begin{aligned}
	A_d^\sigma = -\frac{1}{\pi}\text{Im } \left<\{O_\sigma, O_\sigma^\dagger\} \right>, &&O_\sigma = \frac{J}{2}\left(S^{-\sigma}c_{0,-\sigma} + S^z c_{0,\sigma}\right) 
\end{aligned}\end{equation}
where \(\left<... \right>\) indicates a thermal average, \(\left\{,\right\} \) indicates an anti-commutator and \(c_{0,\sigma} = \sum_{k}c_{k\sigma}\). 

\begin{figure}[!htpb]
	\centering
	\hspace*{-0.8cm}
	\includegraphics[width=0.5\textwidth]{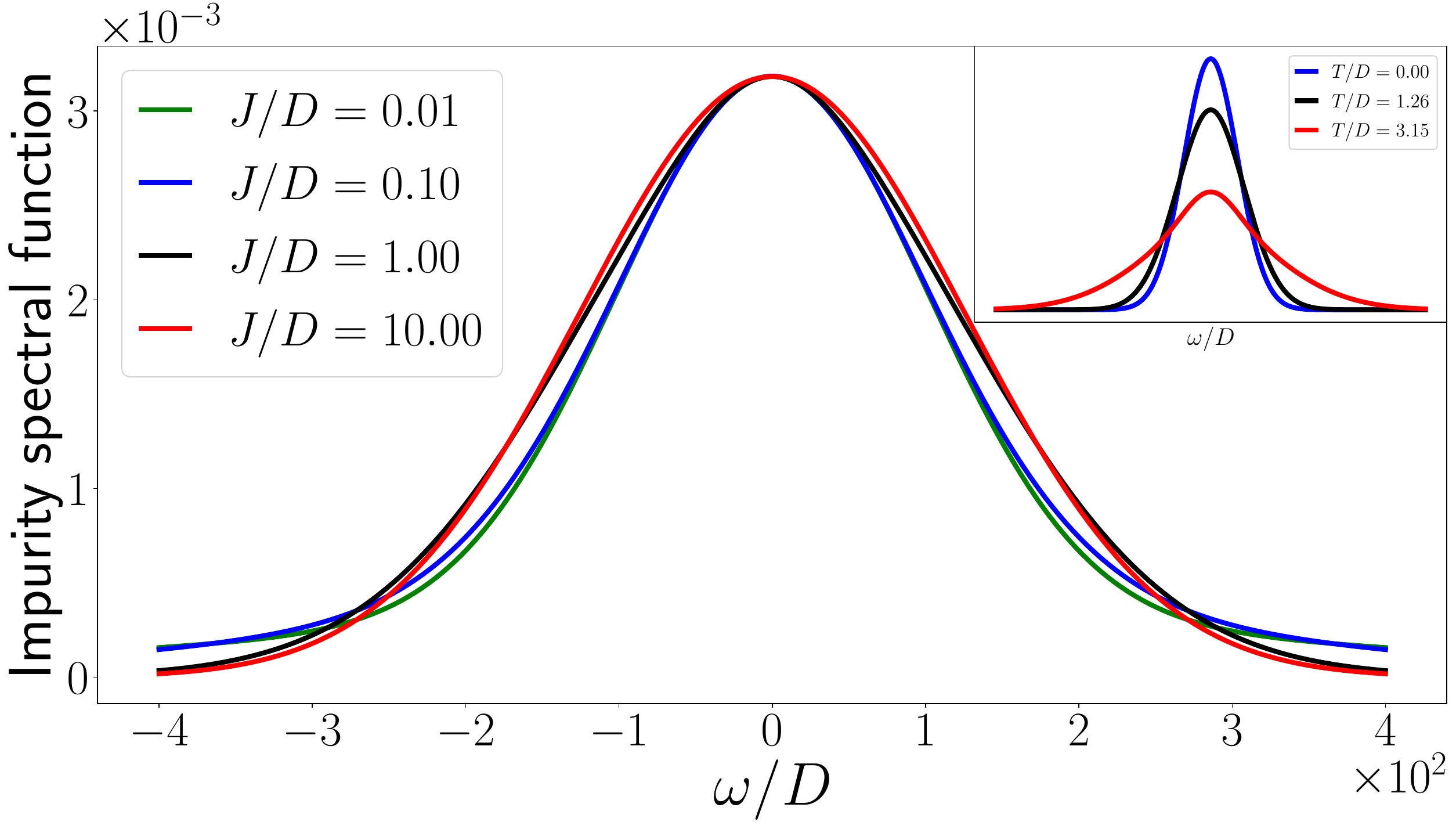}
	\caption{Impurity spectral function as a function of \(\omega/D\) at various values of \(J\), starting from weak-coupling and extending up to strong-coupling. \(D\) is the half-bandwidth, and is set to 1. Inset shows the spectral function at various temperatures. The width increases with increasing temperature (see final paragraph of subsection \ref{cloud-size}).}
	\label{kondo_spec_func}
\end{figure}

In order to plot the spectral function (fig.~\ref{kondo_spec_func}), we chose various values of \(J\) to mimic the crossover from the local moment to the strong-coupling fixed points. The spectral function is normalised by dividing with the total area under the curve.  The evolution of the spectral functions shows the increase in the spectral weight at the zero energy resonance at the cost of that at higher frequencies. This transfer of spectral weight occurs because the high-frequency excitations are slowly getting integrated out and consumed into the IOMS, while the Kondo cloud gets distilled into purely low-energy modes proximate to the Fermi surface. This should be contrasted with the sharpening of the Abrikosov-Suhl resonance in the spectral function of the bare Hamiltonian as obtained from other methods (e.g., NRG), and which corresponds to the excitations of the local Fermi liquid. 

The inset of Fig.~\ref{kondo_spec_func} also shows the change in the spectral function when temperature is introduced. The broadening of the central peak implies a decrease in the relevant length scale and hence a destruction of screening for \(T \gg T_\mathrm{K}\). This was also shown in \cite{costi2000}.
\section{Effective Hamiltonian for excitations of the Kondo cloud}\label{susc2}
In this section we will study the effect of the renormalised Kondo coupling on the self energy of the electrons comprising the Kondo cloud. We integrate out the decoupled electronic states to obtain the effective Hamiltonian $H^{*}$ of the impurity + electronic cloud system. In this Hamiltonian we have additionally kept the electronic dispersion to study the effect of electronic density fluctuation due to the inter-electronic interaction mediated by the impurity spin, 
\begin{eqnarray}
	H^{*}&=&H_{0}^{*}+\frac{J^{*}}{2}\sum_{\substack{j_{1},j_{2}<j^{*},\\ m,m'}}\mathbf{S}\cdot c^{\dagger}_{j_{1},\hat{s}_{m},\alpha}\boldsymbol{\sigma}_{\alpha\beta}c_{j_{2},\hat{s}_{m'},\beta},
\end{eqnarray}
where \(H_{0}^{*}=\sum_{|\Lambda_{j}|<\Lambda^{*},\sigma}\epsilon_{j}\hat{n}_{j,\hat{s},\sigma}\label{density}\). 
In order to study the inter-electronic interaction we need to isolate the quantum impurity from the Kondo cloud. This can be done by first recasting the many body eigenstate $\ket{\Psi}$ of $H^{*}$ in the $\ket{\uparrow_d},\ket{\downarrow_d}$ basis of the impurity and the associated configuration of the rest of the electronic states. We will then perform a perturbative expansion in the small parameter \(\frac{t^2}{J}\) from which we obtain the effective $k$-space Hamiltonian for the Kondo cloud. This is justified by the fact that the Kondo coupling can be made arbitrarily large by performing the expansion as close to the strong-coupling fixed point as desired.
\par\noindent
At the ground state, the wavefunction will be a singlet composed of the impurity spin and the composite spin formed by the conduction electrons at the origin \(\left( \ket{\Uparrow}, \ket{\Downarrow} \right) \): \(|\Psi\rangle = a_{0}\ket{\uparrow_d}\ket{\Downarrow} +a_{1}\ket{\downarrow_d}\ket{\Uparrow}\). With this ground state in mind, we can rewrite the eigenvalue relation \(H^{*}_{K} \ket{\Psi} = E_\text{GS} \ket{\Psi}\) as a set of two coupled equations:
\begin{equation}\begin{aligned}
&a_{0}(H_{0}^{*}+\frac{J^{*}}{2}s_{z})\ket{\Downarrow} + a_{1}\frac{J^{*}}{2}s_{-}\ket{\Uparrow} = a_{0}E_\text{GS}\ket{\Downarrow}, \\
&a_{0}\frac{J^{*}}{2}s_{+}\ket{\Downarrow} +a_{1}(H_{0}^{*}-\frac{J^{*}}{2}s_{z})\ket{\Uparrow}  = a_{1}E_\text{GS}\ket{\Uparrow}~.
\end{aligned}\end{equation}
By combining these two equations and eliminating \(\ket{\Uparrow}\), we obtain the effective Hamiltonian for excitations in the subspace of states \(\ket{\Downarrow}\):
\begin{equation}
\left[H_0^* + \frac{J^{*}}{2}s_{z} + s_-\frac{\frac{(J^{*})^{2}}{4}}{E_\text{GS}+\frac{J^{*}}{2}s_{z}-H_{0}^{*}}s_+\right]\ket{\Downarrow} = E_\text{GS}\ket{\Downarrow}\label{coupled1}
\end{equation}
From here we obtain the form of the effective Hamiltonian accounting for the leading order density-density and off-diagonal four Fermi interaction terms, by expanding the denominator in powers of \(H_0^*/J\).
\begin{align}
	&H_0^* + \frac{J^{*}}{2}s_{z}+ s_-\frac{\frac{(J^{*})^{2}}{4}}{(E_\text{GS}+\frac{J^{*}}{2}s_{z})(1-\frac{1}{E_\text{GS}+\frac{J^{*}}{2}s_{z}}H^{*}_{0})} s_+\nonumber\\
	       &\approx H_0^* + \frac{J^{*}}{2}s_{z}+ s_-\frac{\frac{(J^{*})^{2}}{4}}{E_\text{GS}+\frac{J^{*}}{2}s_{z}}(1+\frac{1}{E_\text{GS}+\frac{J^{*}}{2}s_{z}}H^{*}_{0}\nonumber\\
	       &+\frac{1}{E_\text{GS} +\frac{J^{*}}{2}s_{z}}H^{*}_{0}\frac{1}{E_\text{GS}  + \frac{J^{*}}{2}s_{z}}H^{*}_{0}) s_++O((H_{0}^*)^{3})\nonumber\\
	       &\approx H_0^* + \frac{J^{*}}{2}s_{z}+ s_-\frac{\frac{(J^{*})^{2}}{4}}{E_\text{GS} +\frac{J^{*}}{2}s_{z}}   s_++  s_-\frac{\frac{(J^{*})^{2}}{4}}{(E_\text{GS}+\frac{J^{*}}{2}s_{z})^{2}}H^{*}_{0} s_+\nonumber\\
	       &+ s_-\frac{\frac{(J^{*})^{2}}{4}}{(E_\text{GS}+\frac{J^{*}}{2}s_{z})^{2}}H^{*}_{0}\frac{1}{E_\text{GS} +\frac{J^{*}}{2}s_{z}}H^{*}_{0} s_+, \nonumber\\
	       &\approx H_0^* + \frac{J^{*}}{2}s_{z}+(\frac{1}{2} - s_{z})\frac{\frac{(J^{*})^{2}}{4}}{E_\text{GS}+\frac{J^{*}}{4}}\nonumber\\
	       &+s_{+}\frac{\frac{(J^{*})^{2}}{4}}{(E_\text{GS}+\frac{J^{*}}{2}s_{z})^{2}}(s_{-}H^{*}_{0}+[H^{*}_{0},s_{-}])+\nonumber\\
	       &+s_{+}\frac{\frac{(J^{*})^{2}}{4}}{(E_\text{GS}+\frac{J^{*}}{2}s_{z})^{2}}H^{*}_{0}\frac{1}{E_\text{GS}+\frac{J^{*}}{2}s_{z}}(s_{-}H^{*}_{0}+[H^{*}_{0},s_{-}]).
\end{align}
In the second and third term, we can substitute \(s^z = -\frac{1}{2}\), because \(s^z\ket{\Downarrow} = -\frac{1}{2}\). We will also substitute the singlet binding energy \(-\frac{3J^*}{4}\) as the ground state energy \(E_\text{GS}\) because \(J \gg \epsilon_k\) at the strong-coupling fixed point, and \(H_0^* = E_0^*\) as the kinetic energy part of the ground state \(\ket{\Downarrow}\):
\begin{equation}\begin{aligned}
	H_\text{eff} = -\frac{3J^*}{4} + E_0^* + s_{+}\frac{\frac{(J^{*})^{2}}{4}}{(E_\text{GS}+\frac{J^{*}}{2}s_{z})^{2}}\left(s_{-}H^{*}_{0}+[H^{*}_{0},s_{-}]\right.\\
	\left.+ H^{*}_{0}\frac{1}{E_\text{GS}+\frac{J^{*}}{2}s_{z}}(s_{-}H^{*}_{0}+[H^{*}_{0},s_{-}])\right)~.
\end{aligned}\end{equation}
This effective Hamiltonian has the expected structure; there are constant pieces that represent the diagonal part of the Hamiltonian, and then there are terms that scatter within the subspace. We will henceforth drop the constant terms, and keep only the excitations within the subspace. After simplifying the terms and retaining at most four-fermion interactions, we obtain the following effective Hamiltonian:
\begin{equation}
	H_\text{eff}=2(H^{*}_{0}+ \frac{1}{J^*}{H^{*}_{0}}^2) + \sum_{1234}V_{1234}c^\dagger_{{k_4} \uparrow}c^\dagger_{{k_3} \downarrow}c_{{k_2} \downarrow}c_{{k_1} \uparrow}\label{eff_Ham_Kondo}
\end{equation}
where \(V_{1234} = \left( \epsilon_{k_1} - \epsilon_{k_3} \right)\left[1 - \frac{2}{J^*}\left(\epsilon_{k_3} - \epsilon_{{k_1}} + \epsilon_{{k_2}} + \epsilon_{{k_4}}\right)\right]\). The first term, second and third terms represent a kinetic energy (eq.\eqref{density}), a density-density correlation (see eq.\eqref{localfermiliq} below), and a spin-fluctuation mediated electron-electron scattering process respectively. The presence of the off-diagonal scattering term is expected because the ground state singlet is highly-entangled and integrating out the impurity from this singlet should lead to scattering among the conduction electron states. In other words, it is this off-diagonal interaction that is a signal of the strongly screened local moment.
\section{Many body correlations and entanglement properties of the Kondo cloud}\label{ent2}
\begin{figure}[!htpb]
	\centering
\includegraphics[width=0.45\textwidth]{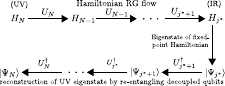}
\caption{The upper line represent the Hamiltonian RG flow via the unitary maps, while the lower line represents the reverse RG flow on the ground state wavefunction obtained at the Kondo IR fixed point. The reverse RG re-entangles decoupled electronic states with the Kondo singlet. This will result in generation of the many-body eigenstates at UV scales.}  \label{URGflowScheme}
\end{figure}

In order to study the effect of the off-diagonal terms  eq.\eqref{eff_Ham_Kondo} on the constituents of the Kondo cloud we perform a reverse URG treatment (shown in Fig.\ref{URGflowScheme}) starting from the Kondo model ground state $|\Psi^{*}\rangle$ at the IR fixed point eq.\eqref{eigState}. For this, we employ the entanglement RG method developed recently by some of us in Refs.\cite{1dhubjhep,siddharthacpi,mukherjee2020}.
\begin{figure}[!htpb]
\centering
\includegraphics[width=0.45\textwidth]{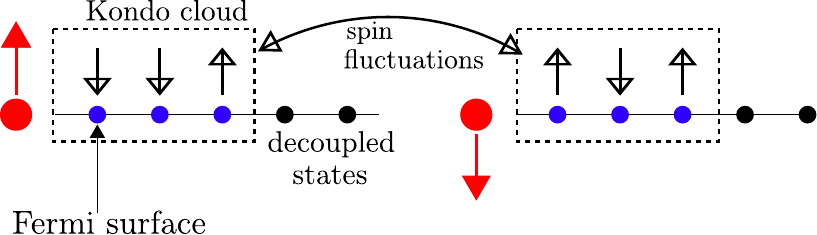}
\caption{Figure describes the electronic Kondo cloud (dashed rectangle) coupled to the impurity spin(red circle and arrow). The blue dots represent momentum+spin states inside the Kondo cloud, while the black dots represent states outside the Kondo cloud (that is, among the IOMS). The ground state singlet is formed by the impurity (red circle) and the Kondo cloud members (blue dots).}\label{ToyKondo}
\end{figure}
For the present study we take a toy model construction of the ground state wavefunction $|\Psi^{*}\rangle$ eq.\eqref{eigState} at IR, the Kondo impurity couples to $12$ electronic states $|\mathbf{k}\sigma\rangle$ of which three are occupied and 9 are unoccupied. The net spin of the electrons comprising the Kondo cloud is oppositely aligned to that the Kondo impurity. The Kondo cloud system is in tensor product with 14 separable electronic states. This construction is represented in Fig.\ref{ToyKondo} At each URG step $U_{j,\uparrow}U_{j,\downarrow}$ two electronic states $|k_{j},\uparrow\rangle$, $|k_{j\downarrow}\rangle$ are disentangled in reaching the IR fixed point. Upon performing reverse RG at each step two electrons are re-entangled into the eigenstates via the inverse unitary maps $U^{\dagger}_{j,\uparrow}U^{\dagger}_{j,\downarrow}$, Fig.\ref{URGflowScheme}, all total we perform seven reverse RG steps. This reverse RG program is numerically implemented using python. Since the forward RG is driven by the unitary transformation \(U = \frac{1}{\sqrt 2}\left( 1 + \eta - \eta^\dagger \right)\), the inverse transformation is \(\frac{1}{\sqrt 2}\left( 1 - \eta + \eta^\dagger \right)\). The \(\eta\) have already been described in appendix-\ref{urg-deriv}. The inverse transformation for re-entangling an electron of spin \(\sigma\) and energy \(\epsilon_q\) can therefore be written as
\begin{equation}\begin{aligned}
	U_{q\sigma}^{-1} = \frac{1}{\sqrt 2}\left[ 1 - \frac{J^2}{2}\frac{1}{2\omega \tau_{q\sigma} - \epsilon_{q}\tau_{q\sigma} - J S^z s^z_q}\left(\hat O + {\hat O}^\dagger\right)\right]
\end{aligned}\end{equation}
where \(\hat O = \sum_{k < \Lambda^*}\sum_{\alpha=\uparrow, \downarrow} \sum_{a={x,y,z}}S^a\sigma^a_{\alpha\sigma}c^\dagger_{k\alpha}c_{q\sigma}\). The wavefunctions for the reverse RG are generated by repeat application of this inverse unitary operator on the fixed point ground state \(\ket{\Psi^*}\). The total operator to re-entangle the energy states \(\epsilon_{q_1}, \epsilon_{q_2}, ..., \epsilon_{q_N}\) is
\begin{equation}
	\ket{\Psi_j} = \prod_{q=q_1}^{q_N}U^{-1}_{q\uparrow} U^{-1}_{q\downarrow}\ket{\Psi^*}
\end{equation}

\begin{figure}[!htpb]
\begin{center}
\includegraphics[width=0.45\textwidth]{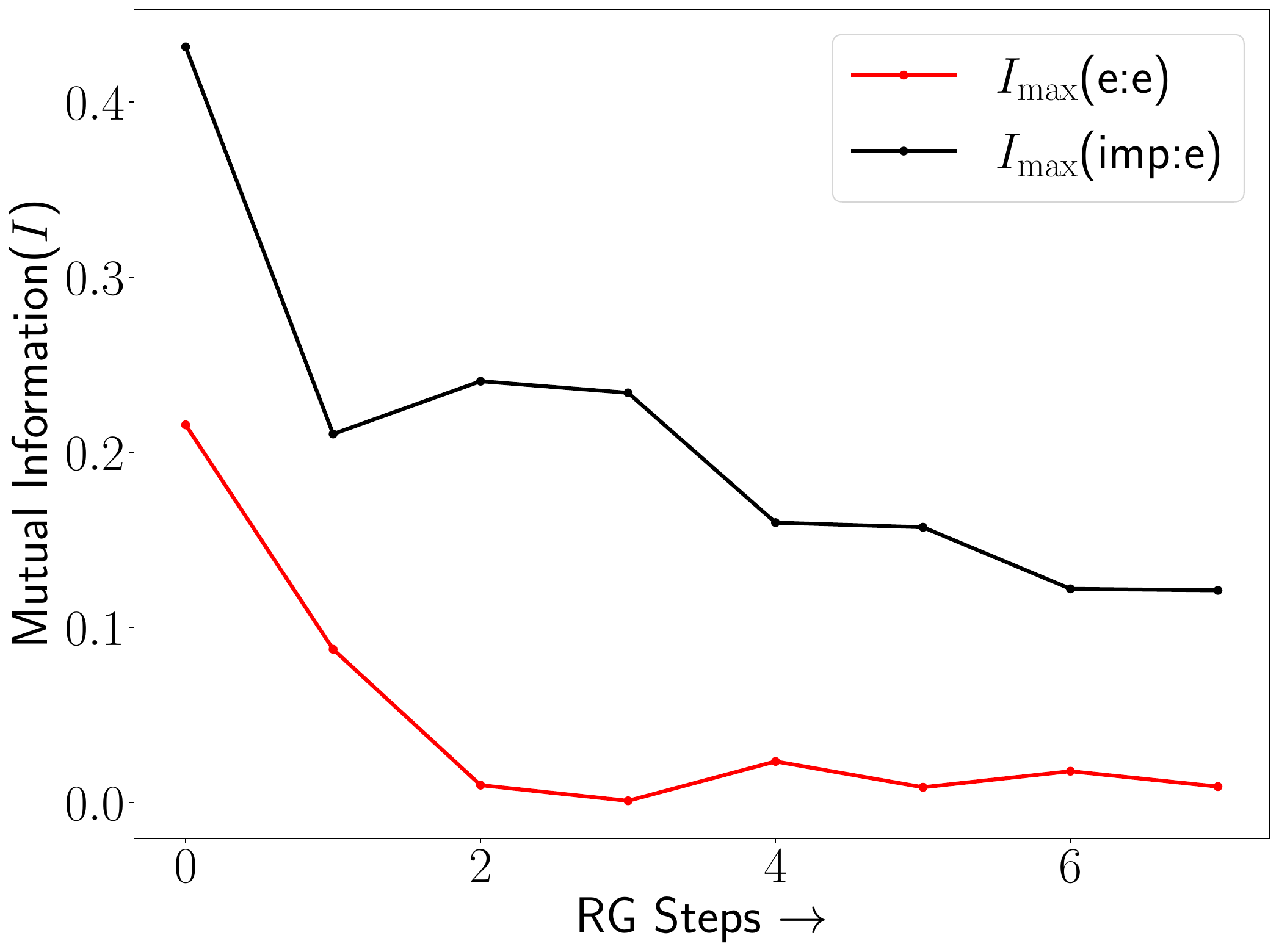}
\includegraphics[width=0.45\textwidth]{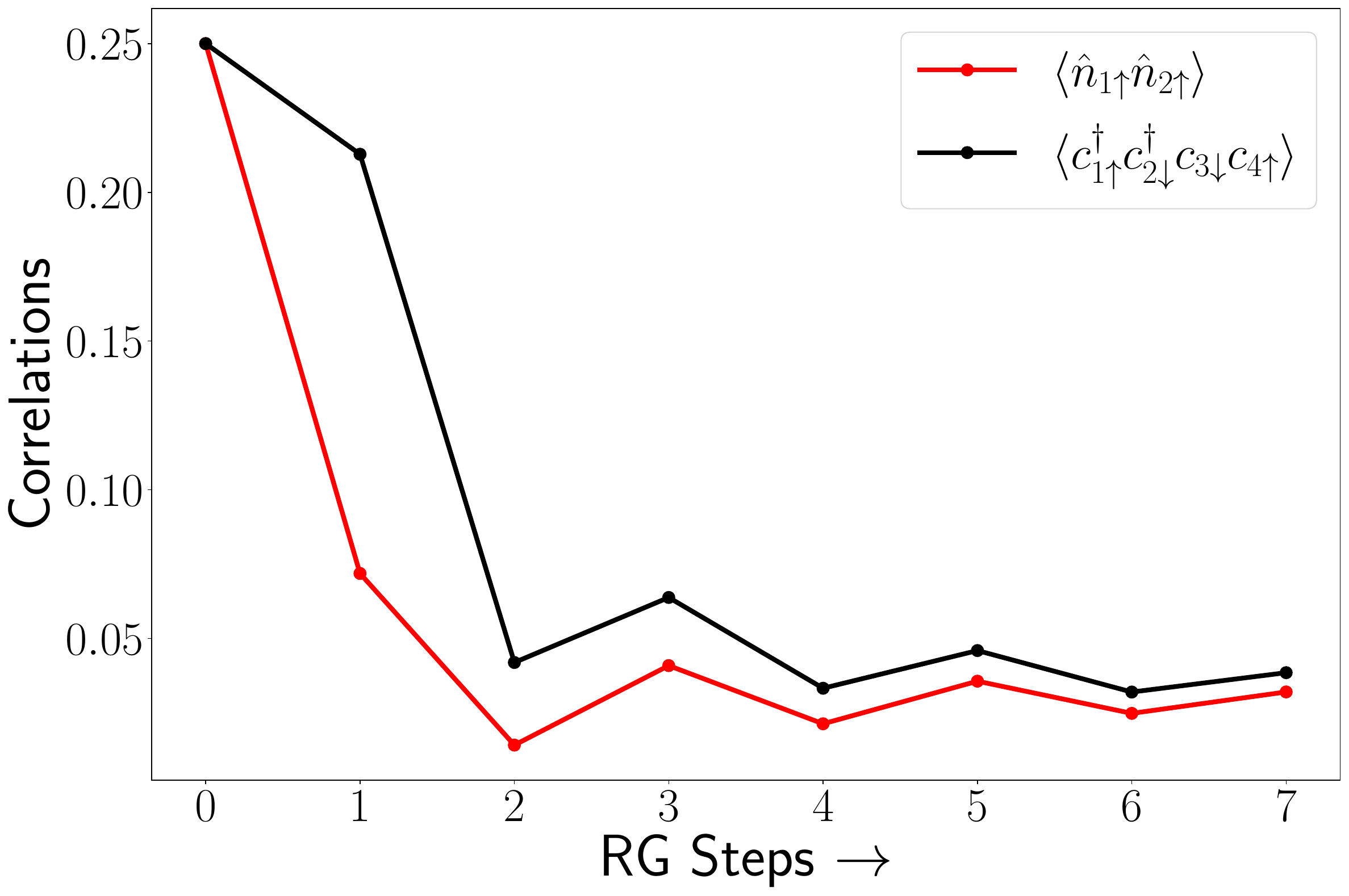}
\end{center}
\caption{Upper panel shows the RG flows for the maximum mutual information between the impurity and electron (black curve) and diagonal correlations (red curve). Lower panel shows the RG flows for the off-diagonal (black curve) and diagonal correlations (red curve).}\label{MI1}
\end{figure}

We use the wavefunctions generated under the reverse RG to compute the mutual information between (a) an electron in the Kondo cloud and the impurity electron, and (b) two electrons within the cloud. MI measures the total amount of quantum and classical correlations in a system~\cite{groisman2005}. The mutual information between two electrons is given by 
\begin{equation}\begin{aligned}
 I(i:j)=-Tr(\rho_{i}\ln\rho_{i})-Tr(\rho_{j}\ln\rho_{j})+Tr(\rho_{ij}\ln\rho_{ij})~,\label{MI}
\end{aligned}\end{equation}
where $\rho_{i}$ or $\rho_{j}$ and $\rho_{ij}$ are the $1$- and $2$-electron reduced density matrices respectively obtained from the wavefunctions obtained at each step of the reverse RG simulation. In Fig.\ref{MI1}, we present the RG flow of both types of mutual information mentioned above. The orange curve in Fig.\ref{MI1}(a) represents the plot for the maximum mutual information $I(e:e)$ (eq.\eqref{MI}) between any two of the electrons comprising the Kondo cloud, and shows that the maximum entanglement content/quantum correlation increases under RG flow from UV to IR. This implies that the electrons within the Kondo cloud is not simply a separable state in momentum space expected of a local Fermi liquid. This is a strong indication of the fact that the two-particle off-diagonal ($c^{\dagger}_{k_{4}\uparrow}c^{\dagger}_{k\downarrow}c_{k_{2}\downarrow}c_{k_{1}\uparrow}$) scattering term in eq.\eqref{eff_Ham_Kondo} is playing a role in the electronic entanglement within the Kondo cloud. Further, the blue curve in Fig.\ref{MI1}(a) shows that the maximum mutual information between the Kondo impurity and any one member of the electronic cloud also increases under RG and to a higher value compared to that between electrons. This originates from the maximally entangled singlet state that is formed between the impurity and the electronic cloud (as can also be observed from the $\ln 2$ entanglement entropy obtained by tracing out the impurity spin from the singlet state). In this manner, the Kondo impurity mediates the entanglement between electrons (orange curve) in the Kondo cloud.

In order to understand this further, we also study (i) the maximum density-density correlations \(\max_{k,k_{1}}\langle\hat{n}_{k\uparrow}\hat{n}_{k_{1}\downarrow}\rangle\) (blue curve in Fig.\ref{MI1}(b)), and (ii) the maximum two-particle off-diagonal correlations $\max_{k,k_{1}}\langle c^{\dagger}_{k\uparrow}c^{\dagger}_{k_{1}\downarrow}c_{k_{2}\downarrow}c_{k_{3}\uparrow}\rangle$ (orange curve in Fig.\ref{MI1}(b)) between electrons within the Kondo cloud. The plots show clearly that both the correlations grow under RG from UV to IR, finally reach the same value at the IR fixed point. On the other hand, the large values of the off-diagonal correlations reinforce our observation of a non-zero mutual information content between the cloud electrons. This implies that the electronic cloud contains, in general, interaction terms beyond the Fermi liquid density-density interaction leading to non-zero entanglement content.

\section{Conclusions}\label{ConcluKondo}
\par\noindent
The Kondo problem~\cite{kondo1964resistance} is one of the oldest and most well-studied problems of electronic correlations in condensed matter physics~\cite{anderson1970poor,wilson1975}. The unitary RG analysis of the Kondo Hamiltonian leads to a zero temperature phase diagram revealing a strong coupling fixed point for an antiferromagnetic Kondo coupling. At the IR fixed point, we obtained the effective Hamiltonian, the ground state wavefunction and the energy eigenspectrum. This enabled the computation of various  thermodynamic quantities such as the impurity susceptibility, specific heat coefficient, Wilson ratio, Wilson number and thermal entropy, all of which  are found to be in good agreement with that obtained from the NRG studies~\cite{bullaNRGreview}. It is also noteworthy that we were able to capture the entire RG (crossover) flow at finite temperatures from the complete effective Hamiltonian (eq.\eqref{kondototham}) obtained upon reaching the IR stable fixed point (as seen, for instance, for the susceptibility $\chi (T)$ in Fig.\ref{suscfig2}). As the URG relies purely on unitary transformations, the eigenspectrum is preserved under the RG flow, and thence so is the partition function. Thus, at each RG fixed point, the effective Hamiltonian eq.\eqref{kondototham} enables the construction of the density matrix for a given temperature scale $k_{B}T$, such that one can compute the finite temperature partition function (eq.\eqref{partfunc}).

Furthermore, we found that the effective Hamiltonian for the Kondo cloud (eq.~\ref{eff_Ham_Kondo}), obtained by integrating out the impurity spin, contains a density-density repulsion (corresponding to a Fermi liquid) as well as a four-fermion interaction term . In order to better understand the roles of the two types of electronic correlations, we performed a comparative study of the RG evolution of four-point number diagonal and number off-diagonal correlators. By using the singlet IR ground state wavefunction obtained from the URG analysis, we also studied the RG evolution of the mutual information (an entanglement based measure) between (a) the impurity and an electron in the cloud and (b) two electrons in the cloud. The results show strong inter-electronic as well as electron-impurity entanglement upon approaching the IR fixed point. This is in agreement with the presence of both types of two-particle correlators at the IR ground state. We find that both the number diagonal and number off-diagonal correlators reach same value at the IR fixed point, indicating that the electronic configuration within the Kondo cloud is not simply a local Fermi liquid. The large entanglement within the Kondo cloud is an indication of the spin singlet it forms together with the impurity spin. This is an example of quantum mechanics at the macro-scale: impurity-bath correlations drive many-particle entanglement between bath electrons, signaling the emergence of a collective spin that binds with the impurity. Given that the Kondo problem represents a simple model for a single qubit coupled to a fermionic environment, our insights can spur further investigations into more complex qubit-bath interactions relevant to the realization and workings of present-day quantum computers in noisy environments.

The calculation of impurity susceptibility in appendix \ref{chi} defines the Kondo temperature as that energy scale below which only the singlet state contributes to the susceptibility and the other states in the entangled 4-dimensional Hilbert space of the impurity spin and the total conduction electron spin drop out.
The subsequent calculation of the wavefunction renormalisation \(Z=1\) in appendix backs up the presence of a local Fermi liquid phase at the fixed point. 
In this way, the low energy physics of the Kondo model can be captured in great depth purely from the URG fixed point Hamiltonian. 
An important point can now be highlighted. 
The physics of the emergent Kondo cloud Hamiltonian (Sec.\ref{susc2}) is found to be in good agreement with that of the local Fermi liquid. 
This includes various thermodynamic quantities (e.g., the impurity susceptibility, specific heat, thermal entropy etc.) and the Wilson ratio (see appendices \ref{chi} and \ref{Cv}). 
We have found that the $k$-space Kondo cloud effective Hamiltonian describes an extended real-space object comprised of electron waves with extent greater than $\xi_{K}$. 
On the other hand, the local Fermi liquid is found to reside at a distance $\xi_{K}$ from the impurity. 
Thus, it appears tempting to conclude that there exists a holographic bulk-boundary relationship between the Kondo cloud system and the local Fermi liquid. 
We speculate that the emergent change in the Luttinger's volume (see eq.\ref{change_L_vol} in Appendix \ref{phase_shift}) of the conduction bath at the strong-coupling fixed point corresponds to the winding number topological quantity~\cite{seki2017topological} that signals such a holography. We do not know at present whether the effective theory for the Kondo cloud found by us can be related to a bulk gravity theory obtained from an AdS-CFT treatment~\cite{kim2019}. We leave this for a future work.

Future studies need to be performed for investigating the nature of the correlated metal comprising the Kondo cloud. 
This work also opens up the prospect of performing a similar study on other variants of the Kondo problem, such as the multi-channel Kondo model~\cite{Gan_Andrei_Coleman_1993,Noz_blandin_1980,emery_kivelson,Gan_mchannel_1994,Tsvelick_Weigmann_mchannel_1984,Tsvelick_weigmann_mchannel_1985,parcollet_olivier_large_N,kimura_taro_Su_N_kondo,PhysRevB.73.224445,cox_jarrell_two_channel_rev,affleck_1991_overscreen,Coleman_tsvelik}. 
The current method can allow one to investigate the critical properties of the over-screened multi-channel Kondo intermediate-coupling fixed point. The low-energy effective Hamiltonian of such models should be helpful in identifying the microscopic origins of the non-Fermi liquid phase of such a fixed point. Entanglement studies of the zero mode will also be helpful in identifying the breakdown of screening, and such studies will lay bare the distinction between various variants of the Kondo impurity models.

\acknowledgments
The authors thank P. Majumdar, A. Mitchell, S. Sen, S. Patra, M. Mahankali and R. K. Singh for several discussions and feedback. Anirban Mukherjee thanks the CSIR, Govt. of India and IISER Kolkata for funding through a research fellowship. Abhirup Mukherjee thanks IISER Kolkata for funding through a research fellowship. AM and SL thank JNCASR, Bangalore for hospitality at the inception of this work. NSV acknowledges funding from JNCASR and a SERB grant (EMR/2017/005398). We thank two anonymous referees for their helpful comments and suggestions.

\appendix
\begin{widetext}
\section{Calculation of effective Hamiltonian from URG}\label{urg-deriv}
Starting from the Kondo Hamiltonian eq.\eqref{KondoH} and using the URG based Hamiltonian RG equation eq.\eqref{HRG}, we obtain the renormalised Hamiltonian :
\begin{align}
\Delta\hat{H}_{(j)} &= \sum_{\substack{m=1,\\ \beta=\uparrow/\downarrow}}^{n_{j}}\frac{(J^{(j)})^{2}\tau_{j,\hat{s}_{m},\beta}}{2(2\omega\tau_{j,\hat{s}_{m},\beta} - \epsilon_{j,l}\tau_{j,\hat{s}_{m},\beta}-J^{(j)}S^{z}s^{z}_{j,\hat{s}_{m}})} \times  \bigg[S^{a}S^{b}\sigma^{a}_{\alpha\beta}\sigma^{b}_{\beta\gamma} \sum_{\substack{(j_{1},j_{2}< j),\\ n,o}}c^{\dagger}_{j_{1},\hat{s}_{n},\alpha}c_{j_{2},\hat{s}_{o},\gamma}(1-\hat{n}_{j,\hat{s}_{m},\beta})\nonumber\\
&+S^{b}S^{a}\sigma^{b}_{\beta\gamma}\sigma^{a}_{\alpha\beta} \sum_{\substack{(j_{1},j_{2}<j),\\ n,o}}c_{j_{2},\hat{s}_{o},\gamma}c^{\dagger}_{j_{1},\hat{s}_{n},\alpha}\hat{n}_{j,\hat{s}_{m},\beta}\bigg] + \sum_{\substack{m=1,\\ \beta=\uparrow/\downarrow}}^{n_{j}}\frac{(J^{(j)})^{2}}{2(2\omega\tau_{j,\hat{s}_{m},\beta} - \epsilon_{j,l}\tau_{j,\hat{s}_{m},\beta}-J^{(j)}S^{z}s^{z}_{j,\hat{s}_{m}})}\nonumber\\
&\times\bigg[S^{x}S^{y}\sigma^{x}_{\alpha\beta}\sigma^{y}_{\beta\alpha}c^{\dagger}_{j,\hat{s}_{m},\alpha}c_{j,\hat{s}_{m},\beta}c^{\dagger}_{j,\hat{s}_{m},\beta}c_{j,\hat{s}_{m},\alpha} + S^{y}S^{x}\sigma^{x}_{\alpha\beta}\sigma^{y}_{\beta\alpha}c^{\dagger}_{j,\hat{s}_{m},\beta}c_{j,\hat{s}_{m},\alpha}c^{\dagger}_{j,\hat{s}_{m},\alpha}c_{j,\hat{s}_{m},\beta}\bigg]~.
\label{major}
\end{align}
The first term in eq.\eqref{major} corresponds to the renormalisation of the Kondo coupling and describes the s-d exchange interactions for the entangled degrees of freedom
\begin{equation}\begin{aligned}
\Delta H^{1}_{(j)} &= \sum_{\substack{m=1,\\ \beta=\uparrow/\downarrow}}^{n_{j}}\frac{(J^{(j)})^{2}\tau_{j,\hat{s}_{m},\beta}}{(2\omega\tau_{j,\hat{s}_{m},\beta} - \epsilon_{j,l}\tau_{j,\hat{s}_{m},\beta}-J^{(j)}S^{z}s^{z}_{j,\hat{s}_{m}})} \sum_{\substack{(j_{1},j_{2}<j),\\ n,o}}\mathbf{S}\cdot c^{\dagger}_{j_{1},\hat{s}_{n},\alpha}\frac{\boldsymbol{\sigma}_{\alpha\beta}}{2}c_{j_{2},\hat{s}_{o},\beta}\\
&=\frac{1}{2}\sum_{\substack{m=1,\\~\beta=\uparrow/\downarrow}}^{n_{j}}\frac{\tau_{j,\hat{s}_{m},\beta}(J^{(j)})^{2}\left[(2\omega\tau_{j,\hat{s}_{m},\beta}-\epsilon_{j,l}\tau_{j,\hat{s}_{m},\beta})+J^{(j)}S^{z}s^{z}_{j,m})\right]}{(\omega - \frac{\epsilon_{j,l}}{2})^{2}-\frac{\left(J^{(j)}\right)^{2}}{16}}\sum_{\substack{(j_{1},j_{2}<j),\\ n,o}}\mathbf{S}\cdot c^{\dagger}_{j_{1},\hat{s}_{n},\alpha}\frac{\boldsymbol{\sigma}_{\alpha\beta}}{2}c_{j_{2},\hat{s}_{o},\beta}\\
&=\frac{1}{2}\left[\sum_{\substack{m=1,\\~\beta=\uparrow/\downarrow}}^{n_{j}}\frac{(J^{(j)})^{2}\left[(\frac{\omega}{2}-\frac{\epsilon_{j,l}}{4})\right]}{(\omega - \frac{\epsilon_{j,l}}{2})^{2}-\frac{\left(J^{(j)}\right)^{2}}{16}} + \frac{1}{2}\sum_{m=1}^{n_{j}}\frac{(J^{(j)})^{3}S^{z}s^{z}_{j,m}(\tau_{j,\hat{s}_{m},\uparrow}+\tau_{j,\hat{s}_{m},\downarrow})}{(\omega - \frac{\epsilon_{j,l}}{2})^{2}-\frac{\left(J^{(j)}\right)^{2}}{16}}\right]\sum_{\substack{(j_{1},j_{2}<j),\\ n,o}}\mathbf{S}\cdot c^{\dagger}_{j_{1},\hat{s}_{n},\alpha}\frac{\boldsymbol{\sigma}_{\alpha\beta}}{2}c_{j_{2},\hat{s}_{o},\beta}\\
&=\frac{n_{j}(J^{(j)})^{2}\left[(\omega - \frac{\epsilon_{j}}{2})\right]}{(\omega - \frac{\epsilon_{j,l}}{2})^{2}-\frac{\left(J^{(j)}\right)^{2}}{16}}\mathbf{S}\cdot \sum_{\substack{(j_{1},j_{2}<j),\\ n,o}} c^{\dagger}_{j_{1},\hat{s}_{n},\alpha}\frac{\boldsymbol{\sigma}_{\alpha\gamma}}{2}c_{j_{2},\hat{s}_{o},\gamma}~.\label{HamRG}
\end{aligned}\end{equation}
In the second last step of the calculation, we have used the result $\tau_{j,\hat{s}_{m},\uparrow}^{2}=\frac{1}{4}$, where $\tau_{j,\hat{s}_{m},\uparrow}=\hat{n}_{j,\hat{s}_{m},\uparrow}-\frac{1}{2}$. In obtaining the last step of the calculation we have assumed $\epsilon_{j,l}=\epsilon_{j}$ for a circular Fermi surface geometry. Further, we have replaced $\tau_{j,\hat{s}_{m},\uparrow}$ and $\tau_{j,\hat{s}_{m},\downarrow}$ by their eigenvalues, $\tau_{j,\hat{s}_{m},\uparrow}=-\tau_{j,\hat{s}_{m},\downarrow}=\frac{1}{2}$, i.e., the resulting decoupled electronic wave vector $|j,\hat{s}_{m}\rangle$ carries a non-zero spin angular momentum. This configuration promotes the spin scattering between the Kondo impurity and the fermionic bath. The second term in eq.\eqref{major} corresponds to the renormalisation of the number diagonal Hamiltonian for the immediately disentangled electronic states $|j,\hat{s}_{m},\sigma\rangle$:
\begin{align}
	\Delta H^{2}_{(j)} &=\sum_{\substack{m=1,\\ \beta=\uparrow/\downarrow}}^{n_{j}}\frac{(J^{(j)})^{2}}{(2\omega\tau_{j,\hat{s}_{m},\beta} - \epsilon_{j,l}\tau_{j,\hat{s}_{m},\beta}-J^{(j)}S^{z}s^{z}_{j,\hat{s}_{m}})} \bigg[S^{x}S^{y}\sigma^{x}_{\alpha\beta}\sigma^{y}_{\beta\alpha}c^{\dagger}_{j,\hat{s}_{m},\alpha}c_{j,\hat{s}_{m},\beta}c^{\dagger}_{j,\hat{s}_{m},\beta}c_{j,\hat{s}_{m},\alpha}\\
&+S^{y}S^{x}\sigma^{x}_{\alpha\beta}\sigma^{y}_{\beta\alpha}c^{\dagger}_{j,\hat{s}_{m},\beta}c_{j,\hat{s}_{m},\alpha}c^{\dagger}_{j,\hat{s}_{m},\alpha}c_{j,\hat{s}_{m},\beta}\bigg]\\
&=\sum_{\substack{m=1,\\ \beta=\uparrow/\downarrow}}^{n_{j}}\frac{(J^{(j)})^{2}}{(2\omega\tau_{j,\hat{s}_{m},\beta} - \epsilon_{j,l}\tau_{j,\hat{s}_{m},\beta}-J^{(j)}S^{z}s^{z}_{j,\hat{s}_{m}})}S^{z}\frac{\sigma^{z}_{\alpha\alpha}}{2}\bigg[\hat{n}_{j,\hat{s}_{m},\alpha}(1-\hat{n}_{j,\hat{s}_{m},\beta})-\hat{n}_{j,\hat{s}_{m},\beta}(1-\hat{n}_{j,\hat{s}_{m},\alpha})\bigg]\\
&=\sum_{m=1}^{n_{j}}\frac{(J^{(j)})^{2}}{(2\omega\tau_{j,\hat{s}_{m},\beta} - \epsilon_{j,l}\tau_{j,\hat{s}_{m},\beta}-J^{(j)}S^{z}s^{z}_{j,\hat{s}_{m}})}S^{z}s^{z}_{j,\hat{s}_{m}}~,
\end{align}
where we have used $\hat{n}_{j,\hat{s}_{m},\alpha}(1-\hat{n}_{j,\hat{s}_{m},\beta})-\hat{n}_{j,\hat{s}_{m},\beta}(1-\hat{n}_{j,\hat{s}_{m},\alpha})=\hat{n}_{j,\hat{s}_{m},\alpha}-\hat{n}_{j,\hat{s}_{m},\beta}$  in the last step, and the spin density for the state $|j,\hat{s}_{m}\rangle$ is given by $s^{z}_{j,\hat{s}_{m}}=\frac{1}{2}(\hat{n}_{j,\hat{s}_{m},\uparrow}-\hat{n}_{j,\hat{s}_{m},\downarrow})$. In obtaining the above RG equation we have replaced  $\hat{\omega}_{(j)}=2\omega\tau_{j,\hat{s}_{m},\beta}$. We set the electronic configuration $\tau_{j,\hat{s}_{m},\uparrow}=-\tau_{j,\hat{s}_{m},\downarrow}=\frac{1}{2}$ to account for the spin scattering between the Kondo impurity and the fermionic bath.  
The operator $\hat{\omega}_{(j)}$ (eq.\eqref{qfOp}) for RG step $j$ is determined by the occupation number diagonal piece of the Hamiltonian  $H^{D}_{(j-1)}$ attained at the next RG step $j-1$. This demands a self-consistent treatment of the RG equation to determine the $\omega$. In this fashion, two-particle and higher order quantum fluctuations are automatically encoded into the RG dynamics of $\hat{\omega}$. In the present work, however, we restrict our study by ignoring the RG contribution in $\omega$. The electron/hole configuration ($|1_{j,\hat{s}_{m},\beta}\rangle$/$|0_{j,\hat{s}_{m},\beta}\rangle$) of the disentangled electronic state (and associated with energy $\pm \epsilon_{j,l}$) is accounted by the fluctuation energy scales $\pm\omega$. 
From the above Hamiltonian RG equations eq.\eqref{HamRG}, we can obtain the form of the Kondo coupling RG equation (eq.\eqref{RGeqn})
\begin{eqnarray}
\Delta J^{(j)}=\frac{n_{j}(J^{(j)})^{2}\left[\omega- \frac{\epsilon_{j,l}}{2}\right]}{(\frac{\epsilon_{j,l}}{2}-\omega)^{2}-\frac{\left(J^{(j)}\right)^{2}}{16}}~.
\end{eqnarray}

It is easy to generalize this method to the more general anisotropic Kondo model with distinct transverse ($J_{\perp}$) and Ising ($J_{z}$) couplings:
\begin{equation}\begin{aligned}
J_z S^z \sum_{kk^\prime}\frac{1}{2}\left(c^\dagger_{k\uparrow}c_{k^\prime \uparrow} - c^\dagger_{k\downarrow}c_{k^\prime \downarrow}\right) + \frac{1}{2}J_{\perp} S^+ \sum_{kk^\prime}c^\dagger_{k \downarrow}c_{k^\prime \uparrow} + \frac{1}{2}J_{\perp} S^- \sum_{kk^\prime}c^\dagger_{k \uparrow}c_{k^\prime \downarrow}
\end{aligned}\end{equation}
We now briefly sketch the calculation.
\begin{itemize}
	\item Since the denominator involves the diagonal \(S^z s^z\) part of the Hamiltonian, only the Ising coupling \(J_z\) will enter the denominator of the RG equation. 
	\item The Ising coupling \(J_z\) will be renormalised only by scattering processes that involve one \(S^+\) operator and one \(S^-\) operator, so that the corresponding scattering vertices that renormalise \(J_z\) are of strength \({J_{\perp}}^2\).
	\item On the other hand, the transverse coupling \(J_{\perp}\) will be renormalised only by processes that involve one \(S^\pm\) operator and one \(S^z\) operator, so that the corresponding vertices are of strength \(J_z J_{\perp}\).
\end{itemize}
With these modifications, the final result of eq.~\ref{HamRG} becomes
\begin{equation}\begin{aligned}
	\frac{n_{j}(\omega - \frac{\epsilon_{j}}{2})}{(\omega - \frac{\epsilon_{j,l}}{2})^{2}-\frac{\left(J_z^{(j)}\right)^{2}}{16}}\left[\left(J^{(j)}_{\perp}\right)^{2}\sum_{\substack{(j_{1},j_{2}<j),\\ n,o,\alpha}} S^z \sigma^z_{\alpha\alpha}c^{\dagger}_{j_{1},\hat{s}_{n},\alpha}c_{j_{2},\hat{s}_{o},\alpha} + J^{(j)}_z J^{(j)}_{\perp}\left(\sum_{\substack{(j_{1},j_{2}<j),\\ n}} S^+ c^{\dagger}_{j_{1},\hat{s}_{n},\downarrow}c_{j_{2},\hat{s}_{o},\uparrow} + \text{h.c.}\right)\right]~.
\end{aligned}\end{equation}

\end{widetext}
The RG equations for the Ising and transverse couplings can now be read off from the renormalised Hamiltonian:
\begin{equation}\begin{aligned}
	\Delta J_z^{(j)}=\frac{n_{j}(J^{(j)}_{\perp})^{2}\left[\omega- \frac{\epsilon_{j,l}}{2}\right]}{(\omega-\frac{\epsilon_{j,l}}{2})^{2}-\frac{\left(J^{(j)}\right)^{2}}{16}}~,\\
	\Delta J_{\perp}^{(j)}=\frac{n_{j}J^{(j)}_z J^{(j)}_{\perp}\left[\omega- \frac{\epsilon_{j,l}}{2}\right]}{(\omega-\frac{\epsilon_{j,l}}{2})^{2}-\frac{\left(J^{(j)}\right)^{2}}{16}}~.
\end{aligned}\end{equation}
These coupled RG equations have the same RG invariant as the Berezinskii-Kosterlitz-Thouless (BKT) RG equations~\cite{berezinskii1971destruction,kosterlitz1978two} as well as the 1D Hubbard model RG equations~\cite{1dhubjhep}:
\begin{equation}\begin{aligned}
	\label{rg_inv}
	\frac{\Delta J_z^{(j)}}{\Delta J_{\perp}^{(j)}} = \frac{J^{(j)}_{\perp}}{J^{(j)}_z} \implies {J_{\perp}^{(j)}}^2 - {J_z^{(j)}}^2 = \text{constant}
\end{aligned}\end{equation}
One can also show that the anisotropy in the couplings is irrelevant by writing the RG equation for the anisotropy parameter \(\kappa = J_{\perp}^{(j)} - J_z^{(j)}\).
\begin{equation}
	\Delta \kappa=\frac{n_{j}J^{(j)}_t\left( J^{(j)}_{\perp} - J^{(j)}_z\right) \left[\omega- \frac{\epsilon_{j,l}}{2}\right]}{(\omega-\frac{\epsilon_{j,l}}{2})^{2}-\frac{\left(J^{(j)}\right)^{2}}{16}}
\end{equation}
Dividing this equation by the RG equation of \(J_z^{(j)}\) gives
\begin{equation}
	\frac{\Delta \kappa}{\Delta J_z^{(j)}} = -\frac{\kappa}{J_{\perp}^{(j)}}~.
\end{equation}
Since both \(\Delta J_z^{(j)}\) and \(J_{\perp}^{(j)}\) are positive in the strong-coupling regime, we find that \(\Delta \kappa / \kappa < 0\). This relation implies that if \(\kappa\) is initially positive (negative), the renormalisation is negative (positive), and \(\kappa\) flow towards zero. This shows that all strong-coupling flows enforce the irrelevance of the anisotropy parameter \(\kappa\).

\section{Effective Hamiltonian for low energy excitations of the Kondo singlet - the local Fermi liquid}
\label{lfl-kc}
The central resonance in the impurity spectral function can be attributed to quasiparticle excitations arising from the local Fermi liquid component of the effective Hamiltonian. Here we will derive an effective Hamiltonian for low-energy excitations in the singlet subspace of the fixed-point theory and hence expose the local Fermi liquid component.

To consider the local states as ground states, we will drop the kinetic energy part and start with the following zeroth Hamiltonian:
\begin{equation}
	\mathcal{H}^*_\text{loc} = \underbrace{J^* \vec{S_d}\cdot\vec{s}}_{H_0} \quad\underbrace{- t^*\sum_{\sigma, \left<0,l\right>} c^\dagger_{0\sigma}c_{l,\sigma} +\text{h.c.}}_{V}
\end{equation}
The four ground states of \(H_0\) are
\begin{equation}\begin{aligned}
	\ket{\phi^{(0)}_i} =& \underbrace{\ket{\chi_i}}_\text{site 1}\otimes\frac{1}{\sqrt 2}\left(\ket{\uparrow, \downarrow} - \ket{\downarrow, \uparrow} \right), i \in \left[1,4\right] ,\\
	\left\{\chi_i\right\} =& \left\{0, \uparrow, \downarrow, 2\right\} , E = -\frac{3J^*}{4}
\end{aligned}\end{equation}
In the singlet part, the first entry is the configuration of the zeroth site while the second entry is the configuration of the impurity site. 

Since \(V\) and \(V^3\) take \(\ket{\phi_{2,4}^{(0)}}\) to a completely orthogonal subspace, the first and third order corrections are 0. The second order shift in the ground state energy for a state \(\phi^{(0)}\) is given by
\begin{equation}\begin{aligned}
	E(2) = \sum_{i\neq \phi^{(0)}}\frac{|\braket{\phi^{(0)}|V|i}^2}{E^{(0)}_\phi - E_i}
\end{aligned}\end{equation}
This second order shift for both the up-spin and double-occupied states (and hence the down and empty states, using SU(2) and p-h symmetries) comes out to be
\begin{equation}\begin{aligned}
	E^{(2)} = -\frac{4t^2}{3J^*}
\end{aligned}\end{equation}
so that the effective Hamiltonian at second order is simply a constant. We move on to the fourth order correction. The general formula is quite formidable, so we only write down the terms that aren't outright zero for this problem:
\begin{equation}\begin{aligned}
	E^{(4)} = \sum_{i^{(0)}\neq \phi, \atop{j^{(0)}\neq \phi, \atop{k^{(0)}\neq\phi}}}\frac{V_{\phi,k}V_{k,j}V_{j,i}V_{i,\phi}}{E_{\phi,_k}E_{\phi,j}E_{\phi,i}} - E_\phi^{(2)}\sum_{m^{(0)}\neq\phi}\frac{|V_{m,\phi}|^2}{\left(E_{\phi,m}\right)^2}
\end{aligned}\end{equation}
where \(V_{x,y} = \bra{x^{(0)}}V\ket{y^{(0)}}\) and \(E_{x,y} = E_{x}^{(0)} - E_{y}^{(0)}\). At fourth order, \(\ket{\phi^{(0)}_2}\)is first excited to \(\ket{0, 2, \uparrow}\) or \(\ket{2, 0, \uparrow}\), and then to the spin-triplet subspace. The total effective Hamiltonian up to fourth order (and up to a constant energy shift) is
\begin{equation}\begin{aligned}
	H_\text{loc}^\text{eff} = - 4\alpha\frac{t^4}{{J^*}^3}\sum_\sigma \hat n_{1\sigma} + 8\alpha \frac{t^4}{{J^*}^3}\hat n_{1 \uparrow}\hat n_{1 \downarrow}
\end{aligned}\end{equation}
The presence of the repulsive correlation term means that electrons that want to occupy the first site will face a repulsion from an electron already present there and the effective  Hamiltonian behaves like a local Fermi liquid \cite{nozieres1974fermi} on the first site.
\section{Scattering phase shift of conduction electrons at strong-coupling and wavefunction renormalisation}
\label{phase_shift}
In general, the conduction electrons will suffer a phase shift as they scatter off the impurity. We will obtain an explicit form for this phase shift in terms of the fixed point coupling \(J^*\), by writing down a simple Hamiltonian that models the low energy theory. We will drop the states of the triplet subspace and retain only the singlet state and the double and hole states of the zeroth site. In order to avoid any two-particle terms, we will drop the impurity site because it is already frozen in the singlet. Also, since the double and hole states are at zero energy, they will drop out of the Hamiltonian, and we will model the zeroth site in the form of a single spin. We define \(\ket{0_\sigma}\) as the state where the zeroth site has a single electron of spin \(\sigma\) and \(\ket{0_0}\) and \(\ket{0_2}\) as the states with 0 and 2 electrons on the zeroth site respectively. Then,
\begin{equation}\begin{aligned}
	\text{spectrum of zeroth site}\to
	\begin{cases}
	\text{state} & \text{energy}\\
	\ket{0_\sigma} & -\frac{3J^*}{4}\\
	\ket{0_0}, \ket{0_2} & 0\\
	\end{cases}
\end{aligned}\end{equation}
 The total simplified Hamiltonian can therefore be written completely in terms of the single-particle states \(\ket{0_\sigma}\) and \(\ket{k_\sigma}\). \(\ket{k_\sigma}\) is the conduction electron state of momentum \(k\) and spin \(\sigma\). This Hamiltonian preserves the spin of the zeroth site and the conduction electrons separately, and the Hamiltonian for a particular spin is
\begin{equation}\begin{aligned}
\mathcal{H}_\sigma = \epsilon_0 \ket{0_\sigma}\bra{0_\sigma} +\sum_{k}\left[- t\left(\ket{0_\sigma}\bra{k_\sigma} + \text{h.c.} \right) + \epsilon_k \ket{k_\sigma}\bra{k_\sigma}\right]
\end{aligned}\end{equation}
where \(\epsilon_0 = -\frac{3J^*}{4}\) is the effective onsite energy of the zeroth site. The matrix elements of the retarded single particle Greens function for spin \(\sigma\), \(G^\sigma\), satisfy the equation \(\sum_\beta \left( \omega - \mathcal{H}_\sigma \right)_{\alpha\beta} G_{\beta \gamma} = \delta_{\alpha \gamma}\), where \(\alpha,\beta\) and \(\gamma\) represent matrix elements between any pair of states among \(0_\sigma\) and \(\left\{ k_\sigma \right\} \), and \(\delta\) is the Kronecker delta~\cite{phillips2012advanced}. Solving the equations for \(G_{00}^\sigma\) gives
\begin{equation}\begin{aligned}
	G_{00}^\sigma(\omega) = \frac{1}{\omega  - \epsilon_0 - \Sigma(\omega)}
\end{aligned}\end{equation}
\(\Sigma(\omega) = \sum_k \frac{t^2}{\omega  - \epsilon_k} = \int d\epsilon \frac{t^2\rho(\epsilon)}{\omega  - \epsilon}\) is the self-energy of the zeroth site of the bath because of hybridisation with the rest of the bath through the hopping term \(t\). The imaginary part is given by \(\Sigma_I(\omega) = -\pi\rho(\omega)t^2\).
Since \(\epsilon_0 \gg t^2\), we can drop the real part of the self energy in the denominator. The \(T-\)matrix of the conduction electrons can now be calculated using this impurity Greens function \cite{coleman2015}:
\begin{equation}\begin{aligned}
	T_{kk^\prime}^\sigma(\omega) = t^2 G_{00}(\omega) = t^2 \frac{\omega - \epsilon_0 + i\Sigma_I}{\left( \omega - \epsilon_0 \right)^2 + \Sigma_I^2 }
\end{aligned}\end{equation}
From scattering theory, it can be shown that the phase shift \(\delta^\text{FL}_{k\sigma}\) of the conduction electron state \(\ket{k\sigma}\) is given by the phase of \(T_{kk}\) \cite{hewson1993}:
\begin{equation}\begin{aligned}
	\label{phase}
	\tan \delta^\text{FL}_{k\sigma}(\omega) = \frac{\Sigma_I}{\omega - \epsilon_0} = \frac{-\pi\rho(\omega)t^2}{\omega + \frac{3J^*}{4}}
\end{aligned}\end{equation}
Note that there is an additional phase shift that we have not accounted for; when we removed the zeroth site from the lattice, all the wavefunctions got shifted by a distance of \(a\): \(\psi_k(x) \sim \sin\left( kx \right)  \to \sin\left(kx - ka\right)\). This is equivalent to a phase shift of \(\delta^0_{k\sigma} = ka\). The total phase shift is therefore
\begin{equation}\begin{aligned}
	\label{delta_tot}
	\delta^\text{tot}_{k\sigma}(\omega) \equiv \delta^0_{k\sigma} + \delta^\text{FL}_{k\sigma}(\omega) = ka + \tan^{-1}\frac{-\pi\rho(\omega)t^2}{\omega + \frac{3J^*}{4}}
\end{aligned}\end{equation}
Further, following the arguments involving the Friedel sum rule given in Ref.\cite{langreth1966}, we know that the total scattering phase shift in the ground state of the single impurity Anderson model (SIAM) at the Fermi surface is equal to \(\pi\) times the impurity occupation \(n_d\). The Kondo model corresponds to the local moment regime of the SIAM, and possesses a value of $n_{d}=1$. It then follows that
\begin{equation}\begin{aligned}
	\label{delta_tot_kf}
	\delta^\text{tot}_{k_F}(0) = \pi \times n_d = \pi~.
\end{aligned}\end{equation}
Further, we know that since $k_{F\sigma}=1/2a$ for a half-filled one-dimensional Fermi volume (for a $s$-wave Kondo exchange coupling), the phase shift $\delta^{0}_{k_{F}}$ is given by
\begin{equation}\begin{aligned}
	\label{delta_zero_kf}
\delta^{0}_{k_{F}} = \sum_{\sigma}\delta^{0}_{k_{F}\sigma} = \sum_{\sigma} k_{F\sigma}a = \pi = \delta^\text{tot}_{k_F}(0)~.
\end{aligned}\end{equation}
Substituting \(\delta^{0}_{k_{F}} = \delta^\text{tot}_{k_F}(0)\) into eq.~\ref{delta_tot} then indicates that the scattering phase shift for the Landau quasiparticle \(\delta^\text{FL}\) vanishes identically at the Fermi surface, and the total phase shift is determined purely by \(\delta^0_{k_{F}}\). 

The phase shift can also be connected to the overlap integral between the ground states \(\ket{\Psi_0}\) and \(\ket{\Psi}\) before and after adding the local interaction respectively. 
The local interaction can either take the form of the Kondo coupling between the impurity and the Kondo cloud, or a hopping between the zeroth site and the rest of the conduction bath. 
In the former case, the ground state \(\ket{\Psi_0}\) is that of a decoupled conduction bath, and \(\ket{\Psi}\) corresponds to the ground state of the conduction electrons in the presence of a strong Kondo coupling. 
In the latter case, \(\ket{\Psi_0}\) is the ground state of a conduction bath from which the zeroth site (i.e., the site strongly coupled with the impurity) is removed, and \(\ket{\Psi}\) is that of the same system but with a small hopping between the zeroth and first sites of the conduction gas. 

Following Anderson's orthogonality theorem~\cite{anderson1967infrared}, it's extension by Yamada and Yosida~\cite{yamada_catastrophe} and the generalized Friedel sum rule due to Langer and Ambegaokar~\cite{langer1961friedel}, the square of the overlap integral (often called the wavefunction renormalisation) is given by
\begin{equation}\begin{aligned}
	Z \equiv |\braket{\Psi|\Psi_0}|^2 = N^{-\frac{1}{\pi^2}\delta_{k_F}(0)^2}~,
\end{aligned}\end{equation}
where \(N\) is the total number of conduction electrons and \(\delta_{k_F}(0)\) is the phase shift produced by the local perturbation at the Fermi surface. 
In the thermodynamic limit, this expression simplifies to 
\begin{equation}\begin{aligned}
Z(N \to \infty) = 1 \text{ if }\delta_{k_F}(0) = 0, \text{ and } 0 \text{ otherwise}.
\end{aligned}\end{equation}
For the first case of the local interaction mentioned above (i.e., for the Kondo coupling), we obtain the wavefunction overlap (\(Z_\text{imp}\)) in terms of the total phase shift: \(Z_\text{imp} = N^{-\frac{1}{\pi^2}\left(\delta^\text{tot}_{k_F}(0)\right)^2} \to 0\) in the limit $N\to\infty$. \textit{This shows the orthogonality catastrophe between the ground states of the conduction bath in the local-moment and strong-coupling phases.} Further, following~\cite{martin-PhysRevLett.48.362}, the orthogonality catastrophe reflects a change in the Luttinger volume ($\Delta N_{L}$) of the conduction bath: 
\begin{equation}
	\label{change_L_vol}
\Delta N_{L} = \frac{\delta_{k_{F}}(0)}{\pi} = 1~.
\end{equation}
This increase in the number of electrons inside the Fermi volume is nothing  but the ``large Fermi surface" effect seen in heavy fermionic systems \cite{fujimoto_satoshi_1997,Shiina1993,lacroix_cyrot_1979,budin_grempel_2002}. Here the effect is infinitesimal because of the local nature of the impurity.

In the second case (i.e., with the zeroth site as the local perturbation), the relevant phase shift is \(\delta^\text{FL}_{k_F}(0) = 0\), and the overlap integral then pertains to the quasiparticle residue of the local Fermi liquid: \(Z_\text{FL} = N^{-\frac{1}{\pi^2}\left(\delta^\text{FL}_{k_F}(0)\right)^2}=1\)~. Further, from Ref.\cite{hewson1993}, the Wilson ratio ($R$) of the local Fermi liquid is given by 
\begin{eqnarray}
R &=& 1 + \sin^{2}(\frac{\delta_{k_{F}}(0)}{2})= 1 + \sin^{2}(\frac{\pi}{2}) = 2~. 
\end{eqnarray}

\section{Calculation of thermodynamic quantities}
\subsection{Impurity contribution to the magnetic susceptibility}
\label{chi}
The complete effective Hamiltonian for the impurity spin ($\mathbf{S}$), Kondo cloud spin ($\mathbf{s}$) and the electrons that comprise the local Fermi liquid has the form,
\begin{equation}
H_{2}=\epsilon_{*}\sum_{m,\sigma}\hat{n}_{*,m,\sigma}+J^{*}\mathbf{S}\cdot\mathbf{s}+J^{*}S^{z}\sum_{m}s^{z}_{*,m}~.
\label{kondototham}
\end{equation}
The Hamiltonian $H_{2}$ has several conserved quantities which we depict below,
\begin{equation}
[H_{2},S^{z}+s^{z}]=0~,~ [H_{2},s^{z}_{*,m}]=0 ~\forall ~1\leq m\leq n_{j}~,
\end{equation}
such that $[H^{*},S^{z}_{tot}]=0$ where $S^{z}_{tot}=S^{z}+s^{z}+\sum_{m}s^{z}_{*,m}$. Therefore, the eigenvalues of $|s^{z}_{*,m}=\uparrow/\downarrow\rangle$  are good quantum numbers; this is simply an outcome of the URG method. For the purposes of computing the impurity magnetization and susceptibility, we keep only the effective impurity - Kondo electron cloud Hamiltonian from $H_{2}$ above
\begin{eqnarray}
H^{*}_{K}=J^{*}\mathbf{S}\cdot\mathbf{s}+BS^{z}~,
\end{eqnarray}
and where we have introduced a local magnetic field $B$ that couples to the impurity magnetic moment through a Zeeman coupling.

We can now obtain the impurity magnetization and susceptibility from the effective Hamiltonian $H^{*}_{K}$. The four state eigenspectrum of $H^{*}_{K}$ is given by
\begin{equation}\begin{aligned}
	E_{1,2}&=\frac{1}{2}(-\frac{J^{*}}{2}\pm\sqrt{B^{2}+J^{*2}}), &E_{3,4}= \frac{1}{4}J^* - \frac{1}{2}B,
\end{aligned}\end{equation}
The partition function for this Hamiltonian (with $\beta=\frac{1}{k_{B}T}$) is given by
\begin{equation}\begin{aligned}
	Z(B)=2e^{-\beta\frac{J^{*}}{4}}\cosh(\beta\frac{B}{2}) + 2e^{\beta\frac{J^{*}}{4}}\cosh(\frac{\beta}{2}(\sqrt{B^{2}+J^{*2}}))~.
\end{aligned}\end{equation}
The susceptibility is then given by,
\begin{equation}
	\chi =\lim_{B\to 0}\frac{d}{dB}\left(\frac{k_{B}T}{Z(B)}\frac{dZ(B)}{dB}\right) = \frac{\frac{\beta}{4}+\frac{1}{2J^{*}}e^{\beta\frac{J^{*}}{2}}\sinh(\frac{\beta}{2}J^{*})}{1+e^{\beta\frac{J^{*}}{2}}\cosh(\frac{\beta}{2}J^{*})}~.
\label{susc}
\end{equation}

\begin{figure}[!htpb]
\begin{center}
\includegraphics[width=0.45\textwidth]{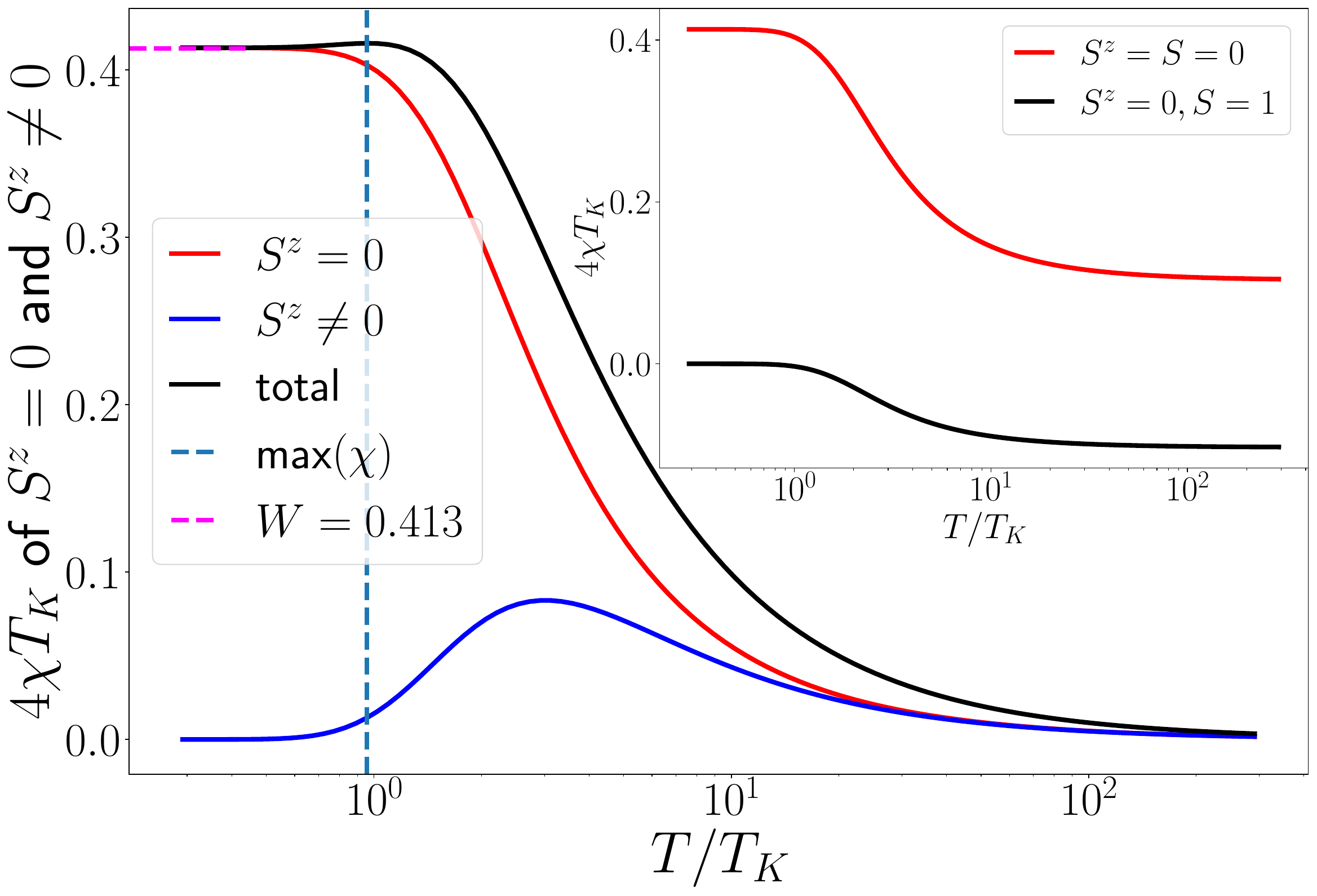}
\end{center}
\caption{Variation of $4T_{K}\chi$ with $T/T_{K}$, for individual spin subspaces as well as the total. $T_{K}$ corresponds to the Kondo temperature. Inset shows the difference in the contributions to \(\chi\) coming from the singlet and the triplet zero states. Dashed line shows the maximum in the \(\chi\), and passes through \(T_K\).}
\label{suscfig1}
\end{figure}

In the plot of \(4\chi T_K\) (fig.~\ref{suscfig1}), the blue and orange lines represent the contributions to the total susceptibility coming from the \(S^z = 0\) and \(S^z \neq 0\) sectors respectively. The green line is the total susceptibility itself. The inset shows a further resolution of the \(S^z = 0\) sector into the \(S = 0\) and \(S = 1\) contributions. We can now make several important observations based on eq.\ref{susc} and the plot. Firstly, at low temperatures \(T \ll T_K\), almost the entirety of the contribution comes from the singlet state. Near \(T_K\), the singlet contribution starts dropping while the \(S^z \neq 0\) contributions start to pick up. In this way, the Kondo temperature scale \(T_K\) arises out of the interplay between symmetry-preserved and the symmetry-broken sectors. Coming down from high temperatures, \(T_K\) can be thought of as that temperature beyond which the susceptibility contribution from the symmetry-broken states vanishes.

\begin{figure}[!htpb]
\begin{center}
\includegraphics[width=0.45\textwidth]{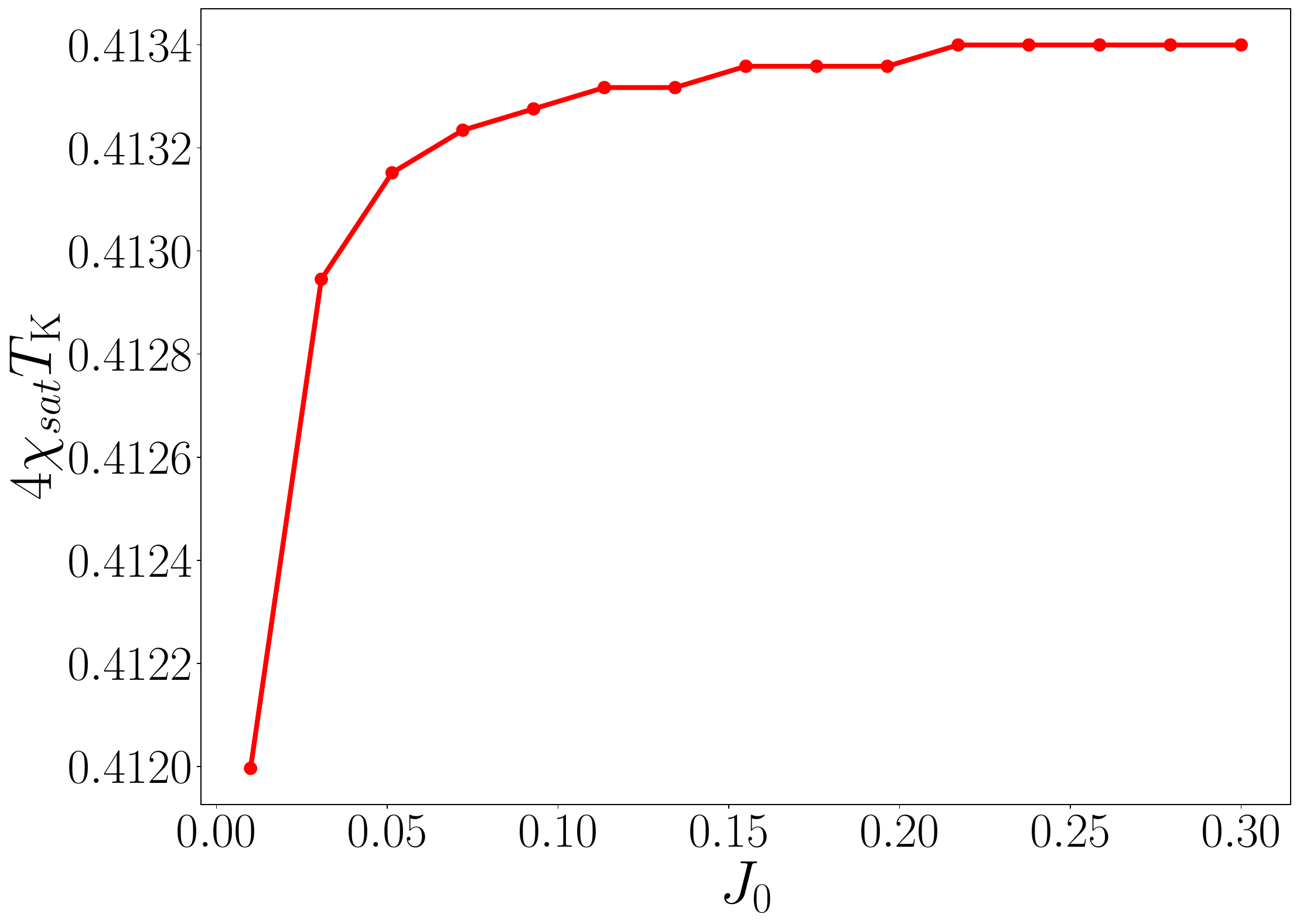}
\end{center}
\caption{Variation of the Wilson number $W = 4T_\text{K}\chi_\text{sat}$ with bare Kondo coupling $J_{0}$. $T_{K}$ corresponds to the Kondo temperature. See discussion in text.}\label{suscfig2}
\end{figure}

Secondly, the saturation value of $4T_{K}\chi$ as $\beta\to \infty$ (i.e., $T\to 0$) is given by
\begin{eqnarray}
	\chi_\text{sat} = \chi(T=0)=\frac{1}{2J^{*}}~.
\end{eqnarray}
We find that the Wilson number $W=4T_{K}\chi_\text{sat}$ takes the form $\frac{2T_{K}}{J^{*}}$. For values of $J^{*}\simeq 16.612t$ and $T_{K}\simeq 3.433t$ we obtain $W=0.413$. This is in good agreement with the value for $W$ obtained from NRG~\cite{bullaNRGreview} and Bethe Ansatz solution of the Kondo problem~\cite{andreiKondoreview,tsvelickKondoreview}. This is shown in Fig.\ref{suscfig1}. Further, we show the variation of $W$ with the bare Kondo coupling $J$ in Fig.\ref{suscfig2}. The figure clearly shows the saturation of $W$ to the value mentioned above as $J$ flows to the strong-coupling fixed point.

Thirdly, the $4T_{K}\chi$ vs temperature curve shown in Fig.\ref{suscfig1} has a non-monotonic behaviour, i.e., we find a maximum obtained from the transcendental equation \(\frac{d\chi}{d\beta} = 0\).
We confirm from our numerical studies that the temperature corresponding to the maximum value of $T_{K}\chi$ tends to $T_{K}$ as $J$ flows to strong coupling. Further, we find that the maximum value of $T_{K}\chi$ does not vary with the bare $J$ in the range $\mathcal{O}(10^{-5})<J<\mathcal{O}(1)$.
As discussed before, this maximum emerges as we go to lower temperatures and indicates that temperature at which the contributions from the \(\ket{\uparrow, \uparrow}\) and \(\ket{\downarrow, \downarrow}\) drop out. Thus, the non-monotonic behaviour in \(\chi\) can be directly attributed to the corresponding non-monotonic behaviour in the contribution coming from these states. 
We note that the impurity susceptibility obtained from NRG treatments of the Kondo problem appear to resemble that obtained by us for the $S_{z}=0$ sector (red curve in Fig.\ref{suscfig1}) rather than the total susceptibility. We do not presently understand the reason for this discrepancy. However, it is also interesting to note that a similar non-monotonic behaviour of the impurity susceptibility is obtained from a Schwinger boson large-N mean-field approach to the fully screened Kondo model \cite{rech_coleman_parcollet_2006_prl}.
\begin{figure}[htp]
\includegraphics[width=0.45\textwidth]{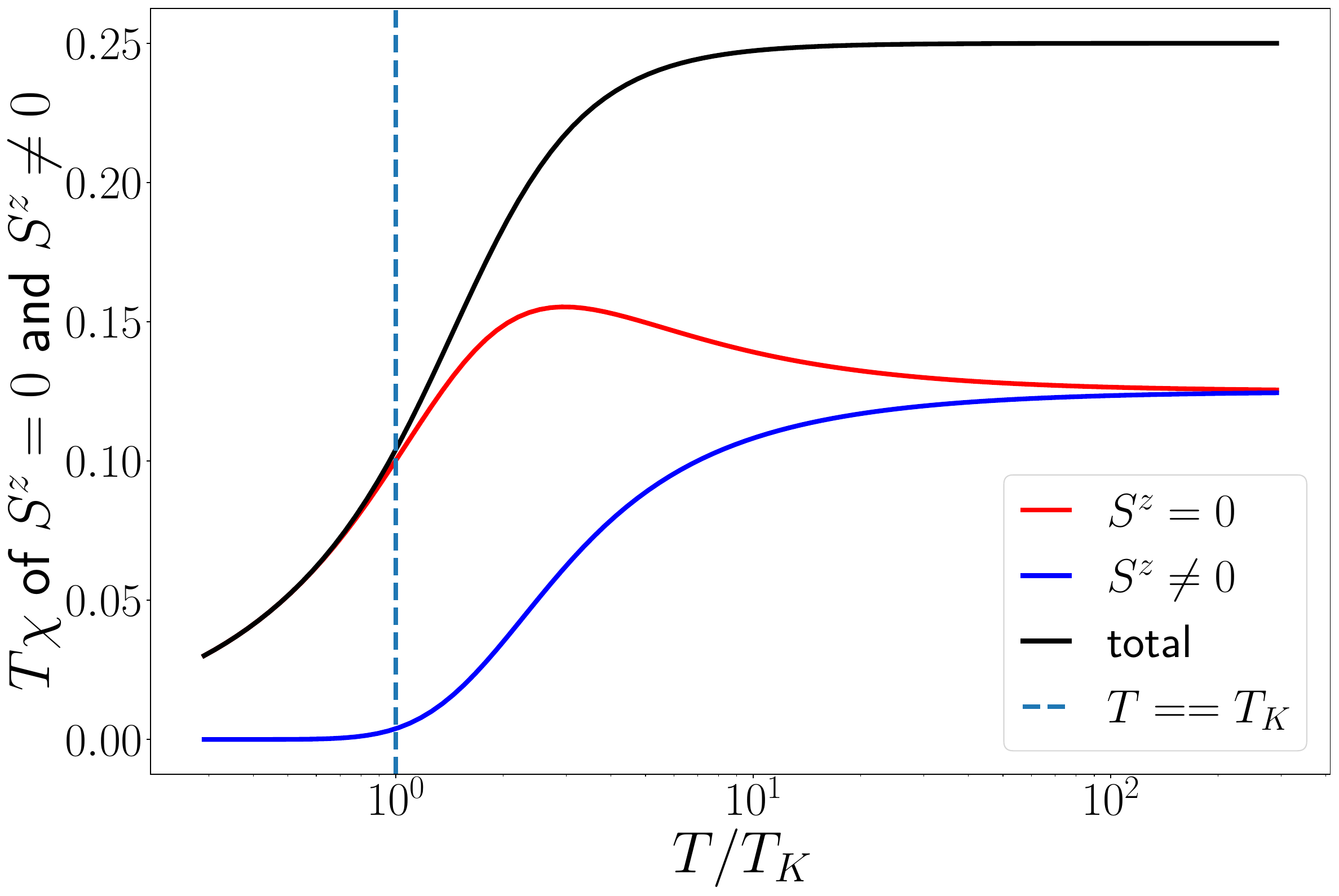}
\caption{Variation of $T\chi$ with $T/T_{K}$. $T_{K}$ corresponds to the Kondo temperature. See discussion in text.}\label{suscfig3}
\end{figure}

Finally, we find that the saturation value of $T\times\chi$ for $\beta\to 0$ is given by the universal value \(k_{B}T\chi(T=\infty)=\frac{1}{4}\), as shown in Fig.\ref{suscfig3}. The saturation value at high-$T$ arises from the two-fold degenerate impurity and reflects the physics of the (almost isolated) local impurity spin moment, while the vanishing value of $T\chi$ originates from the formation of the singlet between the Kondo cloud and the impurity spin (second term in eq.\eqref{kondototham}).

\subsection{Impurity contribution to specific heat and thermal entropy}
\label{Cv}
\begin{figure}[!htpb]
\centering
\includegraphics[width=0.45\textwidth]{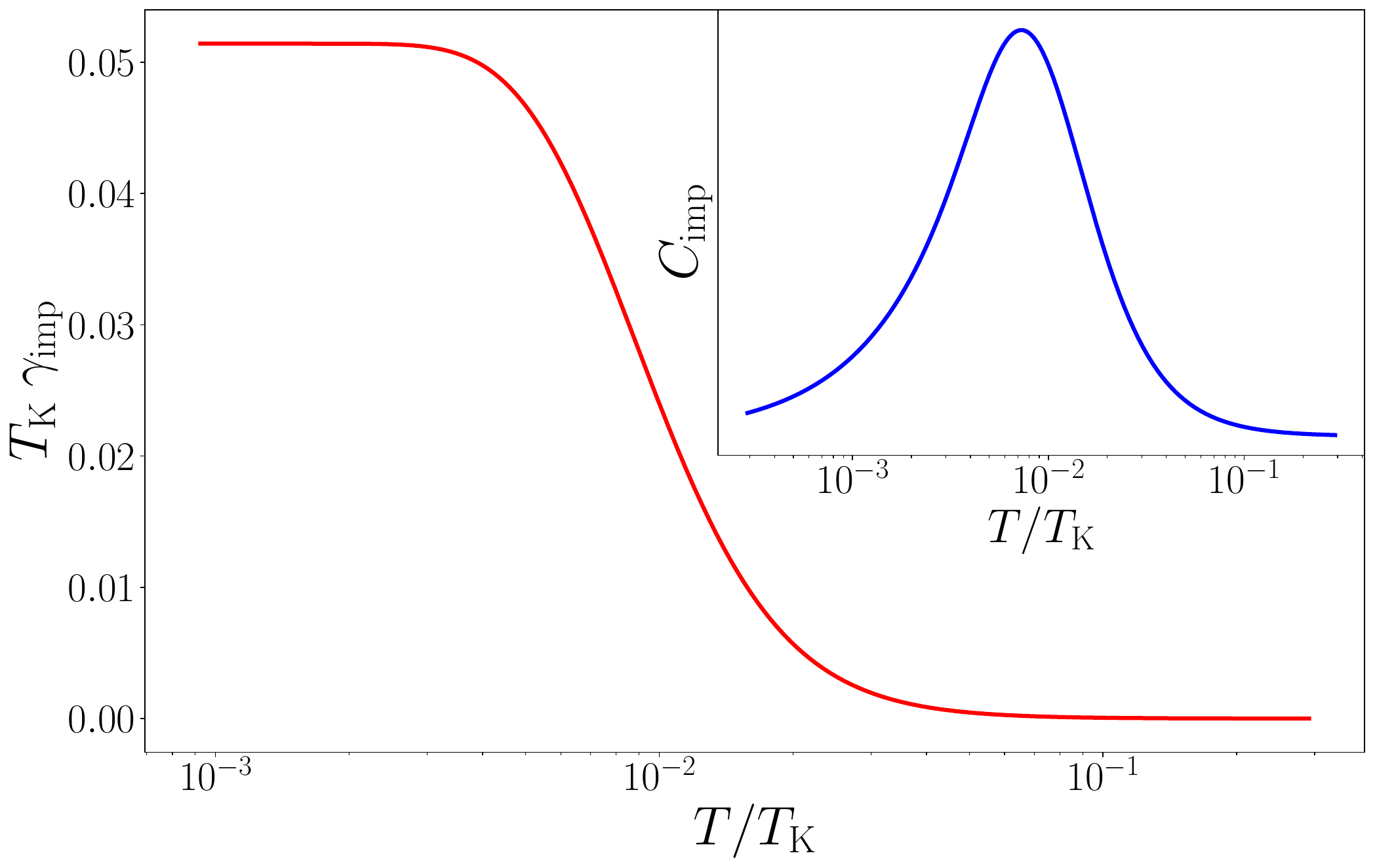}
\caption{Variation of $\gamma_\text{imp}T_{K}$ with $T/T_{K}$. $T_{K}$ corresponds to the Kondo temperature. The saturation to a constant value at low temperatures shows the Fermi liquid-like behaviour. Inset shows the impurity specific heat \(C_\text{imp}\).}
\label{specheat}
\end{figure}
\begin{figure}[!htpb]
\centering
\includegraphics[width=0.45\textwidth]{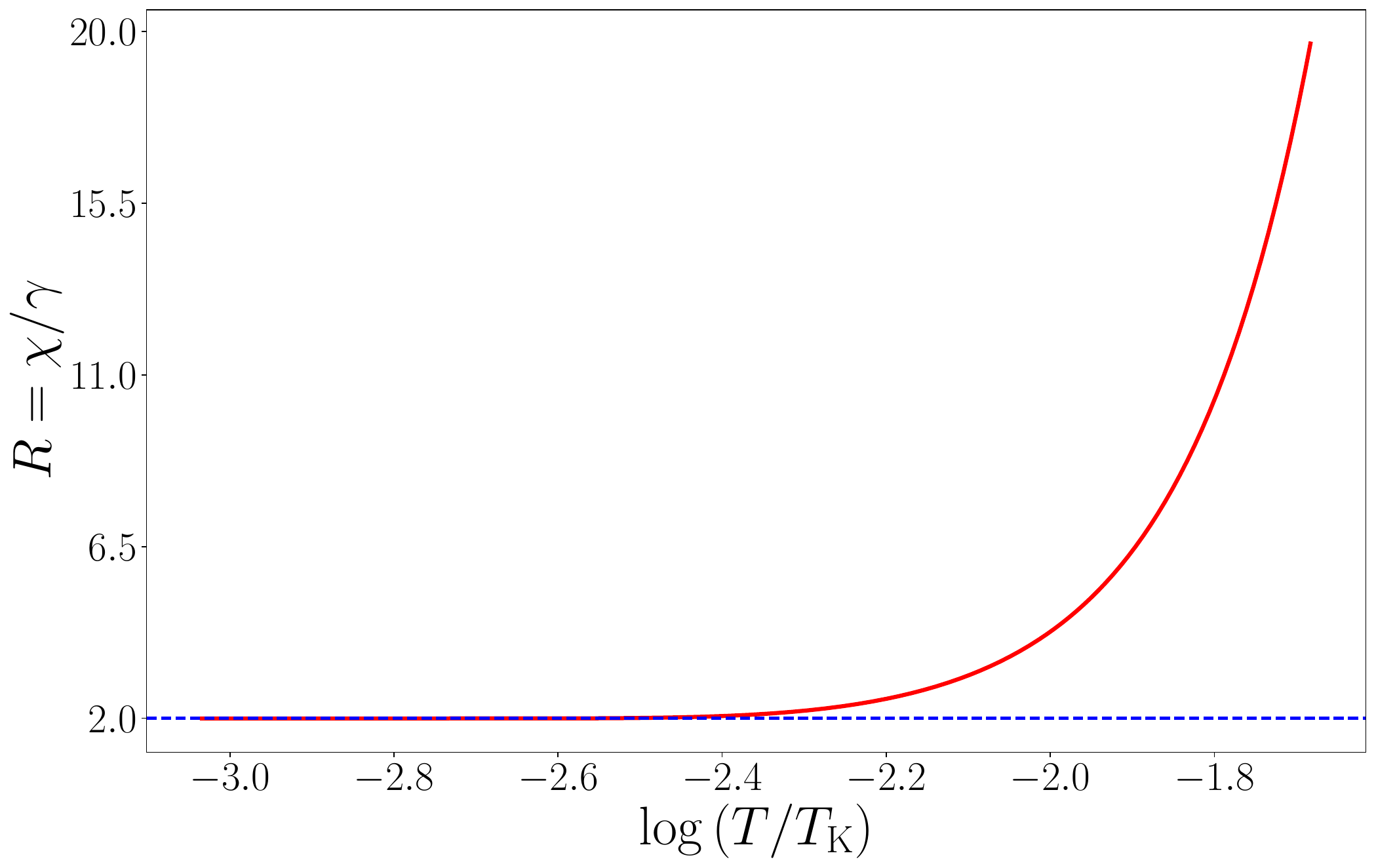}
\caption{Variation of the Wilson ratio $R$ with $T/T_{K}$. \(R\) saturates to a value of 2.009 as \(T \to 0\).}
\label{wilsonratio}
\end{figure}
In order to study the thermodynamic properties of the Fermi liquid we restrict our attention to the density-density terms only. From the density terms we obtain the low excitation energy functional accounting for the quasiparticle interaction
\begin{equation}
E=E_{0}+\sum_{\mathbf{k}_{\Lambda\hat{s}},\Lambda<\Lambda^{*}}\epsilon_{\mathbf{k}}\delta n_{\mathbf{k}\sigma}+\sum_{\mathbf{k},\mathbf{k}'}\frac{\epsilon_{\mathbf{k}}\epsilon_{\mathbf{k}'}}{J^{*}}\delta n_{\mathbf{k}\sigma}\delta n_{\mathbf{k}'\sigma'}~.
\label{localfermiliq}
\end{equation}
This leads to the renormalised one-particle dispersion, where the  self energy term has the following form
\begin{equation}
\bar{\epsilon}_{\Lambda}=\epsilon_{\Lambda}+\Sigma_{\Lambda} ,\Sigma_{\Lambda}=(\sum_{\Lambda',\hat{s},\sigma'}\frac{\epsilon_{\Lambda}\epsilon_{\Lambda'}}{J^{*}}\delta n_{\Lambda',\hat{s},\sigma'}) \text{ for }\Lambda<\Lambda^{*}~. 
\end{equation}
Note that as $\Lambda\to 0$ , $\Sigma_{\Lambda}\to 0$. Next, we compute the specific heat \(C_\text{imp} \left(\equiv C(J^{*})-C(0)\right)\) of the impurity from the Fermi distribution of the renormalised quasiparticles
\begin{equation}\begin{aligned}
C_\text{imp} =\sum_{\Lambda,\hat{s},\sigma}\frac{1}{T^2}\left[\frac{(\bar{\epsilon}_{\Lambda})^{2}e^{\beta\bar{\epsilon}_{\Lambda}}}{(e^{\beta\bar{\epsilon}_{\Lambda}}+1)^{2}}-\frac{(\epsilon_{\Lambda})^{2}e^{\beta\epsilon_{\Lambda}}}{(e^{\beta\epsilon_{\Lambda}}+1)^{2}}\right]~,
\end{aligned}\end{equation}
where $C(J^{*})$ is the specific heat for the electronic system with the Kondo impurity, and $C(0)$ is the specific heat for the free electronic system without coupling to Kondo impurity. The specific heat coefficient is given by $\gamma_\text{imp}=\frac{C_\text{imp}}{k_{B}T}$.

We computed the impurity specific heat using the same set of parameters as used in the computation of the susceptibility, and plotted \(\gamma_\text{imp}T_\text{K}\) in Fig.\ref{specheat}. It is seen that $\gamma_\text{imp}T_{K}$ rises from 0 at high temperatures $T>10^{2}T_{K}$ and saturates at a value $\gamma_\text{imp}(0)=\frac{1}{4J^*}$ for $T<10^{-2}T_{K}$.
The Wilson ratio \(R\) is defined as the ratio of the susceptibility \(\chi_\text{imp}\) and the specific heat coefficient \(\gamma_\text{imp}\). It is found that \(R\) saturates to a value $R=\chi/\gamma_\text{imp}=2.009$ for $T \ll T_\text{K}$.

\begin{figure}
\centering
\includegraphics[width=0.45\textwidth]{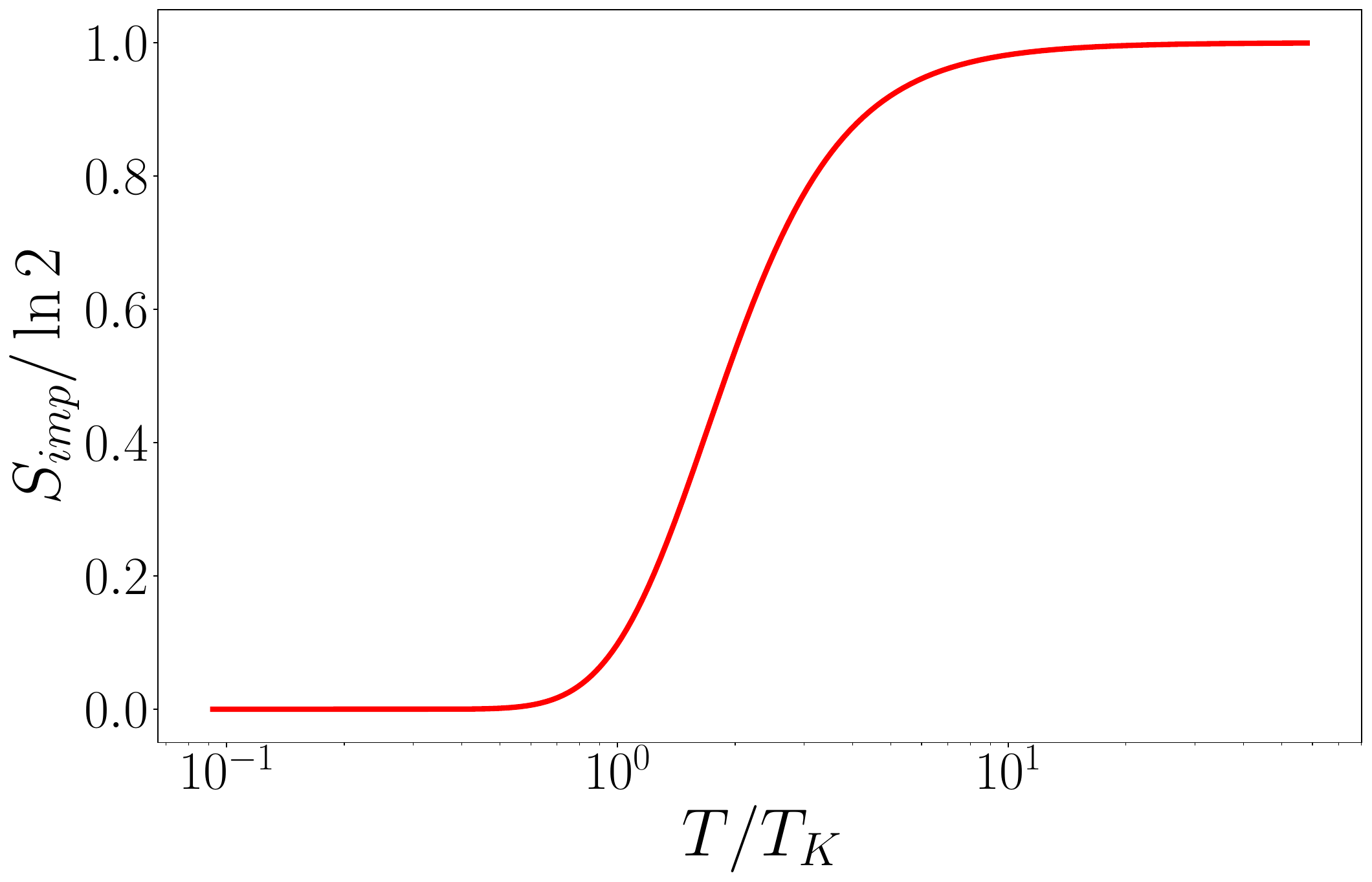}
\caption{Impurity contribution to thermal entropy in units of \(\ln 2\). $T_{K}$ corresponds to the Kondo temperature. The high temperature value of \(\ln 2\) shows is indicative of a doubly-degenerate local moment, while the zero temperature value of 0 shows the ground state is a non-degenerate singlet.}
\label{therm_entr}
\end{figure}
We also calculated the impurity thermal entropy from the zero mode Hamiltonian using the expression
\begin{equation}\begin{aligned}
	S_\text{imp}(T) = -\frac{1}{2}\frac{\partial{F}}{\partial{T}} = \frac{1}{2}\left[\ln Z + \frac{\beta}{Z}\sum_i \epsilon_i e^{-\beta \epsilon_i}\right] 
\end{aligned}\end{equation}
\(Z\) and \(\epsilon_i\) are the partition function and energy eigenvalues of the zero mode Hamiltonian, and the factor of half comes from the fact that since the impurity and the conduction bath site are symmetrical in the zero mode Hamiltonian, the impurity contribution to the entropy will be half of the total contribution. This is shown in Fig.\ref{therm_entr}.
At high temperatures, the impurity behaves like a doubly-degenerate free spin , contributing an entropy of \(\ln 2\). At lower temperatures, the impurity is screened within the unique singlet ground state, leading to a vanishing residual entropy. 
\bibliography{kondo_ms}

\begin{thebibliography}{106}%
\makeatletter
\providecommand \@ifxundefined [1]{%
 \@ifx{#1\undefined}
}%
\providecommand \@ifnum [1]{%
 \ifnum #1\expandafter \@firstoftwo
 \else \expandafter \@secondoftwo
 \fi
}%
\providecommand \@ifx [1]{%
 \ifx #1\expandafter \@firstoftwo
 \else \expandafter \@secondoftwo
 \fi
}%
\providecommand \natexlab [1]{#1}%
\providecommand \enquote  [1]{``#1''}%
\providecommand \bibnamefont  [1]{#1}%
\providecommand \bibfnamefont [1]{#1}%
\providecommand \citenamefont [1]{#1}%
\providecommand \href@noop [0]{\@secondoftwo}%
\providecommand \href [0]{\begingroup \@sanitize@url \@href}%
\providecommand \@href[1]{\@@startlink{#1}\@@href}%
\providecommand \@@href[1]{\endgroup#1\@@endlink}%
\providecommand \@sanitize@url [0]{\catcode `\\12\catcode `\$12\catcode
  `\&12\catcode `\#12\catcode `\^12\catcode `\_12\catcode `\%12\relax}%
\providecommand \@@startlink[1]{}%
\providecommand \@@endlink[0]{}%
\providecommand \url  [0]{\begingroup\@sanitize@url \@url }%
\providecommand \@url [1]{\endgroup\@href {#1}{\urlprefix }}%
\providecommand \urlprefix  [0]{URL }%
\providecommand \Eprint [0]{\href }%
\providecommand \doibase [0]{http://dx.doi.org/}%
\providecommand \selectlanguage [0]{\@gobble}%
\providecommand \bibinfo  [0]{\@secondoftwo}%
\providecommand \bibfield  [0]{\@secondoftwo}%
\providecommand \translation [1]{[#1]}%
\providecommand \BibitemOpen [0]{}%
\providecommand \bibitemStop [0]{}%
\providecommand \bibitemNoStop [0]{.\EOS\space}%
\providecommand \EOS [0]{\spacefactor3000\relax}%
\providecommand \BibitemShut  [1]{\csname bibitem#1\endcsname}%
\let\auto@bib@innerbib\@empty
\bibitem [{\citenamefont {Schrieffer}\ and\ \citenamefont
  {P.A.Wolff}(1966)}]{schrieffer1966}%
  \BibitemOpen
  \bibfield  {author} {\bibinfo {author} {\bibfnamefont {J.}~\bibnamefont
  {Schrieffer}}\ and\ \bibinfo {author} {\bibnamefont {P.A.Wolff}},\
  }\href@noop {} {\bibfield  {journal} {\bibinfo  {journal} {Phys. Rev.}\
  }\textbf {\bibinfo {volume} {149}},\ \bibinfo {pages} {491} (\bibinfo {year}
  {1966})}\BibitemShut {NoStop}%
\bibitem [{\citenamefont {Kondo}(1964)}]{kondo1964resistance}%
  \BibitemOpen
  \bibfield  {author} {\bibinfo {author} {\bibfnamefont {J.}~\bibnamefont
  {Kondo}},\ }\href@noop {} {\bibfield  {journal} {\bibinfo  {journal}
  {Progress of theoretical physics}\ }\textbf {\bibinfo {volume} {32}},\
  \bibinfo {pages} {37} (\bibinfo {year} {1964})}\BibitemShut {NoStop}%
\bibitem [{\citenamefont {hui Zhang}\ \emph {et~al.}(2013)\citenamefont {hui
  Zhang}, \citenamefont {Kahle}, \citenamefont {Herden}, \citenamefont {Stroh},
  \citenamefont {Mayor}, \citenamefont {Schlickum}, \citenamefont {Ternes},
  \citenamefont {Wahl},\ and\ \citenamefont {Kern}}]{Zhang2013}%
  \BibitemOpen
  \bibfield  {author} {\bibinfo {author} {\bibfnamefont {Y.}~\bibnamefont {hui
  Zhang}}, \bibinfo {author} {\bibfnamefont {S.}~\bibnamefont {Kahle}},
  \bibinfo {author} {\bibfnamefont {T.}~\bibnamefont {Herden}}, \bibinfo
  {author} {\bibfnamefont {C.}~\bibnamefont {Stroh}}, \bibinfo {author}
  {\bibfnamefont {M.}~\bibnamefont {Mayor}}, \bibinfo {author} {\bibfnamefont
  {U.}~\bibnamefont {Schlickum}}, \bibinfo {author} {\bibfnamefont
  {M.}~\bibnamefont {Ternes}}, \bibinfo {author} {\bibfnamefont
  {P.}~\bibnamefont {Wahl}}, \ and\ \bibinfo {author} {\bibfnamefont
  {K.}~\bibnamefont {Kern}},\ }\href {\doibase 10.1038/ncomms3110} {\bibfield
  {journal} {\bibinfo  {journal} {Nature Communications}\ }\textbf {\bibinfo
  {volume} {4}} (\bibinfo {year} {2013}),\ 10.1038/ncomms3110}\BibitemShut
  {NoStop}%
\bibitem [{\citenamefont {Anderson}\ and\ \citenamefont
  {Yuval}(1969)}]{anderson1969exact}%
  \BibitemOpen
  \bibfield  {author} {\bibinfo {author} {\bibfnamefont {P.~W.}\ \bibnamefont
  {Anderson}}\ and\ \bibinfo {author} {\bibfnamefont {G.}~\bibnamefont
  {Yuval}},\ }\href@noop {} {\bibfield  {journal} {\bibinfo  {journal}
  {Physical Review Letters}\ }\textbf {\bibinfo {volume} {23}},\ \bibinfo
  {pages} {89} (\bibinfo {year} {1969})}\BibitemShut {NoStop}%
\bibitem [{\citenamefont {Anderson}\ \emph {et~al.}(1970)\citenamefont
  {Anderson}, \citenamefont {Yuval},\ and\ \citenamefont
  {Hamann}}]{anderson1970exact}%
  \BibitemOpen
  \bibfield  {author} {\bibinfo {author} {\bibfnamefont {P.~W.}\ \bibnamefont
  {Anderson}}, \bibinfo {author} {\bibfnamefont {G.}~\bibnamefont {Yuval}}, \
  and\ \bibinfo {author} {\bibfnamefont {D.}~\bibnamefont {Hamann}},\
  }\href@noop {} {\bibfield  {journal} {\bibinfo  {journal} {Physical Review
  B}\ }\textbf {\bibinfo {volume} {1}},\ \bibinfo {pages} {4464} (\bibinfo
  {year} {1970})}\BibitemShut {NoStop}%
\bibitem [{\citenamefont {Anderson}(1970)}]{anderson1970poor}%
  \BibitemOpen
  \bibfield  {author} {\bibinfo {author} {\bibfnamefont {P.}~\bibnamefont
  {Anderson}},\ }\href@noop {} {\bibfield  {journal} {\bibinfo  {journal}
  {Journal of Physics C: Solid State Physics}\ }\textbf {\bibinfo {volume}
  {3}},\ \bibinfo {pages} {2436} (\bibinfo {year} {1970})}\BibitemShut
  {NoStop}%
\bibitem [{\citenamefont {Wilson}(1975{\natexlab{a}})}]{wilson1975}%
  \BibitemOpen
  \bibfield  {author} {\bibinfo {author} {\bibfnamefont {K.~G.}\ \bibnamefont
  {Wilson}},\ }\href@noop {} {\bibfield  {journal} {\bibinfo  {journal} {Rev.
  Mod. Phys.}\ }\textbf {\bibinfo {volume} {47}},\ \bibinfo {pages} {773}
  (\bibinfo {year} {1975}{\natexlab{a}})}\BibitemShut {NoStop}%
\bibitem [{\citenamefont {Bulla}\ \emph {et~al.}(2008)\citenamefont {Bulla},
  \citenamefont {Costi},\ and\ \citenamefont {Pruschke}}]{bullaNRGreview}%
  \BibitemOpen
  \bibfield  {author} {\bibinfo {author} {\bibfnamefont {R.}~\bibnamefont
  {Bulla}}, \bibinfo {author} {\bibfnamefont {T.}~\bibnamefont {Costi}}, \ and\
  \bibinfo {author} {\bibfnamefont {T.}~\bibnamefont {Pruschke}},\ }\href@noop
  {} {\bibfield  {journal} {\bibinfo  {journal} {Rev. Mod. Phys.}\ }\textbf
  {\bibinfo {volume} {80}},\ \bibinfo {pages} {395} (\bibinfo {year}
  {2008})}\BibitemShut {NoStop}%
\bibitem [{\citenamefont {Andrei}\ \emph {et~al.}(1983)\citenamefont {Andrei},
  \citenamefont {Furuya},\ and\ \citenamefont
  {Lowenstein}}]{andreiKondoreview}%
  \BibitemOpen
  \bibfield  {author} {\bibinfo {author} {\bibfnamefont {N.}~\bibnamefont
  {Andrei}}, \bibinfo {author} {\bibfnamefont {K.}~\bibnamefont {Furuya}}, \
  and\ \bibinfo {author} {\bibfnamefont {J.~H.}\ \bibnamefont {Lowenstein}},\
  }\href@noop {} {\bibfield  {journal} {\bibinfo  {journal} {Rev. Mod. Phys.}\
  }\textbf {\bibinfo {volume} {55}},\ \bibinfo {pages} {331} (\bibinfo {year}
  {1983})}\BibitemShut {NoStop}%
\bibitem [{\citenamefont {Tsvelick}\ and\ \citenamefont
  {Wiegmann}(1983)}]{tsvelickKondoreview}%
  \BibitemOpen
  \bibfield  {author} {\bibinfo {author} {\bibfnamefont {A.~M.}\ \bibnamefont
  {Tsvelick}}\ and\ \bibinfo {author} {\bibfnamefont {P.~B.}\ \bibnamefont
  {Wiegmann}},\ }\href@noop {} {\bibfield  {journal} {\bibinfo  {journal} {Adv.
  in Phys.}\ }\textbf {\bibinfo {volume} {32}},\ \bibinfo {pages} {453}
  (\bibinfo {year} {1983})}\BibitemShut {NoStop}%
\bibitem [{\citenamefont {Affleck}(1995)}]{affleck1995conformal}%
  \BibitemOpen
  \bibfield  {author} {\bibinfo {author} {\bibfnamefont {I.}~\bibnamefont
  {Affleck}},\ }\href@noop {} {\bibfield  {journal} {\bibinfo  {journal} {Acta
  Phys.Polon.B}\ }\textbf {\bibinfo {volume} {26}} (\bibinfo {year}
  {1995})}\BibitemShut {NoStop}%
\bibitem [{\citenamefont {Affleck}\ and\ \citenamefont
  {Ludwig}(1993)}]{affleck1993exact}%
  \BibitemOpen
  \bibfield  {author} {\bibinfo {author} {\bibfnamefont {I.}~\bibnamefont
  {Affleck}}\ and\ \bibinfo {author} {\bibfnamefont {A.~W.}\ \bibnamefont
  {Ludwig}},\ }\href@noop {} {\bibfield  {journal} {\bibinfo  {journal}
  {Physical Review B}\ }\textbf {\bibinfo {volume} {48}},\ \bibinfo {pages}
  {7297} (\bibinfo {year} {1993})}\BibitemShut {NoStop}%
\bibitem [{\citenamefont {Nozieres}(1974)}]{nozieres1974fermi}%
  \BibitemOpen
  \bibfield  {author} {\bibinfo {author} {\bibfnamefont {P.}~\bibnamefont
  {Nozieres}},\ }\href@noop {} {\bibfield  {journal} {\bibinfo  {journal}
  {Journal of Low Temperature Physics}\ }\textbf {\bibinfo {volume} {17}},\
  \bibinfo {pages} {31} (\bibinfo {year} {1974})}\BibitemShut {NoStop}%
\bibitem [{\citenamefont {Nozaki}\ \emph {et~al.}(2012)\citenamefont {Nozaki},
  \citenamefont {Ryu},\ and\ \citenamefont {Takayanagi}}]{nozaki2012}%
  \BibitemOpen
  \bibfield  {author} {\bibinfo {author} {\bibfnamefont {M.}~\bibnamefont
  {Nozaki}}, \bibinfo {author} {\bibfnamefont {S.}~\bibnamefont {Ryu}}, \ and\
  \bibinfo {author} {\bibfnamefont {T.}~\bibnamefont {Takayanagi}},\
  }\href@noop {} {\bibfield  {journal} {\bibinfo  {journal} {Journal of High
  Energy Physics}\ }\textbf {\bibinfo {volume} {2012}},\ \bibinfo {pages} {193}
  (\bibinfo {year} {2012})}\BibitemShut {NoStop}%
\bibitem [{\citenamefont {Krishnamurthy}\ \emph {et~al.}(1990)\citenamefont
  {Krishnamurthy}, \citenamefont {Jayaprakash}, \citenamefont {Sarker},\ and\
  \citenamefont {Wenzel}}]{krishnamurthy-physrevlett.64.950}%
  \BibitemOpen
  \bibfield  {author} {\bibinfo {author} {\bibfnamefont {H.}~\bibnamefont
  {Krishnamurthy}}, \bibinfo {author} {\bibfnamefont {C.}~\bibnamefont
  {Jayaprakash}}, \bibinfo {author} {\bibfnamefont {S.}~\bibnamefont {Sarker}},
  \ and\ \bibinfo {author} {\bibfnamefont {W.}~\bibnamefont {Wenzel}},\
  }\href@noop {} {\bibfield  {journal} {\bibinfo  {journal} {Physical review
  letters}\ }\textbf {\bibinfo {volume} {64}},\ \bibinfo {pages} {950}
  (\bibinfo {year} {1990})}\BibitemShut {NoStop}%
\bibitem [{\citenamefont {Hanl}\ and\ \citenamefont
  {Weichselbaum}(2014)}]{andreas_markus_2014}%
  \BibitemOpen
  \bibfield  {author} {\bibinfo {author} {\bibfnamefont {M.}~\bibnamefont
  {Hanl}}\ and\ \bibinfo {author} {\bibfnamefont {A.}~\bibnamefont
  {Weichselbaum}},\ }\href {\doibase 10.1103/PhysRevB.89.075130} {\bibfield
  {journal} {\bibinfo  {journal} {Phys. Rev. B}\ }\textbf {\bibinfo {volume}
  {89}},\ \bibinfo {pages} {075130} (\bibinfo {year} {2014})}\BibitemShut
  {NoStop}%
\bibitem [{\citenamefont {Arai}(1985)}]{tadashi_1985}%
  \BibitemOpen
  \bibfield  {author} {\bibinfo {author} {\bibfnamefont {T.}~\bibnamefont
  {Arai}},\ }\href {\doibase 10.1063/1.335190} {\bibfield  {journal} {\bibinfo
  {journal} {Journal of Applied Physics}\ }\textbf {\bibinfo {volume} {57}},\
  \bibinfo {pages} {3161} (\bibinfo {year} {1985})},\ \Eprint
  {http://arxiv.org/abs/https://doi.org/10.1063/1.335190}
  {https://doi.org/10.1063/1.335190} \BibitemShut {NoStop}%
\bibitem [{\citenamefont {Gaass}\ \emph {et~al.}(2011)\citenamefont {Gaass},
  \citenamefont {H\"uttel}, \citenamefont {Kang}, \citenamefont {Weymann},
  \citenamefont {von Delft},\ and\ \citenamefont {Strunk}}]{gass_kang_2011}%
  \BibitemOpen
  \bibfield  {author} {\bibinfo {author} {\bibfnamefont {M.}~\bibnamefont
  {Gaass}}, \bibinfo {author} {\bibfnamefont {A.~K.}\ \bibnamefont {H\"uttel}},
  \bibinfo {author} {\bibfnamefont {K.}~\bibnamefont {Kang}}, \bibinfo {author}
  {\bibfnamefont {I.}~\bibnamefont {Weymann}}, \bibinfo {author} {\bibfnamefont
  {J.}~\bibnamefont {von Delft}}, \ and\ \bibinfo {author} {\bibfnamefont
  {C.}~\bibnamefont {Strunk}},\ }\href {\doibase
  10.1103/PhysRevLett.107.176808} {\bibfield  {journal} {\bibinfo  {journal}
  {Phys. Rev. Lett.}\ }\textbf {\bibinfo {volume} {107}},\ \bibinfo {pages}
  {176808} (\bibinfo {year} {2011})}\BibitemShut {NoStop}%
\bibitem [{\citenamefont {S\o{}rensen}\ and\ \citenamefont
  {Affleck}(1996)}]{sorensen_erik_affleck_1996}%
  \BibitemOpen
  \bibfield  {author} {\bibinfo {author} {\bibfnamefont {E.~S.}\ \bibnamefont
  {S\o{}rensen}}\ and\ \bibinfo {author} {\bibfnamefont {I.}~\bibnamefont
  {Affleck}},\ }\href {\doibase 10.1103/PhysRevB.53.9153} {\bibfield  {journal}
  {\bibinfo  {journal} {Phys. Rev. B}\ }\textbf {\bibinfo {volume} {53}},\
  \bibinfo {pages} {9153} (\bibinfo {year} {1996})}\BibitemShut {NoStop}%
\bibitem [{\citenamefont {Affleck}\ and\ \citenamefont
  {Simon}(2001)}]{affleck_ian_2001}%
  \BibitemOpen
  \bibfield  {author} {\bibinfo {author} {\bibfnamefont {I.}~\bibnamefont
  {Affleck}}\ and\ \bibinfo {author} {\bibfnamefont {P.}~\bibnamefont
  {Simon}},\ }\href {\doibase 10.1103/PhysRevLett.86.2854} {\bibfield
  {journal} {\bibinfo  {journal} {Phys. Rev. Lett.}\ }\textbf {\bibinfo
  {volume} {86}},\ \bibinfo {pages} {2854} (\bibinfo {year}
  {2001})}\BibitemShut {NoStop}%
\bibitem [{\citenamefont {Simon}\ and\ \citenamefont
  {Affleck}(2003)}]{simon_pascal_2003}%
  \BibitemOpen
  \bibfield  {author} {\bibinfo {author} {\bibfnamefont {P.}~\bibnamefont
  {Simon}}\ and\ \bibinfo {author} {\bibfnamefont {I.}~\bibnamefont
  {Affleck}},\ }\href {\doibase 10.1103/PhysRevB.68.115304} {\bibfield
  {journal} {\bibinfo  {journal} {Phys. Rev. B}\ }\textbf {\bibinfo {volume}
  {68}},\ \bibinfo {pages} {115304} (\bibinfo {year} {2003})}\BibitemShut
  {NoStop}%
\bibitem [{\citenamefont {Busser}\ \emph {et~al.}(2010)\citenamefont {Busser},
  \citenamefont {Martins}, \citenamefont {Ribeiro}, \citenamefont {Vernek},
  \citenamefont {Anda},\ and\ \citenamefont {Dagotto}}]{martin2010}%
  \BibitemOpen
  \bibfield  {author} {\bibinfo {author} {\bibfnamefont {C.~A.}\ \bibnamefont
  {Busser}}, \bibinfo {author} {\bibfnamefont {G.~B.}\ \bibnamefont {Martins}},
  \bibinfo {author} {\bibfnamefont {L.~C.}\ \bibnamefont {Ribeiro}}, \bibinfo
  {author} {\bibfnamefont {E.}~\bibnamefont {Vernek}}, \bibinfo {author}
  {\bibfnamefont {E.~V.}\ \bibnamefont {Anda}}, \ and\ \bibinfo {author}
  {\bibfnamefont {E.}~\bibnamefont {Dagotto}},\ }\href@noop {} {\bibfield
  {journal} {\bibinfo  {journal} {Phys. Rev. 9}\ }\textbf {\bibinfo {volume}
  {81}},\ \bibinfo {pages} {045111} (\bibinfo {year} {2010})}\BibitemShut
  {NoStop}%
\bibitem [{\citenamefont {Ribeiro}\ \emph {et~al.}(2019)\citenamefont
  {Ribeiro}, \citenamefont {Martins}, \citenamefont {Gomez-Silva},\ and\
  \citenamefont {Anda}}]{martin2019}%
  \BibitemOpen
  \bibfield  {author} {\bibinfo {author} {\bibfnamefont {L.~C.}\ \bibnamefont
  {Ribeiro}}, \bibinfo {author} {\bibfnamefont {G.~B.}\ \bibnamefont
  {Martins}}, \bibinfo {author} {\bibfnamefont {G.}~\bibnamefont
  {Gomez-Silva}}, \ and\ \bibinfo {author} {\bibfnamefont {E.~V.}\ \bibnamefont
  {Anda}},\ }\href@noop {} {\bibfield  {journal} {\bibinfo  {journal} {Phys.
  Rev. B}\ }\textbf {\bibinfo {volume} {99}},\ \bibinfo {pages} {085139}
  (\bibinfo {year} {2019})}\BibitemShut {NoStop}%
\bibitem [{\citenamefont {Goldhaber-Gordon}\ \emph {et~al.}(1998)\citenamefont
  {Goldhaber-Gordon}, \citenamefont {Shtrikman}, \citenamefont {Mahalu},
  \citenamefont {Abusch-Magder}, \citenamefont {Meirav},\ and\ \citenamefont
  {Kastner}}]{Goldhaber-Gordon1998}%
  \BibitemOpen
  \bibfield  {author} {\bibinfo {author} {\bibfnamefont {D.}~\bibnamefont
  {Goldhaber-Gordon}}, \bibinfo {author} {\bibfnamefont {H.}~\bibnamefont
  {Shtrikman}}, \bibinfo {author} {\bibfnamefont {D.}~\bibnamefont {Mahalu}},
  \bibinfo {author} {\bibfnamefont {D.}~\bibnamefont {Abusch-Magder}}, \bibinfo
  {author} {\bibfnamefont {U.}~\bibnamefont {Meirav}}, \ and\ \bibinfo {author}
  {\bibfnamefont {M.~A.}\ \bibnamefont {Kastner}},\ }\href {\doibase
  10.1038/34373} {\bibfield  {journal} {\bibinfo  {journal} {Nature}\ }\textbf
  {\bibinfo {volume} {391}},\ \bibinfo {pages} {156} (\bibinfo {year}
  {1998})}\BibitemShut {NoStop}%
\bibitem [{\citenamefont {Cronenwett}\ \emph {et~al.}(1998)\citenamefont
  {Cronenwett}, \citenamefont {Oosterkamp},\ and\ \citenamefont
  {Kouwenhoven}}]{Cronenwett1998}%
  \BibitemOpen
  \bibfield  {author} {\bibinfo {author} {\bibfnamefont {S.~M.}\ \bibnamefont
  {Cronenwett}}, \bibinfo {author} {\bibfnamefont {T.~H.}\ \bibnamefont
  {Oosterkamp}}, \ and\ \bibinfo {author} {\bibfnamefont {L.~P.}\ \bibnamefont
  {Kouwenhoven}},\ }\href {\doibase 10.1126/science.281.5376.540} {\ \textbf
  {\bibinfo {volume} {281}},\ \bibinfo {pages} {540} (\bibinfo {year}
  {1998})}\BibitemShut {NoStop}%
\bibitem [{\citenamefont {Schmid}\ \emph {et~al.}(1998)\citenamefont {Schmid},
  \citenamefont {Weis}, \citenamefont {Eberl},\ and\ \citenamefont
  {v.~Klitzing}}]{Schmid_Weis1998}%
  \BibitemOpen
  \bibfield  {author} {\bibinfo {author} {\bibfnamefont {J.}~\bibnamefont
  {Schmid}}, \bibinfo {author} {\bibfnamefont {J.}~\bibnamefont {Weis}},
  \bibinfo {author} {\bibfnamefont {K.}~\bibnamefont {Eberl}}, \ and\ \bibinfo
  {author} {\bibfnamefont {K.}~\bibnamefont {v.~Klitzing}},\ }\href {\doibase
  10.1016/s0921-4526(98)00533-x} {\ \textbf {\bibinfo {volume} {256-258}},\
  \bibinfo {pages} {182} (\bibinfo {year} {1998})}\BibitemShut {NoStop}%
\bibitem [{\citenamefont {Pustilnik}\ and\ \citenamefont
  {Glazman}(2004)}]{pustilnik_glazman_2004}%
  \BibitemOpen
  \bibfield  {author} {\bibinfo {author} {\bibfnamefont {M.}~\bibnamefont
  {Pustilnik}}\ and\ \bibinfo {author} {\bibfnamefont {L.}~\bibnamefont
  {Glazman}},\ }\href {\doibase 10.1088/0953-8984/16/16/r01} {\ \textbf
  {\bibinfo {volume} {16}},\ \bibinfo {pages} {R513} (\bibinfo {year}
  {2004})}\BibitemShut {NoStop}%
\bibitem [{\citenamefont {V.~Borzenets}\ \emph {et~al.}(2020)\citenamefont
  {V.~Borzenets}, \citenamefont {Shim}, \citenamefont {Chen}, \citenamefont
  {Ludwig}, \citenamefont {Wieck}, \citenamefont {Tarucha}, \citenamefont
  {Sim},\ and\ \citenamefont {Yamamoto}}]{Borzenets2020}%
  \BibitemOpen
  \bibfield  {author} {\bibinfo {author} {\bibfnamefont {I.}~\bibnamefont
  {V.~Borzenets}}, \bibinfo {author} {\bibfnamefont {J.}~\bibnamefont {Shim}},
  \bibinfo {author} {\bibfnamefont {J.~C.~H.}\ \bibnamefont {Chen}}, \bibinfo
  {author} {\bibfnamefont {A.}~\bibnamefont {Ludwig}}, \bibinfo {author}
  {\bibfnamefont {A.~D.}\ \bibnamefont {Wieck}}, \bibinfo {author}
  {\bibfnamefont {S.}~\bibnamefont {Tarucha}}, \bibinfo {author} {\bibfnamefont
  {H.-S.}\ \bibnamefont {Sim}}, \ and\ \bibinfo {author} {\bibfnamefont
  {M.}~\bibnamefont {Yamamoto}},\ }\href {\doibase 10.1038/s41586-020-2058-6}
  {\bibfield  {journal} {\bibinfo  {journal} {Nature}\ }\textbf {\bibinfo
  {volume} {579}},\ \bibinfo {pages} {210} (\bibinfo {year}
  {2020})}\BibitemShut {NoStop}%
\bibitem [{\citenamefont {N\'eel}\ \emph {et~al.}(2008)\citenamefont {N\'eel},
  \citenamefont {Kr\"oger}, \citenamefont {Berndt}, \citenamefont {Wehling},
  \citenamefont {Lichtenstein},\ and\ \citenamefont
  {Katsnelson}}]{neel_berndt_2008}%
  \BibitemOpen
  \bibfield  {author} {\bibinfo {author} {\bibfnamefont {N.}~\bibnamefont
  {N\'eel}}, \bibinfo {author} {\bibfnamefont {J.}~\bibnamefont {Kr\"oger}},
  \bibinfo {author} {\bibfnamefont {R.}~\bibnamefont {Berndt}}, \bibinfo
  {author} {\bibfnamefont {T.~O.}\ \bibnamefont {Wehling}}, \bibinfo {author}
  {\bibfnamefont {A.~I.}\ \bibnamefont {Lichtenstein}}, \ and\ \bibinfo
  {author} {\bibfnamefont {M.~I.}\ \bibnamefont {Katsnelson}},\ }\href
  {\doibase 10.1103/PhysRevLett.101.266803} {\bibfield  {journal} {\bibinfo
  {journal} {Phys. Rev. Lett.}\ }\textbf {\bibinfo {volume} {101}},\ \bibinfo
  {pages} {266803} (\bibinfo {year} {2008})}\BibitemShut {NoStop}%
\bibitem [{\citenamefont {Zhao}\ \emph {et~al.}(2005)\citenamefont {Zhao},
  \citenamefont {Li}, \citenamefont {Chen}, \citenamefont {Xiang},
  \citenamefont {Wang}, \citenamefont {Pan}, \citenamefont {Wang},
  \citenamefont {Xiao}, \citenamefont {Yang}, \citenamefont {Hou},\ and\
  \citenamefont {Zhu}}]{Zhao2005}%
  \BibitemOpen
  \bibfield  {author} {\bibinfo {author} {\bibfnamefont {A.}~\bibnamefont
  {Zhao}}, \bibinfo {author} {\bibfnamefont {Q.}~\bibnamefont {Li}}, \bibinfo
  {author} {\bibfnamefont {L.}~\bibnamefont {Chen}}, \bibinfo {author}
  {\bibfnamefont {H.}~\bibnamefont {Xiang}}, \bibinfo {author} {\bibfnamefont
  {W.}~\bibnamefont {Wang}}, \bibinfo {author} {\bibfnamefont {S.}~\bibnamefont
  {Pan}}, \bibinfo {author} {\bibfnamefont {B.}~\bibnamefont {Wang}}, \bibinfo
  {author} {\bibfnamefont {X.}~\bibnamefont {Xiao}}, \bibinfo {author}
  {\bibfnamefont {J.}~\bibnamefont {Yang}}, \bibinfo {author} {\bibfnamefont
  {J.~G.}\ \bibnamefont {Hou}}, \ and\ \bibinfo {author} {\bibfnamefont
  {Q.}~\bibnamefont {Zhu}},\ }\href {\doibase 10.1126/science.1113449}
  {\bibfield  {journal} {\bibinfo  {journal} {Science}\ }\textbf {\bibinfo
  {volume} {309}},\ \bibinfo {pages} {1542} (\bibinfo {year}
  {2005})}\BibitemShut {NoStop}%
\bibitem [{\citenamefont {Kaminski}\ \emph {et~al.}(2000)\citenamefont
  {Kaminski}, \citenamefont {Nazarov},\ and\ \citenamefont
  {Glazman}}]{kaminski_nazarov2000}%
  \BibitemOpen
  \bibfield  {author} {\bibinfo {author} {\bibfnamefont {A.}~\bibnamefont
  {Kaminski}}, \bibinfo {author} {\bibfnamefont {Y.~V.}\ \bibnamefont
  {Nazarov}}, \ and\ \bibinfo {author} {\bibfnamefont {L.~I.}\ \bibnamefont
  {Glazman}},\ }\href {\doibase 10.1103/PhysRevB.62.8154} {\bibfield  {journal}
  {\bibinfo  {journal} {Phys. Rev. B}\ }\textbf {\bibinfo {volume} {62}},\
  \bibinfo {pages} {8154} (\bibinfo {year} {2000})}\BibitemShut {NoStop}%
\bibitem [{\citenamefont {Krishna-murthy}\ \emph {et~al.}(1980)\citenamefont
  {Krishna-murthy}, \citenamefont {Wilkins},\ and\ \citenamefont
  {Wilson}}]{hrk_wilson_1980}%
  \BibitemOpen
  \bibfield  {author} {\bibinfo {author} {\bibfnamefont {H.~R.}\ \bibnamefont
  {Krishna-murthy}}, \bibinfo {author} {\bibfnamefont {J.~W.}\ \bibnamefont
  {Wilkins}}, \ and\ \bibinfo {author} {\bibfnamefont {K.~G.}\ \bibnamefont
  {Wilson}},\ }\href {\doibase 10.1103/PhysRevB.21.1003} {\bibfield  {journal}
  {\bibinfo  {journal} {Phys. Rev. B}\ }\textbf {\bibinfo {volume} {21}},\
  \bibinfo {pages} {1003} (\bibinfo {year} {1980})}\BibitemShut {NoStop}%
\bibitem [{\citenamefont {Sakai}\ \emph {et~al.}(1989)\citenamefont {Sakai},
  \citenamefont {Shimizu},\ and\ \citenamefont {Kasuya}}]{sakai_osamu_shimizu}%
  \BibitemOpen
  \bibfield  {author} {\bibinfo {author} {\bibfnamefont {O.}~\bibnamefont
  {Sakai}}, \bibinfo {author} {\bibfnamefont {Y.}~\bibnamefont {Shimizu}}, \
  and\ \bibinfo {author} {\bibfnamefont {T.}~\bibnamefont {Kasuya}},\ }\href
  {\doibase 10.1143/JPSJ.58.3666} {\bibfield  {journal} {\bibinfo  {journal}
  {Journal of the Physical Society of Japan}\ }\textbf {\bibinfo {volume}
  {58}},\ \bibinfo {pages} {3666} (\bibinfo {year} {1989})},\ \Eprint
  {http://arxiv.org/abs/https://doi.org/10.1143/JPSJ.58.3666}
  {https://doi.org/10.1143/JPSJ.58.3666} \BibitemShut {NoStop}%
\bibitem [{\citenamefont {Costi}\ and\ \citenamefont
  {Hewson}(1990)}]{costi_hewson_1990}%
  \BibitemOpen
  \bibfield  {author} {\bibinfo {author} {\bibfnamefont {T.}~\bibnamefont
  {Costi}}\ and\ \bibinfo {author} {\bibfnamefont {A.}~\bibnamefont {Hewson}},\
  }\href {\doibase https://doi.org/10.1016/0921-4526(90)90161-M} {\bibfield
  {journal} {\bibinfo  {journal} {Physica B: Condensed Matter}\ }\textbf
  {\bibinfo {volume} {163}},\ \bibinfo {pages} {179} (\bibinfo {year}
  {1990})}\BibitemShut {NoStop}%
\bibitem [{\citenamefont {Costi}\ \emph {et~al.}(1996)\citenamefont {Costi},
  \citenamefont {Kroha},\ and\ \citenamefont {W\"olfle}}]{costi_kroha_wolfle}%
  \BibitemOpen
  \bibfield  {author} {\bibinfo {author} {\bibfnamefont {T.~A.}\ \bibnamefont
  {Costi}}, \bibinfo {author} {\bibfnamefont {J.}~\bibnamefont {Kroha}}, \ and\
  \bibinfo {author} {\bibfnamefont {P.}~\bibnamefont {W\"olfle}},\ }\href
  {\doibase 10.1103/PhysRevB.53.1850} {\bibfield  {journal} {\bibinfo
  {journal} {Phys. Rev. B}\ }\textbf {\bibinfo {volume} {53}},\ \bibinfo
  {pages} {1850} (\bibinfo {year} {1996})}\BibitemShut {NoStop}%
\bibitem [{\citenamefont {Kroha}\ and\ \citenamefont
  {Wölfle}(2005)}]{kroha_wolfle}%
  \BibitemOpen
  \bibfield  {author} {\bibinfo {author} {\bibfnamefont {J.}~\bibnamefont
  {Kroha}}\ and\ \bibinfo {author} {\bibfnamefont {P.}~\bibnamefont
  {Wölfle}},\ }\href {\doibase 10.1143/JPSJ.74.16} {\bibfield  {journal}
  {\bibinfo  {journal} {Journal of the Physical Society of Japan}\ }\textbf
  {\bibinfo {volume} {74}},\ \bibinfo {pages} {16} (\bibinfo {year} {2005})},\
  \Eprint {http://arxiv.org/abs/https://doi.org/10.1143/JPSJ.74.16}
  {https://doi.org/10.1143/JPSJ.74.16} \BibitemShut {NoStop}%
\bibitem [{\citenamefont {Costi}\ and\ \citenamefont
  {Hewson}(1992)}]{costi_hewson_1992}%
  \BibitemOpen
  \bibfield  {author} {\bibinfo {author} {\bibfnamefont {T.~A.}\ \bibnamefont
  {Costi}}\ and\ \bibinfo {author} {\bibfnamefont {A.~C.}\ \bibnamefont
  {Hewson}},\ }\href {\doibase 10.1080/13642819208215080} {\bibfield  {journal}
  {\bibinfo  {journal} {Philosophical Magazine B}\ }\textbf {\bibinfo {volume}
  {65}},\ \bibinfo {pages} {1165} (\bibinfo {year} {1992})},\ \Eprint
  {http://arxiv.org/abs/https://doi.org/10.1080/13642819208215080}
  {https://doi.org/10.1080/13642819208215080} \BibitemShut {NoStop}%
\bibitem [{\citenamefont {{Nozi\`eres, Ph.}}\ and\ \citenamefont {{Blandin,
  A.}}(1980)}]{Noz_blandin_1980}%
  \BibitemOpen
  \bibfield  {author} {\bibinfo {author} {\bibnamefont {{Nozi\`eres, Ph.}}}\
  and\ \bibinfo {author} {\bibnamefont {{Blandin, A.}}},\ }\href {\doibase
  10.1051/jphys:01980004103019300} {\bibfield  {journal} {\bibinfo  {journal}
  {J. Phys. France}\ }\textbf {\bibinfo {volume} {41}},\ \bibinfo {pages} {193}
  (\bibinfo {year} {1980})}\BibitemShut {NoStop}%
\bibitem [{\citenamefont {Gan}\ \emph {et~al.}(1993)\citenamefont {Gan},
  \citenamefont {Andrei},\ and\ \citenamefont
  {Coleman}}]{Gan_Andrei_Coleman_1993}%
  \BibitemOpen
  \bibfield  {author} {\bibinfo {author} {\bibfnamefont {J.}~\bibnamefont
  {Gan}}, \bibinfo {author} {\bibfnamefont {N.}~\bibnamefont {Andrei}}, \ and\
  \bibinfo {author} {\bibfnamefont {P.}~\bibnamefont {Coleman}},\ }\href
  {\doibase 10.1103/PhysRevLett.70.686} {\bibfield  {journal} {\bibinfo
  {journal} {Phys. Rev. Lett.}\ }\textbf {\bibinfo {volume} {70}},\ \bibinfo
  {pages} {686} (\bibinfo {year} {1993})}\BibitemShut {NoStop}%
\bibitem [{\citenamefont {Emery}\ and\ \citenamefont
  {Kivelson}(1992)}]{emery_kivelson}%
  \BibitemOpen
  \bibfield  {author} {\bibinfo {author} {\bibfnamefont {V.~J.}\ \bibnamefont
  {Emery}}\ and\ \bibinfo {author} {\bibfnamefont {S.}~\bibnamefont
  {Kivelson}},\ }\href {\doibase 10.1103/PhysRevB.46.10812} {\bibfield
  {journal} {\bibinfo  {journal} {Phys. Rev. B}\ }\textbf {\bibinfo {volume}
  {46}},\ \bibinfo {pages} {10812} (\bibinfo {year} {1992})}\BibitemShut
  {NoStop}%
\bibitem [{\citenamefont {Gan}(1994)}]{Gan_mchannel_1994}%
  \BibitemOpen
  \bibfield  {author} {\bibinfo {author} {\bibfnamefont {J.}~\bibnamefont
  {Gan}},\ }\href {\doibase 10.1088/0953-8984/6/24/016} {\ \textbf {\bibinfo
  {volume} {6}},\ \bibinfo {pages} {4547} (\bibinfo {year} {1994})}\BibitemShut
  {NoStop}%
\bibitem [{\citenamefont {Tsvelick}\ and\ \citenamefont
  {Wiegmann}(1984)}]{Tsvelick_Weigmann_mchannel_1984}%
  \BibitemOpen
  \bibfield  {author} {\bibinfo {author} {\bibfnamefont {A.~M.}\ \bibnamefont
  {Tsvelick}}\ and\ \bibinfo {author} {\bibfnamefont {P.~B.}\ \bibnamefont
  {Wiegmann}},\ }\href {\doibase 10.1007/BF01319184} {\bibfield  {journal}
  {\bibinfo  {journal} {Zeitschrift f{\"u}r Physik B Condensed Matter}\
  }\textbf {\bibinfo {volume} {54}},\ \bibinfo {pages} {201} (\bibinfo {year}
  {1984})}\BibitemShut {NoStop}%
\bibitem [{\citenamefont {Tsvelick}\ and\ \citenamefont
  {Wiegmann}(1985)}]{Tsvelick_weigmann_mchannel_1985}%
  \BibitemOpen
  \bibfield  {author} {\bibinfo {author} {\bibfnamefont {A.~M.}\ \bibnamefont
  {Tsvelick}}\ and\ \bibinfo {author} {\bibfnamefont {P.~B.}\ \bibnamefont
  {Wiegmann}},\ }\href {\doibase 10.1007/BF01017853} {\bibfield  {journal}
  {\bibinfo  {journal} {Journal of Statistical Physics}\ }\textbf {\bibinfo
  {volume} {38}},\ \bibinfo {pages} {125} (\bibinfo {year} {1985})}\BibitemShut
  {NoStop}%
\bibitem [{\citenamefont {Parcollet}\ and\ \citenamefont
  {Georges}(1997)}]{parcollet_olivier_large_N}%
  \BibitemOpen
  \bibfield  {author} {\bibinfo {author} {\bibfnamefont {O.}~\bibnamefont
  {Parcollet}}\ and\ \bibinfo {author} {\bibfnamefont {A.}~\bibnamefont
  {Georges}},\ }\href {\doibase 10.1103/PhysRevLett.79.4665} {\bibfield
  {journal} {\bibinfo  {journal} {Phys. Rev. Lett.}\ }\textbf {\bibinfo
  {volume} {79}},\ \bibinfo {pages} {4665} (\bibinfo {year}
  {1997})}\BibitemShut {NoStop}%
\bibitem [{\citenamefont {Kimura}\ and\ \citenamefont
  {Ozaki}(2017)}]{kimura_taro_Su_N_kondo}%
  \BibitemOpen
  \bibfield  {author} {\bibinfo {author} {\bibfnamefont {T.}~\bibnamefont
  {Kimura}}\ and\ \bibinfo {author} {\bibfnamefont {S.}~\bibnamefont {Ozaki}},\
  }\href {\doibase 10.7566/JPSJ.86.084703} {\bibfield  {journal} {\bibinfo
  {journal} {Journal of the Physical Society of Japan}\ }\textbf {\bibinfo
  {volume} {86}},\ \bibinfo {pages} {084703} (\bibinfo {year} {2017})},\
  \Eprint {http://arxiv.org/abs/https://doi.org/10.7566/JPSJ.86.084703}
  {https://doi.org/10.7566/JPSJ.86.084703} \BibitemShut {NoStop}%
\bibitem [{\citenamefont {Bensimon}\ \emph {et~al.}(2006)\citenamefont
  {Bensimon}, \citenamefont {Jerez},\ and\ \citenamefont
  {Lavagna}}]{PhysRevB.73.224445}%
  \BibitemOpen
  \bibfield  {author} {\bibinfo {author} {\bibfnamefont {D.}~\bibnamefont
  {Bensimon}}, \bibinfo {author} {\bibfnamefont {A.}~\bibnamefont {Jerez}}, \
  and\ \bibinfo {author} {\bibfnamefont {M.}~\bibnamefont {Lavagna}},\ }\href
  {\doibase 10.1103/PhysRevB.73.224445} {\bibfield  {journal} {\bibinfo
  {journal} {Phys. Rev. B}\ }\textbf {\bibinfo {volume} {73}},\ \bibinfo
  {pages} {224445} (\bibinfo {year} {2006})}\BibitemShut {NoStop}%
\bibitem [{\citenamefont {Cox}\ and\ \citenamefont
  {Jarrell}(1996)}]{cox_jarrell_two_channel_rev}%
  \BibitemOpen
  \bibfield  {author} {\bibinfo {author} {\bibfnamefont {D.~L.}\ \bibnamefont
  {Cox}}\ and\ \bibinfo {author} {\bibfnamefont {M.}~\bibnamefont {Jarrell}},\
  }\href {\doibase 10.1088/0953-8984/8/48/012} {\ \textbf {\bibinfo {volume}
  {8}},\ \bibinfo {pages} {9825} (\bibinfo {year} {1996})}\BibitemShut
  {NoStop}%
\bibitem [{\citenamefont {Affleck}\ and\ \citenamefont
  {Ludwig}(1991)}]{affleck_1991_overscreen}%
  \BibitemOpen
  \bibfield  {author} {\bibinfo {author} {\bibfnamefont {I.}~\bibnamefont
  {Affleck}}\ and\ \bibinfo {author} {\bibfnamefont {A.~W.}\ \bibnamefont
  {Ludwig}},\ }\href {\doibase https://doi.org/10.1016/0550-3213(91)90419-X}
  {\bibfield  {journal} {\bibinfo  {journal} {Nuclear Physics B}\ }\textbf
  {\bibinfo {volume} {360}},\ \bibinfo {pages} {641} (\bibinfo {year}
  {1991})}\BibitemShut {NoStop}%
\bibitem [{\citenamefont {Coleman}\ \emph {et~al.}(1995)\citenamefont
  {Coleman}, \citenamefont {Ioffe},\ and\ \citenamefont
  {Tsvelik}}]{Coleman_tsvelik}%
  \BibitemOpen
  \bibfield  {author} {\bibinfo {author} {\bibfnamefont {P.}~\bibnamefont
  {Coleman}}, \bibinfo {author} {\bibfnamefont {L.~B.}\ \bibnamefont {Ioffe}},
  \ and\ \bibinfo {author} {\bibfnamefont {A.~M.}\ \bibnamefont {Tsvelik}},\
  }\href {\doibase 10.1103/PhysRevB.52.6611} {\bibfield  {journal} {\bibinfo
  {journal} {Phys. Rev. B}\ }\textbf {\bibinfo {volume} {52}},\ \bibinfo
  {pages} {6611} (\bibinfo {year} {1995})}\BibitemShut {NoStop}%
\bibitem [{\citenamefont {Yasui}\ and\ \citenamefont
  {Sudoh}(2013)}]{Yasui_2013}%
  \BibitemOpen
  \bibfield  {author} {\bibinfo {author} {\bibfnamefont {S.}~\bibnamefont
  {Yasui}}\ and\ \bibinfo {author} {\bibfnamefont {K.}~\bibnamefont {Sudoh}},\
  }\href {\doibase 10.1103/PhysRevC.88.015201} {\bibfield  {journal} {\bibinfo
  {journal} {Phys. Rev. C}\ }\textbf {\bibinfo {volume} {88}},\ \bibinfo
  {pages} {015201} (\bibinfo {year} {2013})}\BibitemShut {NoStop}%
\bibitem [{\citenamefont {Hattori}\ \emph {et~al.}(2015)\citenamefont
  {Hattori}, \citenamefont {Itakura}, \citenamefont {Ozaki},\ and\
  \citenamefont {Yasui}}]{hattori_2015}%
  \BibitemOpen
  \bibfield  {author} {\bibinfo {author} {\bibfnamefont {K.}~\bibnamefont
  {Hattori}}, \bibinfo {author} {\bibfnamefont {K.}~\bibnamefont {Itakura}},
  \bibinfo {author} {\bibfnamefont {S.}~\bibnamefont {Ozaki}}, \ and\ \bibinfo
  {author} {\bibfnamefont {S.}~\bibnamefont {Yasui}},\ }\href {\doibase
  10.1103/PhysRevD.92.065003} {\bibfield  {journal} {\bibinfo  {journal} {Phys.
  Rev. D}\ }\textbf {\bibinfo {volume} {92}},\ \bibinfo {pages} {065003}
  (\bibinfo {year} {2015})}\BibitemShut {NoStop}%
\bibitem [{\citenamefont {Fritz}\ and\ \citenamefont
  {Vojta}(2013)}]{fritz_vojta_2013}%
  \BibitemOpen
  \bibfield  {author} {\bibinfo {author} {\bibfnamefont {L.}~\bibnamefont
  {Fritz}}\ and\ \bibinfo {author} {\bibfnamefont {M.}~\bibnamefont {Vojta}},\
  }\href {\doibase 10.1088/0034-4885/76/3/032501} {\ \textbf {\bibinfo {volume}
  {76}},\ \bibinfo {pages} {032501} (\bibinfo {year} {2013})}\BibitemShut
  {NoStop}%
\bibitem [{\citenamefont {Principi}\ \emph {et~al.}(2015)\citenamefont
  {Principi}, \citenamefont {Vignale},\ and\ \citenamefont
  {Rossi}}]{principi_2015}%
  \BibitemOpen
  \bibfield  {author} {\bibinfo {author} {\bibfnamefont {A.}~\bibnamefont
  {Principi}}, \bibinfo {author} {\bibfnamefont {G.}~\bibnamefont {Vignale}}, \
  and\ \bibinfo {author} {\bibfnamefont {E.}~\bibnamefont {Rossi}},\ }\href
  {\doibase 10.1103/PhysRevB.92.041107} {\bibfield  {journal} {\bibinfo
  {journal} {Phys. Rev. B}\ }\textbf {\bibinfo {volume} {92}},\ \bibinfo
  {pages} {041107} (\bibinfo {year} {2015})}\BibitemShut {NoStop}%
\bibitem [{\citenamefont {Mitchell}\ and\ \citenamefont
  {Fritz}(2015)}]{mitchell_lars_2015}%
  \BibitemOpen
  \bibfield  {author} {\bibinfo {author} {\bibfnamefont {A.~K.}\ \bibnamefont
  {Mitchell}}\ and\ \bibinfo {author} {\bibfnamefont {L.}~\bibnamefont
  {Fritz}},\ }\href {\doibase 10.1103/PhysRevB.92.121109} {\bibfield  {journal}
  {\bibinfo  {journal} {Phys. Rev. B}\ }\textbf {\bibinfo {volume} {92}},\
  \bibinfo {pages} {121109} (\bibinfo {year} {2015})}\BibitemShut {NoStop}%
\bibitem [{\citenamefont {Yasui}\ and\ \citenamefont
  {Sudoh}(2017)}]{yasui_kazutaka_2017}%
  \BibitemOpen
  \bibfield  {author} {\bibinfo {author} {\bibfnamefont {S.}~\bibnamefont
  {Yasui}}\ and\ \bibinfo {author} {\bibfnamefont {K.}~\bibnamefont {Sudoh}},\
  }\href {\doibase 10.1103/PhysRevC.95.035204} {\bibfield  {journal} {\bibinfo
  {journal} {Phys. Rev. C}\ }\textbf {\bibinfo {volume} {95}},\ \bibinfo
  {pages} {035204} (\bibinfo {year} {2017})}\BibitemShut {NoStop}%
\bibitem [{\citenamefont {Yasui}(2016)}]{yasui_shigehiro_2016}%
  \BibitemOpen
  \bibfield  {author} {\bibinfo {author} {\bibfnamefont {S.}~\bibnamefont
  {Yasui}},\ }\href {\doibase 10.1103/PhysRevC.93.065204} {\bibfield  {journal}
  {\bibinfo  {journal} {Phys. Rev. C}\ }\textbf {\bibinfo {volume} {93}},\
  \bibinfo {pages} {065204} (\bibinfo {year} {2016})}\BibitemShut {NoStop}%
\bibitem [{\citenamefont {Mukherjee}\ and\ \citenamefont
  {Lal}(2020{\natexlab{a}})}]{anirbanurg1}%
  \BibitemOpen
  \bibfield  {author} {\bibinfo {author} {\bibfnamefont {A.}~\bibnamefont
  {Mukherjee}}\ and\ \bibinfo {author} {\bibfnamefont {S.}~\bibnamefont
  {Lal}},\ }\href@noop {} {\bibfield  {journal} {\bibinfo  {journal} {Nuclear
  Physics B}\ }\textbf {\bibinfo {volume} {960}},\ \bibinfo {pages} {115170}
  (\bibinfo {year} {2020}{\natexlab{a}})}\BibitemShut {NoStop}%
\bibitem [{\citenamefont {Mukherjee}\ and\ \citenamefont
  {Lal}(2020{\natexlab{b}})}]{anirbanurg2}%
  \BibitemOpen
  \bibfield  {author} {\bibinfo {author} {\bibfnamefont {A.}~\bibnamefont
  {Mukherjee}}\ and\ \bibinfo {author} {\bibfnamefont {S.}~\bibnamefont
  {Lal}},\ }\href@noop {} {\bibfield  {journal} {\bibinfo  {journal} {Nuclear
  Physics B}\ }\textbf {\bibinfo {volume} {960}},\ \bibinfo {pages} {115163}
  (\bibinfo {year} {2020}{\natexlab{b}})}\BibitemShut {NoStop}%
\bibitem [{\citenamefont {Mukherjee}\ and\ \citenamefont
  {Lal}(2020{\natexlab{c}})}]{anirbanmott1}%
  \BibitemOpen
  \bibfield  {author} {\bibinfo {author} {\bibfnamefont {A.}~\bibnamefont
  {Mukherjee}}\ and\ \bibinfo {author} {\bibfnamefont {S.}~\bibnamefont
  {Lal}},\ }\href {\doibase 10.1088/1367-2630/ab8831} {\bibfield  {journal}
  {\bibinfo  {journal} {New Journal of Physics}\ }\textbf {\bibinfo {volume}
  {22}},\ \bibinfo {pages} {063007} (\bibinfo {year}
  {2020}{\natexlab{c}})}\BibitemShut {NoStop}%
\bibitem [{\citenamefont {Mukherjee}\ and\ \citenamefont
  {Lal}(2020{\natexlab{d}})}]{anirbanmott2}%
  \BibitemOpen
  \bibfield  {author} {\bibinfo {author} {\bibfnamefont {A.}~\bibnamefont
  {Mukherjee}}\ and\ \bibinfo {author} {\bibfnamefont {S.}~\bibnamefont
  {Lal}},\ }\href {\doibase 10.1088/1367-2630/ab890c} {\bibfield  {journal}
  {\bibinfo  {journal} {New Journal of Physics}\ }\textbf {\bibinfo {volume}
  {22}},\ \bibinfo {pages} {063008} (\bibinfo {year}
  {2020}{\natexlab{d}})}\BibitemShut {NoStop}%
\bibitem [{\citenamefont {Pal}\ \emph {et~al.}(2019)\citenamefont {Pal},
  \citenamefont {Mukherjee},\ and\ \citenamefont {Lal}}]{santanukagome}%
  \BibitemOpen
  \bibfield  {author} {\bibinfo {author} {\bibfnamefont {S.}~\bibnamefont
  {Pal}}, \bibinfo {author} {\bibfnamefont {A.}~\bibnamefont {Mukherjee}}, \
  and\ \bibinfo {author} {\bibfnamefont {S.}~\bibnamefont {Lal}},\ }\href
  {\doibase 10.1088/1367-2630/ab05ff} {\bibfield  {journal} {\bibinfo
  {journal} {New Journal of Physics}\ }\textbf {\bibinfo {volume} {21}},\
  \bibinfo {pages} {023019} (\bibinfo {year} {2019})}\BibitemShut {NoStop}%
\bibitem [{\citenamefont {Patra}\ and\ \citenamefont
  {Lal}(2021)}]{siddharthacpi}%
  \BibitemOpen
  \bibfield  {author} {\bibinfo {author} {\bibfnamefont {S.}~\bibnamefont
  {Patra}}\ and\ \bibinfo {author} {\bibfnamefont {S.}~\bibnamefont {Lal}},\
  }\href {\doibase 10.1103/PhysRevB.104.144514} {\bibfield  {journal} {\bibinfo
   {journal} {Phys. Rev. B}\ }\textbf {\bibinfo {volume} {104}},\ \bibinfo
  {pages} {144514} (\bibinfo {year} {2021})}\BibitemShut {NoStop}%
\bibitem [{\citenamefont {Anirban~Mukherjee}\ and\ \citenamefont
  {Lal}(2021)}]{1dhubjhep}%
  \BibitemOpen
  \bibfield  {author} {\bibinfo {author} {\bibfnamefont {S.~P.}\ \bibnamefont
  {Anirban~Mukherjee}}\ and\ \bibinfo {author} {\bibfnamefont {S.}~\bibnamefont
  {Lal}},\ }\href {\doibase 10.1007/JHEP04(2021)148} {\bibfield  {journal}
  {\bibinfo  {journal} {Journal of High Energy Physics}\ }\textbf {\bibinfo
  {volume} {2021}} (\bibinfo {year} {2021}),\
  10.1007/JHEP04(2021)148}\BibitemShut {NoStop}%
\bibitem [{\citenamefont {Yoo}\ \emph {et~al.}(2018)\citenamefont {Yoo},
  \citenamefont {Lee},\ and\ \citenamefont {Sim}}]{yoo_gwangshu_2018}%
  \BibitemOpen
  \bibfield  {author} {\bibinfo {author} {\bibfnamefont {G.}~\bibnamefont
  {Yoo}}, \bibinfo {author} {\bibfnamefont {S.-S.~B.}\ \bibnamefont {Lee}}, \
  and\ \bibinfo {author} {\bibfnamefont {H.-S.}\ \bibnamefont {Sim}},\ }\href
  {\doibase 10.1103/PhysRevLett.120.146801} {\bibfield  {journal} {\bibinfo
  {journal} {Phys. Rev. Lett.}\ }\textbf {\bibinfo {volume} {120}},\ \bibinfo
  {pages} {146801} (\bibinfo {year} {2018})}\BibitemShut {NoStop}%
\bibitem [{\citenamefont {Mukherjee}\ and\ \citenamefont
  {Lal}(2020{\natexlab{e}})}]{mukherjee2020}%
  \BibitemOpen
  \bibfield  {author} {\bibinfo {author} {\bibfnamefont {A.}~\bibnamefont
  {Mukherjee}}\ and\ \bibinfo {author} {\bibfnamefont {S.}~\bibnamefont
  {Lal}},\ }\href@noop {} {\bibfield  {journal} {\bibinfo  {journal} {arXiv
  preprint arXiv:2003.06118}\ } (\bibinfo {year}
  {2020}{\natexlab{e}})}\BibitemShut {NoStop}%
\bibitem [{\citenamefont {E.~S.~Sorensen}\ and\ \citenamefont
  {Affleck}(2007)}]{sorensen2007}%
  \BibitemOpen
  \bibfield  {author} {\bibinfo {author} {\bibfnamefont {N.~L.}\ \bibnamefont
  {E.~S.~Sorensen}, \bibfnamefont {M.-S.~Chang}}\ and\ \bibinfo {author}
  {\bibfnamefont {I.}~\bibnamefont {Affleck}},\ }\href@noop {} {\bibfield
  {journal} {\bibinfo  {journal} {Journal of Statistical Mechanics: Theory and
  Experiment}\ }\textbf {\bibinfo {volume} {2007}},\ \bibinfo {pages} {P08003}
  (\bibinfo {year} {2007})}\BibitemShut {NoStop}%
\bibitem [{\citenamefont {Eriksson}\ and\ \citenamefont
  {Johannesson}(2011)}]{eriksson_2011}%
  \BibitemOpen
  \bibfield  {author} {\bibinfo {author} {\bibfnamefont {E.}~\bibnamefont
  {Eriksson}}\ and\ \bibinfo {author} {\bibfnamefont {H.}~\bibnamefont
  {Johannesson}},\ }\href {\doibase 10.1103/PhysRevB.84.041107} {\bibfield
  {journal} {\bibinfo  {journal} {Phys. Rev. B}\ }\textbf {\bibinfo {volume}
  {84}},\ \bibinfo {pages} {041107} (\bibinfo {year} {2011})}\BibitemShut
  {NoStop}%
\bibitem [{\citenamefont {Lee}\ \emph {et~al.}(2015)\citenamefont {Lee},
  \citenamefont {Park},\ and\ \citenamefont {Sim}}]{Lee_park_2015}%
  \BibitemOpen
  \bibfield  {author} {\bibinfo {author} {\bibfnamefont {S.-S.~B.}\
  \bibnamefont {Lee}}, \bibinfo {author} {\bibfnamefont {J.}~\bibnamefont
  {Park}}, \ and\ \bibinfo {author} {\bibfnamefont {H.-S.}\ \bibnamefont
  {Sim}},\ }\href {\doibase 10.1103/PhysRevLett.114.057203} {\bibfield
  {journal} {\bibinfo  {journal} {Phys. Rev. Lett.}\ }\textbf {\bibinfo
  {volume} {114}},\ \bibinfo {pages} {057203} (\bibinfo {year}
  {2015})}\BibitemShut {NoStop}%
\bibitem [{\citenamefont {Sorensen}\ and\ \citenamefont
  {Affleck}(1996)}]{sorensen1996}%
  \BibitemOpen
  \bibfield  {author} {\bibinfo {author} {\bibfnamefont {E.~S.}\ \bibnamefont
  {Sorensen}}\ and\ \bibinfo {author} {\bibfnamefont {I.}~\bibnamefont
  {Affleck}},\ }\href@noop {} {\bibfield  {journal} {\bibinfo  {journal} {Phys.
  Rev. B}\ }\textbf {\bibinfo {volume} {53}},\ \bibinfo {pages} {9153}
  (\bibinfo {year} {1996})}\BibitemShut {NoStop}%
\bibitem [{\citenamefont {Barzykin}\ and\ \citenamefont
  {Affleck}(1998)}]{barzykin1998}%
  \BibitemOpen
  \bibfield  {author} {\bibinfo {author} {\bibfnamefont {V.}~\bibnamefont
  {Barzykin}}\ and\ \bibinfo {author} {\bibfnamefont {I.}~\bibnamefont
  {Affleck}},\ }\href@noop {} {\bibfield  {journal} {\bibinfo  {journal} {Phys.
  Rev. B}\ }\textbf {\bibinfo {volume} {57}},\ \bibinfo {pages} {432} (\bibinfo
  {year} {1998})}\BibitemShut {NoStop}%
\bibitem [{\citenamefont {Barzykin}\ and\ \citenamefont
  {Affleck}(1999)}]{barzykin1999}%
  \BibitemOpen
  \bibfield  {author} {\bibinfo {author} {\bibfnamefont {V.}~\bibnamefont
  {Barzykin}}\ and\ \bibinfo {author} {\bibfnamefont {I.}~\bibnamefont
  {Affleck}},\ }\href@noop {} {\bibfield  {journal} {\bibinfo  {journal} {J.
  Phys. A: Math. Gen.}\ }\textbf {\bibinfo {volume} {32}},\ \bibinfo {pages}
  {867} (\bibinfo {year} {1999})}\BibitemShut {NoStop}%
\bibitem [{\citenamefont {Yang}\ and\ \citenamefont
  {Feiguin}(2017)}]{yangfeigun2017}%
  \BibitemOpen
  \bibfield  {author} {\bibinfo {author} {\bibfnamefont {C.}~\bibnamefont
  {Yang}}\ and\ \bibinfo {author} {\bibfnamefont {A.~E.}\ \bibnamefont
  {Feiguin}},\ }\href@noop {} {\bibfield  {journal} {\bibinfo  {journal} {Phys.
  Rev. B}\ }\textbf {\bibinfo {volume} {95}},\ \bibinfo {pages} {115106}
  (\bibinfo {year} {2017})}\BibitemShut {NoStop}%
\bibitem [{\citenamefont {Vidal}\ and\ \citenamefont
  {Werner}(2002)}]{Vidal_2002}%
  \BibitemOpen
  \bibfield  {author} {\bibinfo {author} {\bibfnamefont {G.}~\bibnamefont
  {Vidal}}\ and\ \bibinfo {author} {\bibfnamefont {R.~F.}\ \bibnamefont
  {Werner}},\ }\href {\doibase 10.1103/PhysRevA.65.032314} {\bibfield
  {journal} {\bibinfo  {journal} {Phys. Rev. A}\ }\textbf {\bibinfo {volume}
  {65}},\ \bibinfo {pages} {032314} (\bibinfo {year} {2002})}\BibitemShut
  {NoStop}%
\bibitem [{\citenamefont {Bayat}\ \emph {et~al.}(2010)\citenamefont {Bayat},
  \citenamefont {Sodano},\ and\ \citenamefont {Bose}}]{bayat_2010}%
  \BibitemOpen
  \bibfield  {author} {\bibinfo {author} {\bibfnamefont {A.}~\bibnamefont
  {Bayat}}, \bibinfo {author} {\bibfnamefont {P.}~\bibnamefont {Sodano}}, \
  and\ \bibinfo {author} {\bibfnamefont {S.}~\bibnamefont {Bose}},\ }\href
  {\doibase 10.1103/PhysRevB.81.064429} {\bibfield  {journal} {\bibinfo
  {journal} {Phys. Rev. B}\ }\textbf {\bibinfo {volume} {81}},\ \bibinfo
  {pages} {064429} (\bibinfo {year} {2010})}\BibitemShut {NoStop}%
\bibitem [{\citenamefont {Andrei}\ and\ \citenamefont
  {Rajan}(1982)}]{andrei_rajan_1982}%
  \BibitemOpen
  \bibfield  {author} {\bibinfo {author} {\bibfnamefont {N.}~\bibnamefont
  {Andrei}}\ and\ \bibinfo {author} {\bibfnamefont {V.~T.}\ \bibnamefont
  {Rajan}},\ }\href {\doibase 10.1063/1.330233} {\bibfield  {journal} {\bibinfo
   {journal} {Journal of Applied Physics}\ }\textbf {\bibinfo {volume} {53}},\
  \bibinfo {pages} {7933} (\bibinfo {year} {1982})},\ \Eprint
  {http://arxiv.org/abs/https://doi.org/10.1063/1.330233}
  {https://doi.org/10.1063/1.330233} \BibitemShut {NoStop}%
\bibitem [{\citenamefont
  {Wilson}(1975{\natexlab{b}})}]{wilson1975renormalization}%
  \BibitemOpen
  \bibfield  {author} {\bibinfo {author} {\bibfnamefont {K.~G.}\ \bibnamefont
  {Wilson}},\ }\href@noop {} {\bibfield  {journal} {\bibinfo  {journal}
  {Reviews of Modern Physics}\ }\textbf {\bibinfo {volume} {47}},\ \bibinfo
  {pages} {773} (\bibinfo {year} {1975}{\natexlab{b}})}\BibitemShut {NoStop}%
\bibitem [{\citenamefont {Oliveira}\ and\ \citenamefont
  {Wilkins}(1981)}]{Wilkins_oliveira_1981}%
  \BibitemOpen
  \bibfield  {author} {\bibinfo {author} {\bibfnamefont {L.~N.}\ \bibnamefont
  {Oliveira}}\ and\ \bibinfo {author} {\bibfnamefont {J.~W.}\ \bibnamefont
  {Wilkins}},\ }\href {\doibase 10.1103/PhysRevLett.47.1553} {\bibfield
  {journal} {\bibinfo  {journal} {Phys. Rev. Lett.}\ }\textbf {\bibinfo
  {volume} {47}},\ \bibinfo {pages} {1553} (\bibinfo {year}
  {1981})}\BibitemShut {NoStop}%
\bibitem [{\citenamefont {Boyce}\ and\ \citenamefont
  {Slichter}(1974)}]{boyce_slichter_1974}%
  \BibitemOpen
  \bibfield  {author} {\bibinfo {author} {\bibfnamefont {J.~B.}\ \bibnamefont
  {Boyce}}\ and\ \bibinfo {author} {\bibfnamefont {C.~P.}\ \bibnamefont
  {Slichter}},\ }\href {\doibase 10.1103/PhysRevLett.32.61} {\bibfield
  {journal} {\bibinfo  {journal} {Phys. Rev. Lett.}\ }\textbf {\bibinfo
  {volume} {32}},\ \bibinfo {pages} {61} (\bibinfo {year} {1974})}\BibitemShut
  {NoStop}%
\bibitem [{\citenamefont {Nyg{\aa}rd}\ \emph {et~al.}(2000)\citenamefont
  {Nyg{\aa}rd}, \citenamefont {Cobden},\ and\ \citenamefont
  {Lindelof}}]{Nygrd_cobden_2000}%
  \BibitemOpen
  \bibfield  {author} {\bibinfo {author} {\bibfnamefont {J.}~\bibnamefont
  {Nyg{\aa}rd}}, \bibinfo {author} {\bibfnamefont {D.~H.}\ \bibnamefont
  {Cobden}}, \ and\ \bibinfo {author} {\bibfnamefont {P.~E.}\ \bibnamefont
  {Lindelof}},\ }\href {\doibase 10.1038/35042545} {\bibfield  {journal}
  {\bibinfo  {journal} {Nature}\ }\textbf {\bibinfo {volume} {408}},\ \bibinfo
  {pages} {342} (\bibinfo {year} {2000})}\BibitemShut {NoStop}%
\bibitem [{\citenamefont {Babi\ifmmode~\acute{c}\else \'{c}\fi{}}\ \emph
  {et~al.}(2004)\citenamefont {Babi\ifmmode~\acute{c}\else \'{c}\fi{}},
  \citenamefont {Kontos},\ and\ \citenamefont
  {Sch\"onenberger}}]{babi_kantos_2004}%
  \BibitemOpen
  \bibfield  {author} {\bibinfo {author} {\bibfnamefont {B.}~\bibnamefont
  {Babi\ifmmode~\acute{c}\else \'{c}\fi{}}}, \bibinfo {author} {\bibfnamefont
  {T.}~\bibnamefont {Kontos}}, \ and\ \bibinfo {author} {\bibfnamefont
  {C.}~\bibnamefont {Sch\"onenberger}},\ }\href {\doibase
  10.1103/PhysRevB.70.235419} {\bibfield  {journal} {\bibinfo  {journal} {Phys.
  Rev. B}\ }\textbf {\bibinfo {volume} {70}},\ \bibinfo {pages} {235419}
  (\bibinfo {year} {2004})}\BibitemShut {NoStop}%
\bibitem [{\citenamefont {Chorley}\ \emph {et~al.}(2012)\citenamefont
  {Chorley}, \citenamefont {Galpin}, \citenamefont {Jayatilaka}, \citenamefont
  {Smith}, \citenamefont {Logan},\ and\ \citenamefont
  {Buitelaar}}]{chorley_galpin_2012}%
  \BibitemOpen
  \bibfield  {author} {\bibinfo {author} {\bibfnamefont {S.~J.}\ \bibnamefont
  {Chorley}}, \bibinfo {author} {\bibfnamefont {M.~R.}\ \bibnamefont {Galpin}},
  \bibinfo {author} {\bibfnamefont {F.~W.}\ \bibnamefont {Jayatilaka}},
  \bibinfo {author} {\bibfnamefont {C.~G.}\ \bibnamefont {Smith}}, \bibinfo
  {author} {\bibfnamefont {D.~E.}\ \bibnamefont {Logan}}, \ and\ \bibinfo
  {author} {\bibfnamefont {M.~R.}\ \bibnamefont {Buitelaar}},\ }\href {\doibase
  10.1103/PhysRevLett.109.156804} {\bibfield  {journal} {\bibinfo  {journal}
  {Phys. Rev. Lett.}\ }\textbf {\bibinfo {volume} {109}},\ \bibinfo {pages}
  {156804} (\bibinfo {year} {2012})}\BibitemShut {NoStop}%
\bibitem [{\citenamefont {Manoharan}\ \emph {et~al.}(2000)\citenamefont
  {Manoharan}, \citenamefont {Lutz},\ and\ \citenamefont
  {Eigler}}]{Manoharan2000}%
  \BibitemOpen
  \bibfield  {author} {\bibinfo {author} {\bibfnamefont {H.~C.}\ \bibnamefont
  {Manoharan}}, \bibinfo {author} {\bibfnamefont {C.~P.}\ \bibnamefont {Lutz}},
  \ and\ \bibinfo {author} {\bibfnamefont {D.~M.}\ \bibnamefont {Eigler}},\
  }\href {\doibase 10.1038/35000508} {\bibfield  {journal} {\bibinfo  {journal}
  {Nature}\ }\textbf {\bibinfo {volume} {403}},\ \bibinfo {pages} {512}
  (\bibinfo {year} {2000})}\BibitemShut {NoStop}%
\bibitem [{\citenamefont {Fiete}\ and\ \citenamefont
  {Heller}(2003)}]{gregory_fiete_2003}%
  \BibitemOpen
  \bibfield  {author} {\bibinfo {author} {\bibfnamefont {G.~A.}\ \bibnamefont
  {Fiete}}\ and\ \bibinfo {author} {\bibfnamefont {E.~J.}\ \bibnamefont
  {Heller}},\ }\href {\doibase 10.1103/RevModPhys.75.933} {\bibfield  {journal}
  {\bibinfo  {journal} {Rev. Mod. Phys.}\ }\textbf {\bibinfo {volume} {75}},\
  \bibinfo {pages} {933} (\bibinfo {year} {2003})}\BibitemShut {NoStop}%
\bibitem [{\citenamefont {Rossi}\ and\ \citenamefont
  {Morr}(2006)}]{rossi_enrico_2006}%
  \BibitemOpen
  \bibfield  {author} {\bibinfo {author} {\bibfnamefont {E.}~\bibnamefont
  {Rossi}}\ and\ \bibinfo {author} {\bibfnamefont {D.~K.}\ \bibnamefont
  {Morr}},\ }\href {\doibase 10.1103/PhysRevLett.97.236602} {\bibfield
  {journal} {\bibinfo  {journal} {Phys. Rev. Lett.}\ }\textbf {\bibinfo
  {volume} {97}},\ \bibinfo {pages} {236602} (\bibinfo {year}
  {2006})}\BibitemShut {NoStop}%
\bibitem [{\citenamefont {Pr\"{u}ser}\ \emph {et~al.}(2011)\citenamefont
  {Pr\"{u}ser}, \citenamefont {Wenderoth}, \citenamefont {Dargel},
  \citenamefont {Weismann}, \citenamefont {Peters}, \citenamefont {Pruschke},\
  and\ \citenamefont {Ulbrich}}]{Prser2011}%
  \BibitemOpen
  \bibfield  {author} {\bibinfo {author} {\bibfnamefont {H.}~\bibnamefont
  {Pr\"{u}ser}}, \bibinfo {author} {\bibfnamefont {M.}~\bibnamefont
  {Wenderoth}}, \bibinfo {author} {\bibfnamefont {P.~E.}\ \bibnamefont
  {Dargel}}, \bibinfo {author} {\bibfnamefont {A.}~\bibnamefont {Weismann}},
  \bibinfo {author} {\bibfnamefont {R.}~\bibnamefont {Peters}}, \bibinfo
  {author} {\bibfnamefont {T.}~\bibnamefont {Pruschke}}, \ and\ \bibinfo
  {author} {\bibfnamefont {R.~G.}\ \bibnamefont {Ulbrich}},\ }\href {\doibase
  10.1038/nphys1876} {\bibfield  {journal} {\bibinfo  {journal} {Nature
  Physics}\ }\textbf {\bibinfo {volume} {7}},\ \bibinfo {pages} {203} (\bibinfo
  {year} {2011})}\BibitemShut {NoStop}%
\bibitem [{\citenamefont {Mitchell}\ \emph {et~al.}(2011)\citenamefont
  {Mitchell}, \citenamefont {Becker},\ and\ \citenamefont
  {Bulla}}]{mitchell_bulla_2011}%
  \BibitemOpen
  \bibfield  {author} {\bibinfo {author} {\bibfnamefont {A.~K.}\ \bibnamefont
  {Mitchell}}, \bibinfo {author} {\bibfnamefont {M.}~\bibnamefont {Becker}}, \
  and\ \bibinfo {author} {\bibfnamefont {R.}~\bibnamefont {Bulla}},\ }\href
  {\doibase 10.1103/PhysRevB.84.115120} {\bibfield  {journal} {\bibinfo
  {journal} {Phys. Rev. B}\ }\textbf {\bibinfo {volume} {84}},\ \bibinfo
  {pages} {115120} (\bibinfo {year} {2011})}\BibitemShut {NoStop}%
\bibitem [{\citenamefont {Costi}(2000)}]{costi2000}%
  \BibitemOpen
  \bibfield  {author} {\bibinfo {author} {\bibfnamefont {T.~A.}\ \bibnamefont
  {Costi}},\ }\href {\doibase 10.1103/PhysRevLett.85.1504} {\bibfield
  {journal} {\bibinfo  {journal} {Phys. Rev. Lett.}\ }\textbf {\bibinfo
  {volume} {85}},\ \bibinfo {pages} {1504} (\bibinfo {year}
  {2000})}\BibitemShut {NoStop}%
\bibitem [{\citenamefont {Rosch}\ \emph {et~al.}(2003)\citenamefont {Rosch},
  \citenamefont {Costi}, \citenamefont {Paaske},\ and\ \citenamefont
  {W\"olfle}}]{rosch_costi2003}%
  \BibitemOpen
  \bibfield  {author} {\bibinfo {author} {\bibfnamefont {A.}~\bibnamefont
  {Rosch}}, \bibinfo {author} {\bibfnamefont {T.~A.}\ \bibnamefont {Costi}},
  \bibinfo {author} {\bibfnamefont {J.}~\bibnamefont {Paaske}}, \ and\ \bibinfo
  {author} {\bibfnamefont {P.}~\bibnamefont {W\"olfle}},\ }\href {\doibase
  10.1103/PhysRevB.68.014430} {\bibfield  {journal} {\bibinfo  {journal} {Phys.
  Rev. B}\ }\textbf {\bibinfo {volume} {68}},\ \bibinfo {pages} {014430}
  (\bibinfo {year} {2003})}\BibitemShut {NoStop}%
\bibitem [{\citenamefont {Groisman}\ \emph {et~al.}(2005)\citenamefont
  {Groisman}, \citenamefont {Popescu},\ and\ \citenamefont
  {Winter}}]{groisman2005}%
  \BibitemOpen
  \bibfield  {author} {\bibinfo {author} {\bibfnamefont {B.}~\bibnamefont
  {Groisman}}, \bibinfo {author} {\bibfnamefont {S.}~\bibnamefont {Popescu}}, \
  and\ \bibinfo {author} {\bibfnamefont {A.}~\bibnamefont {Winter}},\
  }\href@noop {} {\bibfield  {journal} {\bibinfo  {journal} {Phys. Rev. A}\
  }\textbf {\bibinfo {volume} {72}},\ \bibinfo {pages} {032317} (\bibinfo
  {year} {2005})}\BibitemShut {NoStop}%
\bibitem [{\citenamefont {Seki}\ and\ \citenamefont
  {Yunoki}(2017)}]{seki2017topological}%
  \BibitemOpen
  \bibfield  {author} {\bibinfo {author} {\bibfnamefont {K.}~\bibnamefont
  {Seki}}\ and\ \bibinfo {author} {\bibfnamefont {S.}~\bibnamefont {Yunoki}},\
  }\href@noop {} {\bibfield  {journal} {\bibinfo  {journal} {Physical Review
  B}\ }\textbf {\bibinfo {volume} {96}},\ \bibinfo {pages} {085124} (\bibinfo
  {year} {2017})}\BibitemShut {NoStop}%
\bibitem [{\citenamefont {Kim}\ \emph {et~al.}(2019)\citenamefont {Kim},
  \citenamefont {Chung}, \citenamefont {Park},\ and\ \citenamefont
  {Han}}]{kim2019}%
  \BibitemOpen
  \bibfield  {author} {\bibinfo {author} {\bibfnamefont {K.-S.}\ \bibnamefont
  {Kim}}, \bibinfo {author} {\bibfnamefont {S.~B.}\ \bibnamefont {Chung}},
  \bibinfo {author} {\bibfnamefont {C.}~\bibnamefont {Park}}, \ and\ \bibinfo
  {author} {\bibfnamefont {J.-H.}\ \bibnamefont {Han}},\ }\href {\doibase
  10.1103/PhysRevD.99.105012} {\bibfield  {journal} {\bibinfo  {journal} {Phys.
  Rev. D}\ }\textbf {\bibinfo {volume} {99}},\ \bibinfo {pages} {105012}
  (\bibinfo {year} {2019})}\BibitemShut {NoStop}%
\bibitem [{\citenamefont {Berezinskii}(1971)}]{berezinskii1971destruction}%
  \BibitemOpen
  \bibfield  {author} {\bibinfo {author} {\bibfnamefont {V.}~\bibnamefont
  {Berezinskii}},\ }\href@noop {} {\bibfield  {journal} {\bibinfo  {journal}
  {Soviet Journal of Experimental and Theoretical Physics}\ }\textbf {\bibinfo
  {volume} {32}},\ \bibinfo {pages} {493} (\bibinfo {year} {1971})}\BibitemShut
  {NoStop}%
\bibitem [{\citenamefont {Kosterlitz}\ and\ \citenamefont
  {Thouless}(1978)}]{kosterlitz1978two}%
  \BibitemOpen
  \bibfield  {author} {\bibinfo {author} {\bibfnamefont {J.}~\bibnamefont
  {Kosterlitz}}\ and\ \bibinfo {author} {\bibfnamefont {D.~J.}\ \bibnamefont
  {Thouless}},\ }\href@noop {} {\bibfield  {journal} {\bibinfo  {journal}
  {Progress in low temperature physics}\ }\textbf {\bibinfo {volume} {7}},\
  \bibinfo {pages} {371} (\bibinfo {year} {1978})}\BibitemShut {NoStop}%
\bibitem [{\citenamefont {Phillips}(2012)}]{phillips2012advanced}%
  \BibitemOpen
  \bibfield  {author} {\bibinfo {author} {\bibfnamefont {P.}~\bibnamefont
  {Phillips}},\ }\href@noop {} {\emph {\bibinfo {title} {Advanced solid state
  physics}}}\ (\bibinfo  {publisher} {Cambridge University Press},\ \bibinfo
  {year} {2012})\BibitemShut {NoStop}%
\bibitem [{\citenamefont {Coleman}(2015)}]{coleman2015}%
  \BibitemOpen
  \bibfield  {author} {\bibinfo {author} {\bibfnamefont {P.}~\bibnamefont
  {Coleman}},\ }\href@noop {} {\emph {\bibinfo {title} {Introduction to
  many-body physics}}}\ (\bibinfo  {publisher} {Cambridge University Press},\
  \bibinfo {year} {2015})\ \bibinfo {note} {chapter:18}\BibitemShut {NoStop}%
\bibitem [{\citenamefont {Hewson}(1993)}]{hewson1993}%
  \BibitemOpen
  \bibfield  {author} {\bibinfo {author} {\bibfnamefont {A.~C.}\ \bibnamefont
  {Hewson}},\ }\href@noop {} {\emph {\bibinfo {title} {The Kondo Problem to
  Heavy Fermions}}}\ (\bibinfo  {publisher} {Cambridge University Press},\
  \bibinfo {year} {1993})\BibitemShut {NoStop}%
\bibitem [{\citenamefont {Langreth}(1966)}]{langreth1966}%
  \BibitemOpen
  \bibfield  {author} {\bibinfo {author} {\bibfnamefont {D.~C.}\ \bibnamefont
  {Langreth}},\ }\href {\doibase 10.1103/PhysRev.150.516} {\bibfield  {journal}
  {\bibinfo  {journal} {Phys. Rev.}\ }\textbf {\bibinfo {volume} {150}},\
  \bibinfo {pages} {516} (\bibinfo {year} {1966})}\BibitemShut {NoStop}%
\bibitem [{\citenamefont {Anderson}(1967)}]{anderson1967infrared}%
  \BibitemOpen
  \bibfield  {author} {\bibinfo {author} {\bibfnamefont {P.~W.}\ \bibnamefont
  {Anderson}},\ }\href@noop {} {\bibfield  {journal} {\bibinfo  {journal}
  {Physical Review Letters}\ }\textbf {\bibinfo {volume} {18}},\ \bibinfo
  {pages} {1049} (\bibinfo {year} {1967})}\BibitemShut {NoStop}%
\bibitem [{\citenamefont {Yamada}\ and\ \citenamefont
  {Yosida}(1978)}]{yamada_catastrophe}%
  \BibitemOpen
  \bibfield  {author} {\bibinfo {author} {\bibfnamefont {K.}~\bibnamefont
  {Yamada}}\ and\ \bibinfo {author} {\bibfnamefont {K.}~\bibnamefont
  {Yosida}},\ }\href {\doibase 10.1143/PTP.59.1061} {\bibfield  {journal}
  {\bibinfo  {journal} {Progress of Theoretical Physics}\ }\textbf {\bibinfo
  {volume} {59}},\ \bibinfo {pages} {1061} (\bibinfo {year} {1978})},\ \Eprint
  {http://arxiv.org/abs/https://academic.oup.com/ptp/article-pdf/59/4/1061/5315849/59-4-1061.pdf}
  {https://academic.oup.com/ptp/article-pdf/59/4/1061/5315849/59-4-1061.pdf}
  \BibitemShut {NoStop}%
\bibitem [{\citenamefont {Langer}\ and\ \citenamefont
  {Ambegaokar}(1961)}]{langer1961friedel}%
  \BibitemOpen
  \bibfield  {author} {\bibinfo {author} {\bibfnamefont {J.}~\bibnamefont
  {Langer}}\ and\ \bibinfo {author} {\bibfnamefont {V.}~\bibnamefont
  {Ambegaokar}},\ }\href@noop {} {\bibfield  {journal} {\bibinfo  {journal}
  {Physical Review}\ }\textbf {\bibinfo {volume} {121}},\ \bibinfo {pages}
  {1090} (\bibinfo {year} {1961})}\BibitemShut {NoStop}%
\bibitem [{\citenamefont {Martin}(1982)}]{martin-PhysRevLett.48.362}%
  \BibitemOpen
  \bibfield  {author} {\bibinfo {author} {\bibfnamefont {R.~M.}\ \bibnamefont
  {Martin}},\ }\href@noop {} {\bibfield  {journal} {\bibinfo  {journal}
  {Physical Review Letters}\ }\textbf {\bibinfo {volume} {48}},\ \bibinfo
  {pages} {362} (\bibinfo {year} {1982})}\BibitemShut {NoStop}%
\bibitem [{\citenamefont {Fujimoto}\ and\ \citenamefont
  {Kawakami}(1997)}]{fujimoto_satoshi_1997}%
  \BibitemOpen
  \bibfield  {author} {\bibinfo {author} {\bibfnamefont {S.}~\bibnamefont
  {Fujimoto}}\ and\ \bibinfo {author} {\bibfnamefont {N.}~\bibnamefont
  {Kawakami}},\ }\href {\doibase 10.1143/JPSJ.66.2157} {\bibfield  {journal}
  {\bibinfo  {journal} {Journal of the Physical Society of Japan}\ }\textbf
  {\bibinfo {volume} {66}},\ \bibinfo {pages} {2157} (\bibinfo {year}
  {1997})},\ \Eprint
  {http://arxiv.org/abs/https://doi.org/10.1143/JPSJ.66.2157}
  {https://doi.org/10.1143/JPSJ.66.2157} \BibitemShut {NoStop}%
\bibitem [{\citenamefont {Shiina}\ and\ \citenamefont
  {Ishii}(1993)}]{Shiina1993}%
  \BibitemOpen
  \bibfield  {author} {\bibinfo {author} {\bibfnamefont {R.}~\bibnamefont
  {Shiina}}\ and\ \bibinfo {author} {\bibfnamefont {C.}~\bibnamefont {Ishii}},\
  }\href {\doibase 10.1007/BF00132091} {\bibfield  {journal} {\bibinfo
  {journal} {Journal of Low Temperature Physics}\ }\textbf {\bibinfo {volume}
  {91}},\ \bibinfo {pages} {105} (\bibinfo {year} {1993})}\BibitemShut
  {NoStop}%
\bibitem [{\citenamefont {Lacroix}\ and\ \citenamefont
  {Cyrot}(1979)}]{lacroix_cyrot_1979}%
  \BibitemOpen
  \bibfield  {author} {\bibinfo {author} {\bibfnamefont {C.}~\bibnamefont
  {Lacroix}}\ and\ \bibinfo {author} {\bibfnamefont {M.}~\bibnamefont
  {Cyrot}},\ }\href {\doibase 10.1103/PhysRevB.20.1969} {\bibfield  {journal}
  {\bibinfo  {journal} {Phys. Rev. B}\ }\textbf {\bibinfo {volume} {20}},\
  \bibinfo {pages} {1969} (\bibinfo {year} {1979})}\BibitemShut {NoStop}%
\bibitem [{\citenamefont {Burdin}\ \emph {et~al.}(2002)\citenamefont {Burdin},
  \citenamefont {Grempel},\ and\ \citenamefont {Georges}}]{budin_grempel_2002}%
  \BibitemOpen
  \bibfield  {author} {\bibinfo {author} {\bibfnamefont {S.}~\bibnamefont
  {Burdin}}, \bibinfo {author} {\bibfnamefont {D.~R.}\ \bibnamefont {Grempel}},
  \ and\ \bibinfo {author} {\bibfnamefont {A.}~\bibnamefont {Georges}},\ }\href
  {\doibase 10.1103/PhysRevB.66.045111} {\bibfield  {journal} {\bibinfo
  {journal} {Phys. Rev. B}\ }\textbf {\bibinfo {volume} {66}},\ \bibinfo
  {pages} {045111} (\bibinfo {year} {2002})}\BibitemShut {NoStop}%
\bibitem [{\citenamefont {Rech}\ \emph {et~al.}(2006)\citenamefont {Rech},
  \citenamefont {Coleman}, \citenamefont {Zarand},\ and\ \citenamefont
  {Parcollet}}]{rech_coleman_parcollet_2006_prl}%
  \BibitemOpen
  \bibfield  {author} {\bibinfo {author} {\bibfnamefont {J.}~\bibnamefont
  {Rech}}, \bibinfo {author} {\bibfnamefont {P.}~\bibnamefont {Coleman}},
  \bibinfo {author} {\bibfnamefont {G.}~\bibnamefont {Zarand}}, \ and\ \bibinfo
  {author} {\bibfnamefont {O.}~\bibnamefont {Parcollet}},\ }\href {\doibase
  10.1103/PhysRevLett.96.016601} {\bibfield  {journal} {\bibinfo  {journal}
  {Phys. Rev. Lett.}\ }\textbf {\bibinfo {volume} {96}},\ \bibinfo {pages}
  {016601} (\bibinfo {year} {2006})}\BibitemShut {NoStop}%
\end{thebibliography}%
\end{document}